\documentclass[prb,singlecolumn,a4paper,amsmath,amssymb]{revtex4}

\usepackage{geometry}
\usepackage{amsfonts}
\usepackage{latexsym}
\usepackage{graphicx}
\usepackage[usenames]{color}
\usepackage[utf8]{inputenc}
\usepackage[bulgarian]{babel}
\usepackage{hyperref}
\usepackage{caption}
\usepackage{morefloats}

\begin{document}

\title{Volt-Ampere characteristic of ``black box'' with a negative resistance}
\author{Stojan~G.~Manolev}
\email[E-mail: ]{manolest@yahoo.com}
\affiliation{Middle School ``Goce Delchev'',\\
Purvomaiska str. 3, MKD-2460 Valandovo}
\author{Vasil~G.~Yordanov}
\email[E-mail: ]{vasil.yordanov@gmail.com, peshista@gmail.com}
\affiliation{Faculty of Physics,\\
St.~Clement of Ohrid University at Sofia,\\
5 James Bourchier Blvd., BG-1164 Sofia, Bulgaria}
\author{Nikolay~N.~Tomchev}
\email[E-mail: ]{ntomchev@mail.bg}
\affiliation{ Foreign Language School  "Ekzarch Jossiff I",\\
7 Shishman Str., Lovech, Bulgaria}
\author{Todor~M.~Mishonov}
\email[E-mail: ]{mishonov@gmail.com}
\affiliation{Department of Theoretical Physics, Faculty of Physics,\\
St.~Clement of Ohrid University at Sofia,\\
5 James Bourchier Blvd., BG-1164 Sofia, Bulgaria}

\pacs{negative resistor, black box, student problem, Open Experimental Physics Olympiad}

\date{25 February, 2016}

\begin{abstract}
This problem was given at Third Experimental Physics Olympiad ``The day of the resistor'', 31 October 2015, Kumanovo, organized by  the Regional Society of Physicists of Strumica, Macedonia and the Sofia Branch of the Union of Physicists in Bulgaria.
\end{abstract}

\maketitle

\newcommand*\ground{\includegraphics[width=0.4cm]{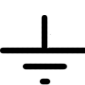}}
\captionsetup{labelfont={normalsize},textfont={small},justification=centerlast}
\maketitle

\section{Experimental problem}
\begin{figure}[h]
\includegraphics[width=8.8cm]{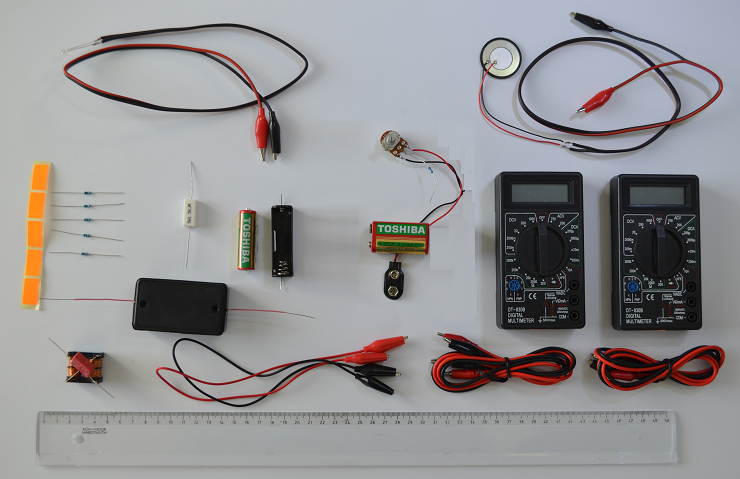}
\caption{Description of the experimental set. Two multimeters, one battery 1.5~V (AA), one white ceramic resistor, 5 other resistors, yellow 
sticker one plastic line 50~cm, one capacitor with piezo element, LED with long cables, 4~cables for the multimeters (black and red), potentiometer with 9~V battery connector, parallel connected capacitor and inductor (resonant circuit) and the most important ``black box''.}
\label{setup}
\end{figure}

With the experimental set~\ref{setup}, analyze the dependence between current and voltage of the ``black box'', that must not be opened during the Olympiad.

\subsubsection{Description of the problems}

Here follow two quality problems, described in section~\ref{quatity_tasks}, which aim to verify whether the device, hidden in the ``black box'', works.
Section~\ref{black_box_investigation} describes in detail the experiments that must be carried out with the set to analyse VAC of the ``black box''.
Then in section~\ref{the_world_is_not_perfect} will make a more detailed analysis and
you will see that the physical properties which are found up to now on this subject are valid within certain limits.
Then follows the static analysis section~\ref{dynamics}, which will explore the dynamic characteristics of the ``black box'' by mechanical experiments.
There is also the purely theoretical section~\ref{Theoretical Problem} connected to the theoretical description of the proposed experiments.
We suggest the students who are not very confident in the experiment,
but better cope with mathematics,
to concentrate on the theoretical problem.
For the tireless there is a homework assignment, described in section~\ref{Homework}.

\section{Two quality problems}

\label{quatity_tasks}
\subsection{Lighting the LED with the``black box''.}
In the set of Фиг.~\ref{setup}, there is one ``black box'' and one LED with long cables and two alligator clips.
Connect the LEDs to the ``black box'' and if  the LED does not light, swap the ``alligator clips''.
At least in one of the polarities the LED lights.
Upon further quality problems will have to measure current and voltage in LEDs and explain how these variables are related to the current-voltage characteristics of the elements.
\begin{figure}[h]
\includegraphics[width=3.3cm]{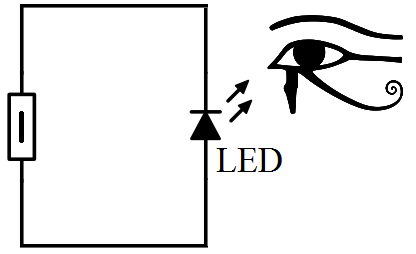}
\caption{In one of the two possible polarities of the LED lights up when connected to electrodes of the ``black box''. What are the properties of that box and what is inside it? This is the aim of the Olympiad.}
\label{LEDNR}
\end{figure}

\subsection{Generating electrical oscillations of the ``black box''}
\begin{figure}[h]
\includegraphics[width=8.8cm]{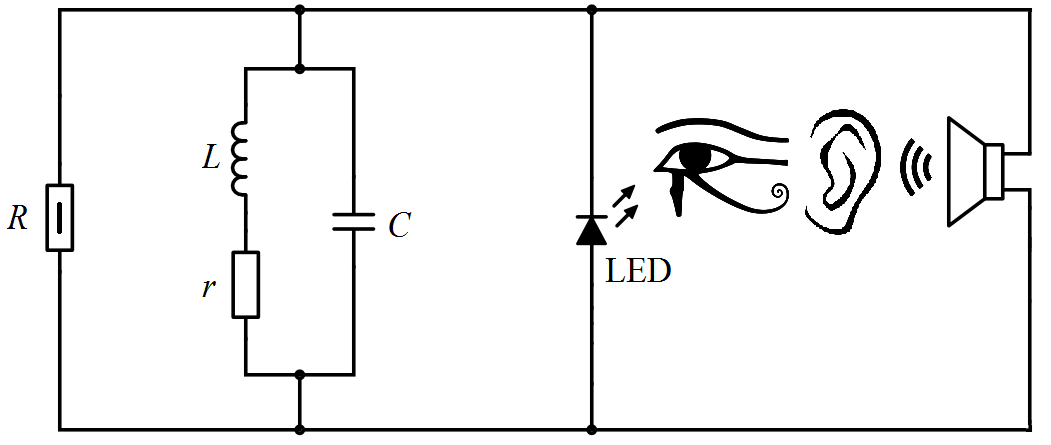}
\caption{Generating electrical oscillations in a resonant circuit using a ``black box''. If moving the LED the eye sees the light pulse
and the ear hears the hum of the piezo element, excited by AC voltage.
}
\label{LCNR}
\end{figure}

Parallel to the ``black box'' and the LED connect a resonant circuit.
Move the LED and you will see that the light pulsates now.
Moreover, if you include parallel and piezo element, as shown in Фиг.~\ref{LCNR}, you will hear a faint buzz.
Confirm whether the LED flashes and the piezo element buzzes. 
The following tasks are related to the detailed quantitative analysis of these oscillations and their theoretical explanation.

\begin{figure}[h]
\includegraphics[width=8.8cm]{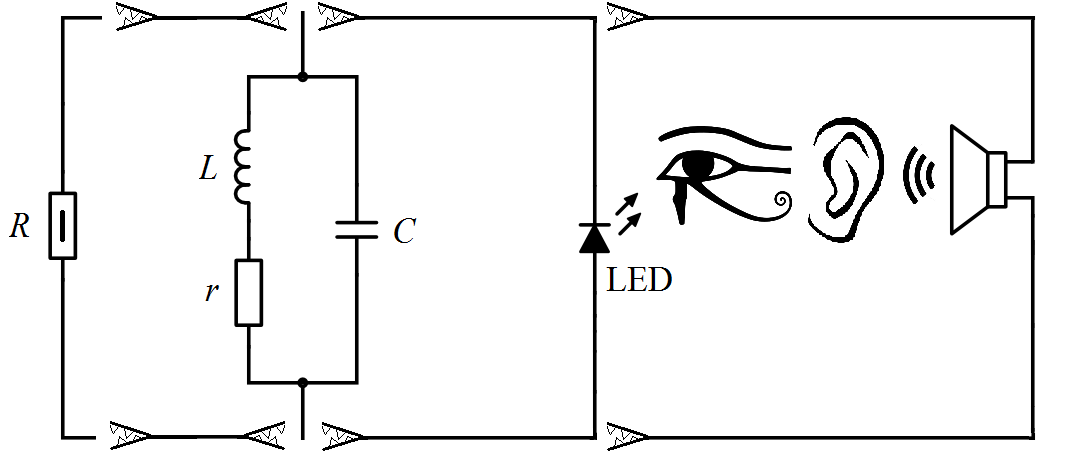}
\caption{Fig~\ref{LCNR} with alligator clips.}
\label{LCNR-with-crocodiles}
\end{figure}

In which of the two quality problems the LED is brighter --  constantly or pulsating emission?

\pagebreak
\section{Experimental problem, 100~points}
\subsection{Analysis of the static behavior of the ``black box''. Section for younger students.}
\label{black_box_investigation}

\textit{The subpoints of the section~\ref{black_box_investigation}, are addressed to younger students. The problems are easier and give less points.}

\begin{enumerate}
\item \textbf{Measure the voltage and the current through the ``black box'', related to the LED and without it. (7~points)}

Connect the circuit in Figure~\ref{Hypothesis_reject} and measure the voltage $U$ on the ``black box'' and the current $I$, which flows through it.
If the LED of the first or the last circuit does not light up, change its polarity.
The results of the measurements write down in the table, as shown in the exemplary table~\ref{template_4_setups}.

What causes the small difference between the voltages $U_a$ and $U^*$?
\begin{figure}[h]
\includegraphics[width=16.2cm]{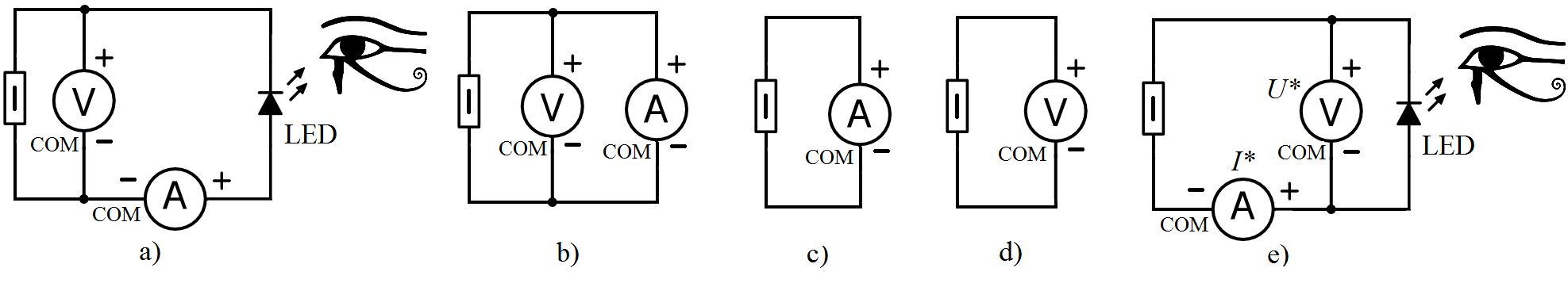}
\caption{Five circuits for analysis of the ``black box'': (a) To the circuit with LED of Figure~\ref{LEDNR} add an amperemeter and a voltmeter.
Their values are $U_a$ and $I_a$. (b) Replace the LED with wire and the values are $U_b$ and $I_b$. (c) We leave the the amperemeter and the value is $I_c$. (d) Replace the amperemeter and the voltmeter and the value is $U_d$. (e) Measure the voltage $U^*$ and the current $I^*$ according to the following scheme.
}
\label{Hypothesis_reject}
\end{figure}

\begin{table}[h]
\caption{Template for a table for entering experimental data for the experiment shown in Fig.~\ref{Hypothesis_reject}.}
\begin{tabular}{| c| c | c | }
\tableline
$\#$ & $I$ [$\mu A$] & $U$ [$V$ ]  \\
\tableline
a) & $I_a=\qquad \qquad$ &  $U_a=\qquad \qquad$   \\
b) & $I_b=\qquad \qquad$ &  $U_b=\qquad \qquad$   \\
c) &  $I_c=\qquad \qquad$ &    \\
d) &   &  $U_d=\qquad \qquad$   \\
e) & $I^*=\qquad \qquad$ &  $U^*=\qquad \qquad$   \\
\tableline
\end{tabular}
\label{template_4_setups}
\end{table}

\item \textbf{Measure the voltage of the battery $\mathcal{E}=$1.5 V. (1 point)}

Use the multimeter as a voltmeter and measure the voltage of the battery. 
The device displays the sign of the voltage.
You can remember the rules for characters and that the black lead is connected to the input of the device marked with ``ground'' (~\ground) or COM and red to one of the other inputs.

\item \textbf{Measure the resistance of the big white resistor. (1 point)}

Turn the multimeter as an ohmmeter and measure and note the resistance $R_\mathrm{WR}$ of the big white resistor with thermal lining of white cement.
Work with an accuracy of 1~$\Omega.$ $R_\mathrm{WR}$=?

\item  \textbf{Measure the resistance of the five small resistors. (3~points)}

On the connecting wires of the five resistors, that are given to you, stick yellow label.
Switch the multimeter to operate as an ohmmeter.
Write the numerical values of the resistances on their labels.
Arrange them in size and write numbers on the stickers.
Represent the resistances in a table
$r_1 < r_2 < r_3 < r_4<r_5$, work with an accuracy of 1~$\Omega.$

\item  \textbf{Using the five small resistors and battery of $\mathcal{E}=$1.5 V measure the relationship between current and voltage of the big white resistor with resistance $R_\mathrm{WR}$. (7~points)}

An electrical circuit to measure the relationship between current
and voltage is shown on Fig.~\ref{resistance-measurement} and Fig.~\ref{R_neg_R_with_crocodiles}.
Use one multimeter as an amperemeter and plug it consecutively to the white resistor.
Watch for the signs -- the current has direction!
The other multimeter turn as a voltmeter, parallel to the white resistor and again watch out for signs and polarity.
For the big white resistor acts the Ohm's law $U/I=R_\mathrm{WR}$. 
If the voltage is positive, the current is positive, otherwise the voltage is negative and the current is negative.
If the signs of voltage and current are opposite see where you went wrong in linking of the devices.
\begin{figure}[h]
\includegraphics[width=11.5cm]{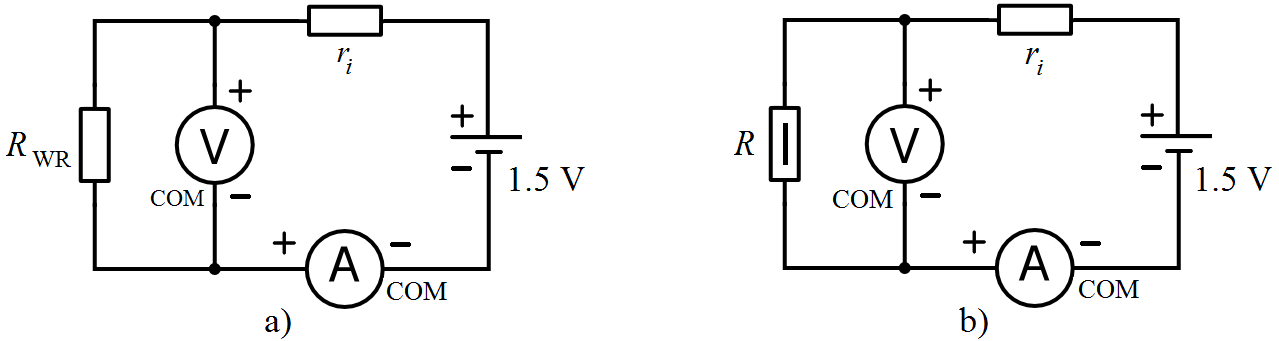}
\caption{The electric circuit is for analysis of a part of the volt-ampere characteristic (VAC) of: (a) the white resistor and (b) the ``black box''.
The voltmeter (V) is connected parallel to the measuring element, and measure the voltage $U$ and the ampermeter (A) is connected in series and measure the current $I$. 
When the circuit is closed with different resistances $r_i\in(0,\;600\,\Omega)$
current and voltage are different.
Thus are obtained in several points of VAC.
For small voltages ratio  $R=U/I$ is a constant and this is one possible way to verify Ohm's law
and signs of the measured current and voltage.
Resistor $R_\mathrm{WR}$ on the scheme on the left is replaced by the ``black box'' of the figure to the right and this is the only difference in the two circuits.
}
\label{resistance-measurement}
\end{figure}

\begin{figure}[h]
\includegraphics[width=15cm]{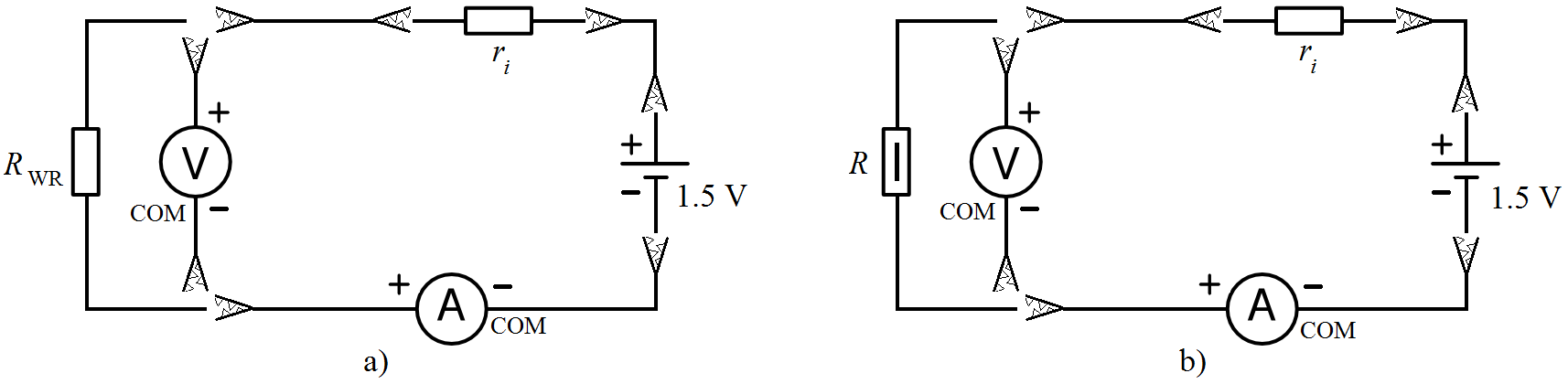}
\caption{Circuit diagram of the electric circuit of Fig.~\ref{resistance-measurement} with alligator clips.}
\label{R_neg_R_with_crocodiles}
\end{figure}

Ampermeter connect in series with the battery of 1.5~V. 
Close the circuit consistent with resistances
0~$\Omega$ separately the 5 $r_i.$ 
For each measurement note in a table 5 columns and 6 lines in the form given in Table ~\ref{template}:
1)~number of resistors $i$,
2)~resistance $r_i$,
3)~current $I_i$ and
4)~the voltage of the voltmeter $U_i$ 
5)~the calculated resistance of the measuring element $U_i/I_i$.

\begin{table}[ht]
\caption{Model of a table for processing of experimental data for the experiment shown in Fig.~\ref{resistance-measurement}.
Columns indicate:
1)~the number of the resistor $i$,
2)~resistances of resistors  $r_i$ and the short circuit $r_0=0\;\Omega$ is the zero line of the table,
3)~the current $I_i$, which flows through the circuit for various resistances $r_i$ series-connected with battery with voltage $\mathcal{E}$,
4)~voltage $U_i$ on the white resistor or the ``black box'',
5)~the ratio of the voltage $U_i$ and current $I_i$.
}
\begin{tabular}{| r | r | r | r | r | r | r |}
\tableline
 i& $r_i \, [\Omega]$ & $I_i \,[\mu \mathrm{A}]$ & $U_i\,[\mathrm{V}]$ & $U_i/I_i\,[\Omega]$ \\
\tableline
0 & 0 &  & & \\
1 &   &  &  &\\
2 &   &  &  &\\
3 &   &  &  &\\
4 &   &  &  &\\
5 &   &  & &\\
\tableline
\end{tabular}
\label{template}
\end{table}

\item  \textbf{Using the five small resistors and a battery of $\mathcal{E}=$1.5 V measure the relationship between the current and the voltage of the ``black box''. (7~points)}

For this task repeat the same measurements as in the previous subpoint, replacing the white resistor with the ``black box''.

\item  \textbf{Draw a volt-ampere characteristic (VAC) of the ``black box'', the white resistor and a common graphic, using data from the tables of the previous two subpoints. (7~points)}

Present the data from the tables graphically. 
Put on the ~\textit{x} axis -- voltage $U_i$ and on the ~\textit{y} axis -- current $I_i$.
Scale the graphic with appropriate axis ranges and beginning from $U=0$ and $I=0$.
This is a small part of the volt-ampere characteristic (VAC) of the ``black box'' and of the white resistor, 
and the dependence of $I(U)$ have only 6 points for each of the analyzed elements.
Through the points on the graph draw a straight line that passes closest to them.
For each element put a label on its corresponding straight line.

\item \textbf{Determine the slope $\Delta U / \Delta I$ of VAC drawn in the previous subpoint. (9~points)}

The symbol $\Delta$ means difference $\Delta U=U_2-U_1$ and $\Delta I=I_2-I_1$. Choose two points from the drawn straight lines.
What kind of characteristics of the analyzed elements connected slope  $\Delta U / \Delta I$?
Do you find anything unusual for this characteristic of the ``black box''?

\subsection{The world is not perfect. Detailed analysis of VAC. Section for older students.}
\label{the_world_is_not_perfect}

At low voltages, the VAC of the ``black box'' is a part of a straight line but how does the VAC look at higher voltage levels, you will discover shortly.

\begin{figure}[h]
\includegraphics[width=8cm]{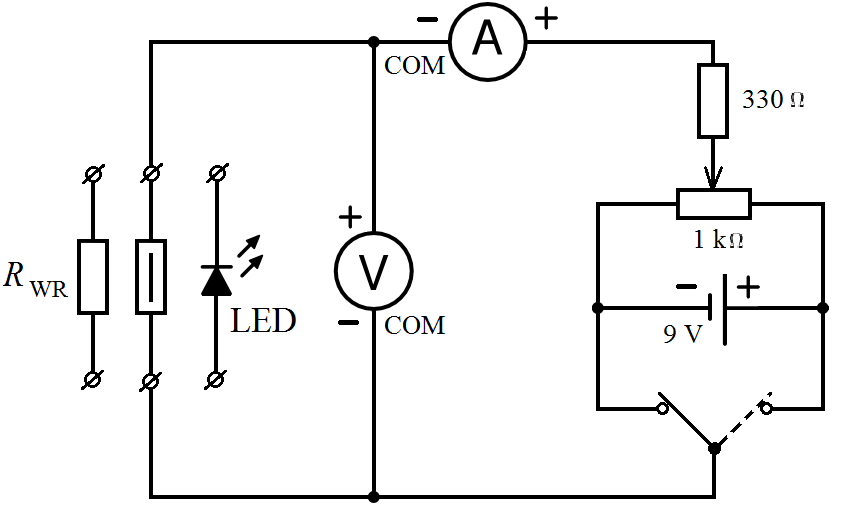}
\caption{Circuit for the analysis of VAC $I(U)$. Firstly, the white resistor $R_\mathrm{WR}$ is analysed, then it is replaced with the ``black box'' and finally the LED.
The voltmeter (V) measures the voltage $U$ and the amperemeter (A) measures the current $I$.
The voltage is created by the battery with a voltage $\mathcal{E}=9\,\mathrm{V}$. 
When we rotate with an arm the shaft of the potentiometer with resistance
$1\,\mathrm{k}\Omega$
the voltage $U$ is altered from 0 to $+\mathcal{E}$. The symbolically drawn switch can be implemented by switching of the alligator clip.
When you switch it from the one end terminal of the potentiometer to the other, the voltage $U$ is altered from $-\mathcal{E}$ to 0.
So we have voltage $-\mathcal{E}<U<+\mathcal{E}$. 
The resistance of $330\,\Omega$
limits the current scheme and protects the the LED from damage.
}
\label{resistance-measurement-I-V-curve}
\end{figure}

\item \textbf{Analyze in detail the dependence between current and voltage of the white resistor. (7~points)}
 
Plug a 9~V battery to the connector contacts connected to the potentiometer.
See the description of Fig.~\ref{resistance-measurement-I-V-curve} for explanation the scheme.
On amendment of the voltage use the potentiometer.|
First scroll quickly the potentiometer between its extreme positions and find out in what ranges of the multimeters it works. 
Then explore the dependence $I_\mathrm{WR}(U)$ the pairs of variables $(I_\mathrm{WR},U)$ note in a table.
For the white resistor the Ohm's law is in effect. To draw a linear dependence $I_\mathrm{WR}(U)$ five points are enough, two of which are at voltages $U_\mathrm{min}$ and $U_\mathrm{max}$.
Again, according to Ohm's law, the signs of voltage and current must be the same.
If the signs of voltage and current are opposite see where you went wrong.

\item \textbf{Analyze in detail the dependence between current and voltage of the ``black box''. (5~points)}

In the circuit of the previous subpoint replace the white resistor with the ``black box'' without making other changes.
Rotating the shaft of the potentiometer change the voltage $U$ between $U_\mathrm{min}$ and $U_\mathrm{max}$ in approximately 1 $V$ and note in a table pairs of variables $(I, U)$.
Note in the table also the voltages in which current has a maximum or minimum.

\item \textbf{Analyze in detail the dependence between current and voltage of the LED. (3~points)}

In the circuit of the previous subpoint replace the ``black box'' with the LED without making other changes.
See in what end position of the potentiometer the LED lights the brightest.
Now rotate the shaft of the potentiometer and watch the amperemeter values.
When the current is less than 6~mA  note pairs of numbers $(I_\mathrm{LED}, U)$ in 1~mA.
When current is less than 2~mA=2000~$\mu$A note pairs of numbers in 200~$\mu$A, 
until it reaches the current less than 200~$\mu$A.

\textit{When you have finished this third VAC disconnect the battery from 9~V contacts because it will become exhausted very quickly.}

\item \textbf{Draw on a common graphic VAC of: the white resistor $I_\mathrm{WR}(U)$, the ``black box''  $I(U)$ and the LED $I_\mathrm{LED}(U)$. (10~points)}

First analyze the smallest and the largest value of current and voltage. These parameters define the rectangle, which will be drawn VAC of three elements.
We recommend that scale on ~\textit{x} axis
$\mathrm{1\,V=1\,cm}$
and on ~\textit{y} axis $\mathrm{1\,mA=1\,cm}$.
And for the three elements $I(U)$ VACs are continuous curves. 
Draw guiding the eyes straight lines (or curves) through the experimental points.

\item \textbf{Why are VACs important? (5~points)}

Draw a mirror image of a VAC LED where $I$ is replaced by $-I$, and rotate VAC around the horizontal axis.
Оn VAC LED mark with a small circle the values of the table~\ref{template_4_setups}.
Take note that this point is near the intersection of the VAC ``black box'' and the VAC of the LED.
Is this accidental?

\item \textbf{Analysing the VAC: the white resistor, the ``black box'' and the LED. (7~points)}

If the VAC $ (U) $ consists of separate straight lines, determine the corresponding resistances $R=\Delta U/\Delta I$ by their slopes.
The symbol $\Delta$ means difference $\Delta U=U_2-U_1$ and $\Delta I=I_2-I_1$. Choose two points from the drawn straight lines.
When the considered in section curvature is negligible, instead of $\Delta$ write $\mathrm{d}.$ The resistance produced by the slope VAC is called differential resistance, and the reciprocal value differential conductivity $\sigma_\mathrm{diff}=\mathrm{d}I/\mathrm{d}U$.

Find: a) the resistance of the white resistor, b) the resistance of the central part of the ``black box'', 
c) the resistance of the left part of the VAC of the ``black box'', 
d) the resistance of the right part of the VAC of the ``black box'', 
and e) the resistance of the LED when it is on~(light) in the range of 2-5 mA.

\item \textbf{What characteristic of VAC of the ``black box'' is essential for permanent lighting or flashing of the LED in both quality problems. (4~points)}

\textit{The ``black box'' is the most important central part of VAC, which includes a zero voltage.
This part of VAC leads to constant light of the LED and the generation of alternating current in the resonance circuit that is seen in the quality problems shown in Fig.~\ref{LEDNR} and Fig.~\ref{LCNR} and the many possible applications of technical devices, hidden in the ``black box''.}

Any difference in VAC of the ``black box'' and the white resistor reveals the cause of light of the LED.
What is that difference?

\item \textbf{Describe qualitatively how to create a current that causes permanent lighting or flashing of the LED in both quality problems. (10~points)}

\subsection{Measurement of the frequency of electric oscillations generated by ``black box''}
\label{dynamics}
Let us connect the electric circuit from the second quantitative problem as it is shown on Figure~\ref{LCNR}, where all elements of the circuit are connected in parallel: the ``black box'', $LC$, piezo element and LED.
If one use a GSM earphone instead of piezo element then an additional resistor greater than 10~$\Omega$ has to be added in sequence.
\item \textbf{Mechanical measurement of the frequency of electric oscillations. (7~points)}

The electric measurements are easier and more accurate.
In order to measure the frequency of the oscillations without frequency meter in the given setup it is necessary to show a little dexterity.

With the stick tape stick the LED to one of the ends of the elastic ruler.
With one of your hands press the other end of the elastic ruler to the table, and with the other generate oscillation on the ruler.
You can observe static (not moving) spots created by the LED oscillation on the ruler.
If the frequency of the oscillation $f_\mathrm{res}$ is a multiple of the mechanical frequency $f_\mathrm{mech}$, then the light spots are immovable.
Count the number of the light spots $N$.
Keep disturbing the elastic ruler to make it oscillating with a constant frequency and amplitude.
Change the length of the free part of the ruler to achieve unmovable light spots.
Count the number of the oscillation of the ruler for 10 seconds and thus determine the mechanical frequency of the oscillations $f_\mathrm{mech}$.
Express the frequency of the electric oscillation $f_\mathrm{res}$ as a function of the number of the light spots $N$ and the frequency of the mechanical oscillations $f_\mathrm{res}(N,f_\mathrm{mech})$.

\item \textbf{Calculate the frequency of the oscillations if you know the values of $L$ and $C$. (2~points)}

An alternative method to determine the frequency of the oscillation is using the Thomson-formula $f_\mathrm{res}=1/(2\pi\sqrt{LC}).$ 
On the capacitor you can see its value 4.7~$\mu F.$ On the toroidal coils it is written that it is composed of 2 coils with inductance of 100~mH, which means that the inductance of the two sequentially connected inductors is $L$=400~mH.
How does the calculated value for the frequency agree with the experimental measured value?

\item \textbf{Determine by hearing the sound from the piezo element what is the frequency of its oscillations. (1~point)}

If you have an absolute music hearing and remember the frequency of the frequencies of musical tones you can determine, which musical tone corresponds to the frequency of the oscillation of the buzzer. 
What is the value of that frequency?
It is not expected to achieve accuracy and error of 50\% would be satisfactory good agreement of these 3 methods to determine the frequency without frequency meter.

\end{enumerate}%

\section{Theoretical problem, 35~points}
\label{Theoretical Problem}

\subsection{The problem}
\begin{figure}[h]
\includegraphics[width=9cm]{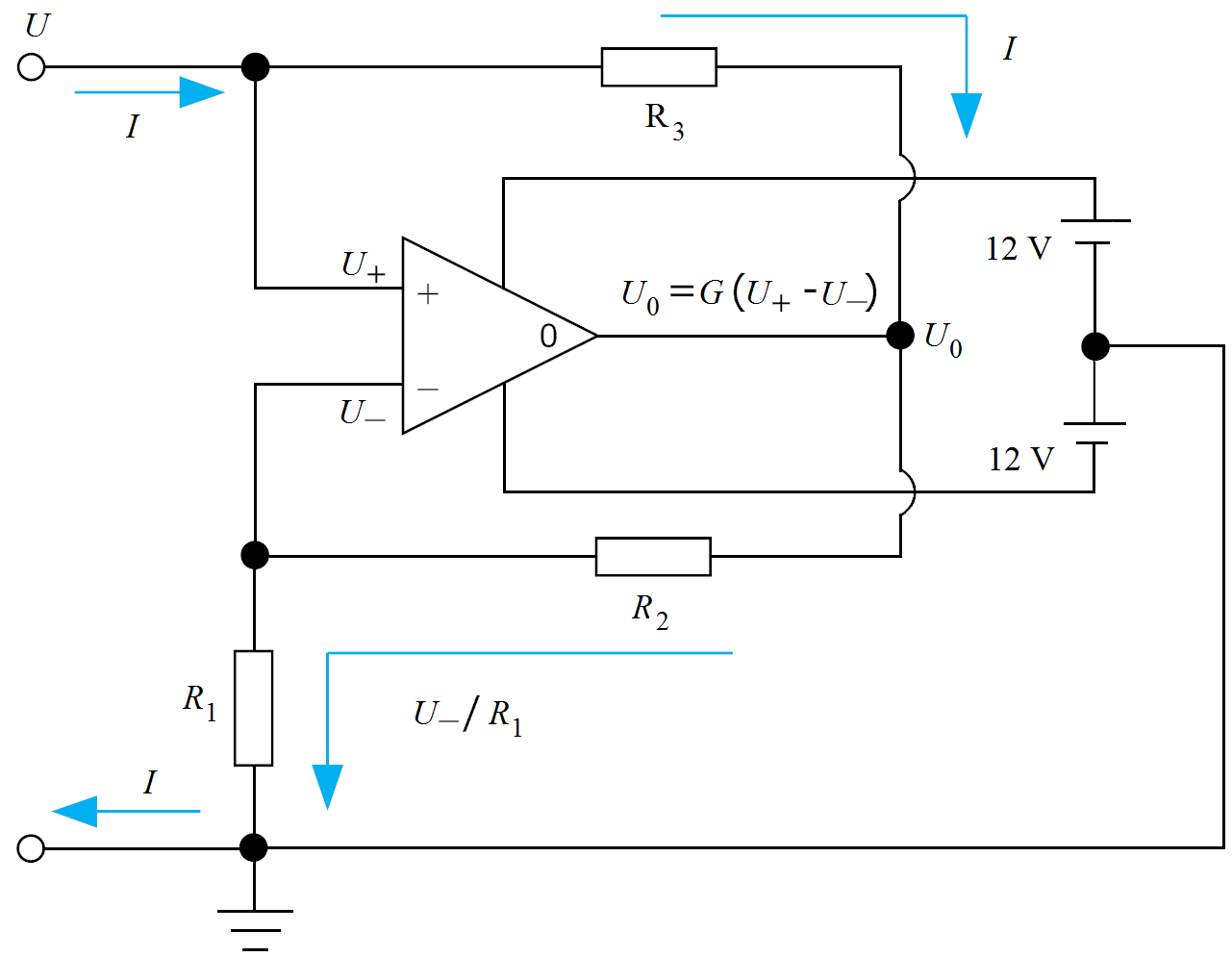}
\caption{Find within percent accuracy the effective resistance $R=U/I$ of the circuit with the three resistors $R_1,$ $R_2$ и $R_3$ and the triangle shown is a voltage amplifier with an amplification $G=10^5$ that is powered from two batteries with a voltage $V_S$. The voltages $U_0$, $U_{-}$, as well as the current $I$  are unknown. The notation $U_0$ originates from $U_\mathrm{output}=(U_+-U_-)G$.}
\label{circuit}
\end{figure}

On the electric circuit shown in Figure~\ref{circuit},
the input currents in the nodes (+) and (--) are zero, and the output currect in the node (0) is such that 
the corresponding voltages are related $U_0=(U_+ - U_-)G$,  where the gain coefficient $G=10^5 \gg 1$ has very big value.
The triangle sign notes an electric element (called amplifier) that is sourced with two batteries with electromotive force $\mathcal{E}_\mathrm{B}=12\;\mathrm{V}.$ 
The node between the two batteries is connected with wire to the resistor $R_1$ and one of the inputs of the schematics.
It is very convinient to choose the potential of this node to be zero $U_\mathrm{CP}=0$.
The engineers call this node ``ground'' (\ground).
The index $\mathrm{CP}$ originates from the english words Common Point.
On the multimeter this node is noted as COM.
In the opposite case if the $U_\mathrm{CP} \neq 0$, then the equation for the voltage amplification becomes $U_0=(U_+ - U_-)G+U_\mathrm{CP}$. 
The current that flows through the ground node is equal to zero.
Calculate the ratio between the input voltage $U$  and the current $I$ that flows through the entire circuit with accuracy 1\%.
Find the effective resistance $R=U/I=R(R_1,R_2, R_3)$ as a function of the three resistors in the circuit.
For simplicity you can assume that the gain $G$ goes to infinity $G \rightarrow \infty$. 
Replace the following example values of the resistors $R_1=R_2=10\;\mathrm{k} \Omega,\; R_3=1.5 \;\mathrm{k} \Omega$ in the found expression.
In short: find the
(1) final expression for the resistance of the entire circuit
(2) calculate the value of the resistance with accuracy 1\%.
What is the sign of the expression $R=U/I$ and what is the module of its value. 
How the resistance light on the LED?

\subsection{Riddle, 2~points} 
Red and robust flying. Guess it?

\section{Homework problem, $\mathbf{137\,\$}$}
\label{Homework}
\textit{After the Olympiad, find a screwdriver and remove the screws from the cover of the ``black box''.
Take out the batteries or turn one of the switches from position ``On'' to ``Off''.}

The ``black box'' VAC (Volt-Ampere Curve) for small voltages is a straight line with a constant relation 
 $U/I$ with the dimension of resistance, just like the Ohm’s law. Try to measure the resistance of the
 ``black box'' with an ohmmeter. 
Compare the readings of the ommeter and the determined resistance via an examination of the VAC.

Explain why the determined resistance via the VAC could not be measured with an ohmmeter?
What modification has to be done in the ``black box'' circuit shown in Fig.~\ref{circuit} and why will it make possible the measurement with an ohmmeter, for instance with the multimeter DT-830B that you were given at the Olympiad in the range 20~k$\Omega$?

The first participant that finds the answer of at least one of both questions and sends it from his/her registered for the Olympiad email address to \texttt{epo@bgphysics.eu} by 07:00 on November 1 2015 earns the prize of 137~\$. 
You can work in teams, read books, use \textit{Google} and consult radio engineers and university professors in electronics worldwide through Internet. 
At midnight in Kumanovo, it is late afternoon in California, while the working day in Japan starts – there are always working colleagues around the globe.

\acknowledgments
The authors are thankful to Albert Varonov for the critical reading of the manuscript.

\section{Original Bulgarian Text: Условие на задачата}

\begin{figure}[h]
\includegraphics[width=8.8cm]{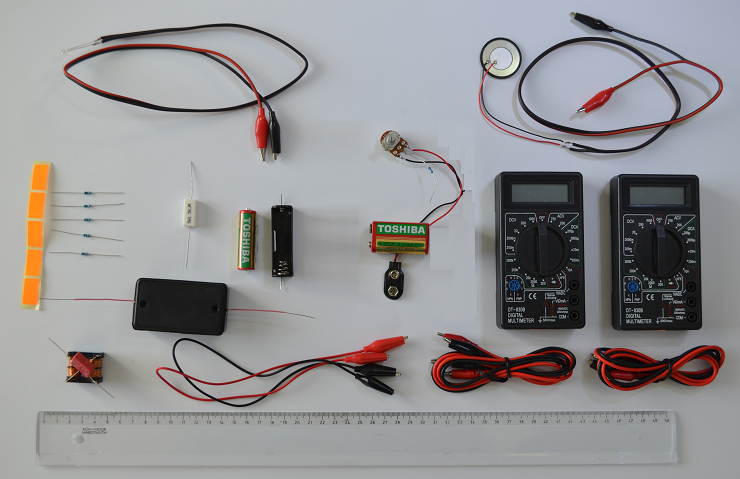}
\caption{Описание на експерименталния набора. 
Вие носите един мултиметър заедно със съединяващите го кабели 
и ще Ви бъде даден още един за временно ползване. 
И така разполагате с два милиметъра, 
една батерия от 1.5~V~(AA), 
1~батерия от 9~V,
1~бял керамичен резистор със съпротивление 1.5~k$\Omega$, 
5~други резистора, 
жълти етикети, 
1~пластмасова линия с дължина 50~cm,
един кондензатор с пиезо-пластинка,
светодиод запоен на дълги жици,
4 съединителни проводника за мултиметрите (червени и черни), 
потенциометър свързан  с контакти за батерия от 9~V,
запоени и успоредно свързани кондензатор и индуктивност 
(това е един резонансен $LC$ контур), 
и най-важното една ``черна кутия''.
Проверете дали имате пълния набор елементи показан на снимката.}
\label{BG_setup}
\end{figure}

С помощта на дадения Ви експериментален набор, показан на Фиг.~\ref{BG_setup}, изследвайте зависимостта между тока и напрежението $I(U)$ или още както се нарича волт-амперна характеристика (ВАХ) на ``черната кутия'', която не бива да се отваря по време на олимпиадата.

\subsubsection{Описание на задачите на Олимпиадата}

Следват две качествените задачи описани в раздел~\ref{BG_quatity_tasks}, които имат за цел да проверят дали устройството
скрито в ``черната кутия'' работи. 
В секция~\ref{BG_black_box_investigation} подробно са описани експериментите, които трябва да извършите с дадения Ви набор за да изследвате ВАХ на ``черната кутия''. 
След това в секция~\ref{BG_the_world_is_not_perfect} ще направите едно по-детайлно изследване и 
ще видите, че физичните свойства, които сте открили до сега за този обект, са валидни в определени граници.
След статичния анализ следва една секция~\ref{BG_dynamics}, в която ще изследвате динамичните свойства на  ``черната кутия'' чрез механични опити.
Има и чисто теоретичен раздел~\ref{BG_TheoreticalProblem} свързан с теоретичното описание на предложените експерименти.
За учениците, които не са много уверени в експеримента, 
но по-добре се справят с математиката 
предлагаме да концентрират усилията си върху теоретичната задача. 
А за неуморните има и задача за домашно описана в секция~\ref{BG_Homework} с парична премия от 137~\$; краен срок утре (01.11.2015) сутрин до 7:00.

След приключването на олимпиадата експерименталния набор остава подарък за
кабинета по физика, моля предайте го на учителя си
за да демонстрирате експеримента пред учениците в училището.
На организаторите трябва да върнете само дадения Ви за състезанието мултиметър.
Желаем добро настроение, интересно изследване, забавление и успех.

\section{Две качествени задачи}

\label{BG_quatity_tasks}
\subsection{Запалване на светодиод с ``черна кутия''.  Ако светодиода не свети обърнете се към учителите}
В дадения Ви набор, показан на Фиг.~\ref{BG_setup}, имате една ``черна кутия'' и един светодиод, запоен на края на дълъг кабел с 2 ``крокодила''. 
Свържете светодиода към ``черната кутия'' и ако не светне разменете ``крокодилите''. 
Поне при една от полярностите диода светва.
При по-нататъшните количествени задачи ще трябва да се измерят тока и напрежението през светещия диод и да се обясни как тези величини са свързани с волт-амперните характеристики на елементите. 
\begin{figure}[h]
\includegraphics[width=3.3cm]{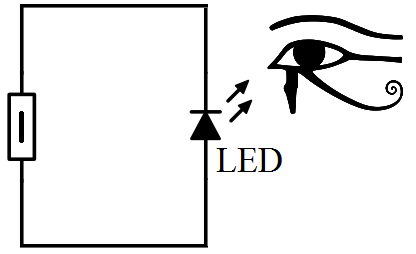}
\caption{При една от двете възможни полярности на светодиода той светва, когато е свързан с електродите на ``черната кутия''. Какви са свойствата на тази кутия и какво има вътре -- това е задачата на Олимпиадата.}
\label{BG_LEDNR}
\end{figure}

\subsection{Възбуждане на електрични трептения с ``черната кутия''. Ако светодиодът не мига и пиезо-пластинката не бръмчи, обърнете се към учителите}
Успоредно към ``черната кутия''  и светещия диод включвате резонансния $LC$ контур от пак успоредно свързани кондензатор $C$ и индуктивност $L$ с феритна сърцевина и вътрешно съпротивление $r$. 
Раздвижете светодиода и ще видите, че сега светлината пулсира. 
Нещо повече, ако включите успоредно и пиезо-пластинката, както е показано на Фиг.~\ref{BG_LCNR}, ще чуете слабо бръмчене. 
Потвърдете дали светодиода мига и пиезо-пластинката бръмчи.
Следващите задачи са свързани с детайлното количествено изследване на тези трептения и тяхното теоретично обяснение. 
\begin{figure}[h]
\includegraphics[width=8.8cm]{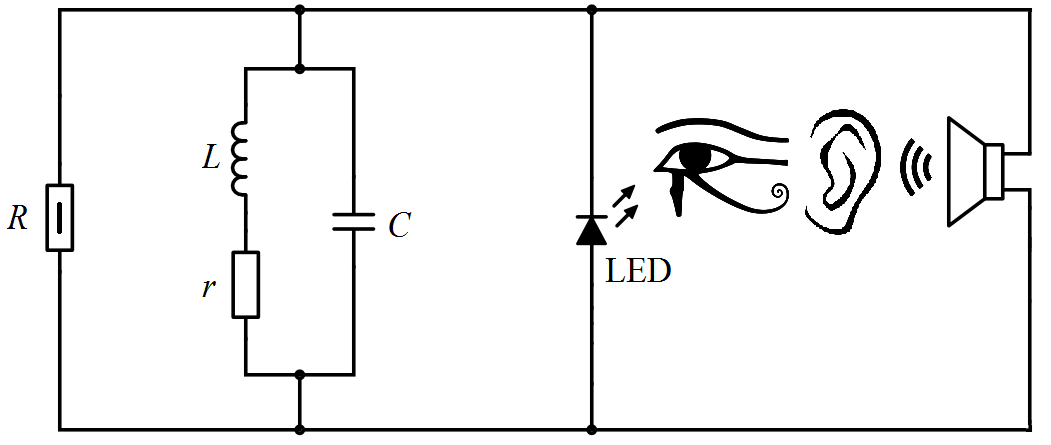}
\caption{Възбуждане на електрични трептения в $LC$ резонансен кръг с помощта на ``черната кутия''. Ако движим светодиода окото вижда, че светлината пулсира, 
а ухото чува бръмченето на пиезо-пластинката възбудено от променливото напрежение.
}
\label{BG_LCNR}
\end{figure}

\begin{figure}[h]
\includegraphics[width=8.8cm]{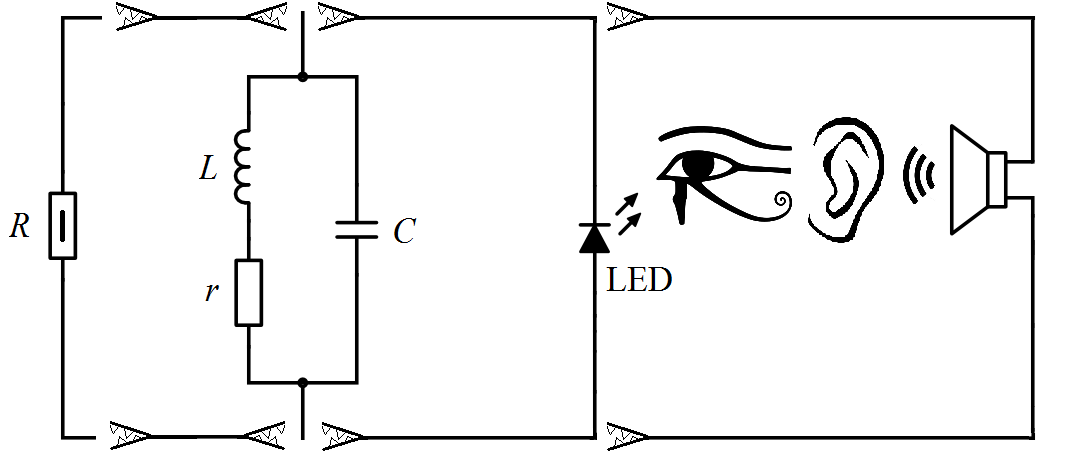}
\caption{Схема на свързване на електричната верига от фигура~\ref{BG_LCNR}, чрез наличните съединителни проводници, завършващи с накрайници 
-- щипки (``крокодили'').}
\label{BG_LCNR-with-crocodiles}
\end{figure}

При коя от двете качествени задачи диода свети по-ярко, при постоянното светене или при пулсиращото?

\pagebreak
\section{Експериментална задача, 101~точки}
\subsection{Изследване на статичното поведение на ``черната кутия''. Раздел за по-малките ученици}
\label{BG_black_box_investigation}

\textit{Подусловията от този раздел~\ref{BG_black_box_investigation}, са адресирани за по-малките ученици. Задачите в тях (1--8) са по-прости и носят по-малко точки. 
Разбира се, ако Ви остане време, продължете и с останалите задачи.}

\begin{enumerate}%
\item \textbf{Измерете напрежението и тока през ``черната кутия'' свързана със светодиод и без него. (7~точки)}

Свържете схемите от Фиг.~\ref{BG_Hypothesis_reject} и измерете напрежението $U$ върху ``черната кутия'' и тока $I$, който протича през нея. 
Ако светодиодът от първата или последната верига не свети сменете полярността му.
Резултатите от измерванията представете таблично, както е показано на примерна 
Табл.~\ref{BG_template_4_setups}.

На какво се дължи малката разлика между напреженията $U_a$ и $U^*$?
\begin{figure}[h]
\includegraphics[width=16.2cm]{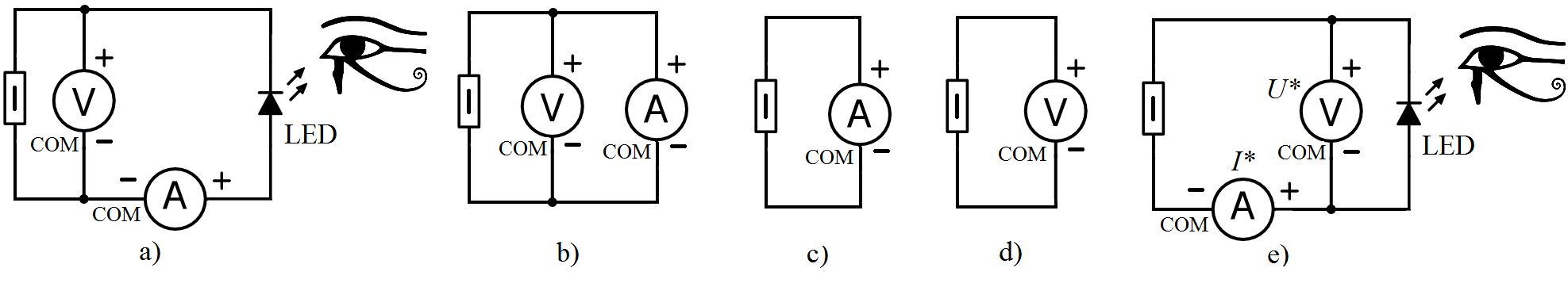}
\caption{Пет постановки за изследване на ``черната кутия'': (a) Към постановката със светещия диод от
Фиг.~\ref{BG_LEDNR} се добавят амперметър и волтметър. Показанията им са $U_a$ и $I_a$. (b) Заменяме светодиода с проводник и показанията на уредите са $U_b$ и $I_b$. (c) Махаме волтметъра и амперметъра показва $I_c$. (d) Заменете амперметъра с волтметър и запишете напрежението $U_d$. (e) Измерете напреженията $U^*$ и $I^*$ по последната схема.  Запишете измерванията в примерна 
Табл.~\ref{BG_template_4_setups}.
}
\label{BG_Hypothesis_reject}
\end{figure}

\begin{table}[h]
\caption{Образец за таблица за обработка на експериментални данни за експеримента показан на 
Фиг.~\ref{BG_Hypothesis_reject}.}
\begin{tabular}{| c| c | c | }
\tableline
$\#$ & $I$ [$\mu A$] & $U$ [$V$ ]  \\
\tableline
a) & $I_a=\qquad \qquad$ &  $U_a=\qquad \qquad$   \\
b) & $I_b=\qquad \qquad$ &  $U_b=\qquad \qquad$   \\
c) &  $I_c=\qquad \qquad$ &    \\
d) &   &  $U_d=\qquad \qquad$   \\
e) & $I^*=\qquad \qquad$ &  $U^*=\qquad \qquad$   \\
\tableline
\end{tabular}
\label{BG_template_4_setups}
\end{table}

\item \textbf{Измерете напрежението на батерията от $\mathcal{E}=$1.5 V. (1 точка)}

Включете комбинирания уред (мултиметъра, мултицета) като волтметър и измерете напрежението на батерията $\mathcal{E}$ от 1.5 V.
Уредът отчита и знакът на напрежението. Можете  да запомните правилата за знаците и условно да включвате 
червения проводник към входа на мултиметъра, а черния към другия вход, който на повечето прибори е означен като ``земя'' (~\ground) или с надпис COM. 

\item \textbf{Измерете съпротивлението на големия бял резистор. (1 точка)}

Включете мултиметъра като омметър и измерете и запишете съпротивлението $R_\mathrm{WR}$ на големия бял резистор с топлинна облицовка от бял цимент. 
Работете с точност от 1~$\Omega.$ $R_\mathrm{WR}$=?

\item  \textbf{Измерете съпротивлението на петте малки резистора. (3~точки)}

Върху съединителните проводници на дадените Ви пет резистора залепете жълто етикетче. 
Мултиметърърт продължава да бъде включен като омметър.
Напишете върху етикетчетата числената стойност на съпротивленията.
Подредете ги по големина и напишете номерата им върху етикетчетата.
Представете съпротивленията в таблица 
$r_1 < r_2 < r_3 < r_4<r_5$, като работите с точност от 1~$\Omega.$

\item  \textbf{С помощта на петте малки резистора и батерийката от $\mathcal{E}=$1.5 V измерете зависимостта между тока и напрежението на големия бял резистор със съпротивление $R_\mathrm{WR}$. (7~точки)}

Електричната схема за измерване на зависимостта между тока 
и напрежението е дадена на Фиг.~\ref{BG_resistance-measurement} и 
Фиг.~\ref{BG_R_neg_R_with_crocodiles}.
Mултицетът, който носите, използвайте като амперметър 
и го включете последователно на белия резистор. 
Внимавайте за знаците -- токът има своя посока!
Другия мултиметър включете като волтметър успоредно на изследвания бял резистор
и пак внимавайте за знаците и полярността на включването.
За големия бял резистор е в сила закона на Ом $U/I=R_\mathrm{BG_WR}$. 
Ако напрежението е положително, токът е положителен, ако напрежението е отрицателно, токът е отрицателен. 
Ако знаците на напрежението и тока са противоположни вижте къде сте сбъркали в свързването на уредите.
\begin{figure}[h]
\includegraphics[width=11.5cm]{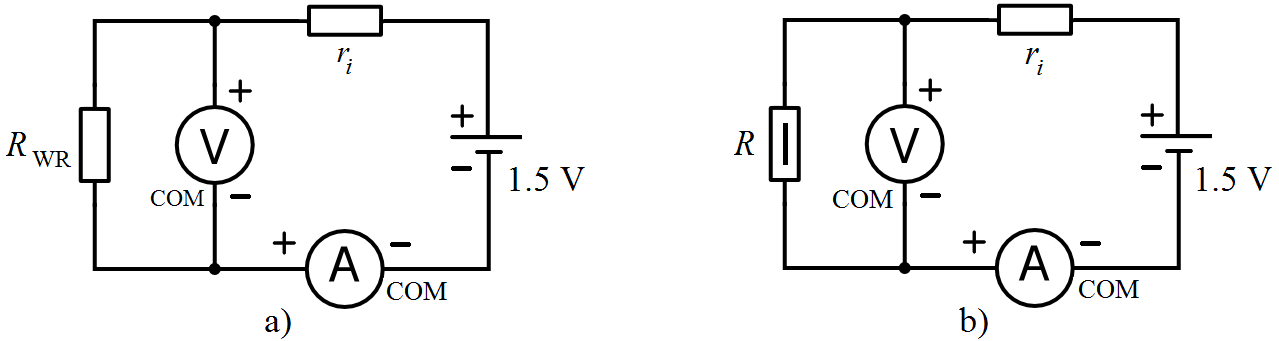}
\caption{Електрична схема за изследване на част от
волт-амперната характеристика (ВАХ) на: (a) белия резистор и (b) ``черната кутия''
и чрез нея техните съпротивления.
Волтметърът (V) е свързан успоредно на изследвания елемент и
измерва напрежението $U$, 
а амперметърът (A) е свързан последователно и измерва тока $I$. 
Когато веригата се затваря с различни съпротивления $r_i\in(0,\;600\,\Omega)$
токът и напрежението са различни.
Така се получават няколко точки от ВАХ. За малки напрежения отношението  $R=U/I$ е постоянно
и това е един възможен начин за проверка на закона на Ом 
и знаците на измерваните ток и напрежение.
Резистора $R_\mathrm{WR}$ на схемата от ляво е заменен с ``черната кутия'' от фигурата в дясно и това е единственото различие на двете постановки.
}
\label{BG_resistance-measurement}
\end{figure}

\begin{figure}[h]
\includegraphics[width=15cm]{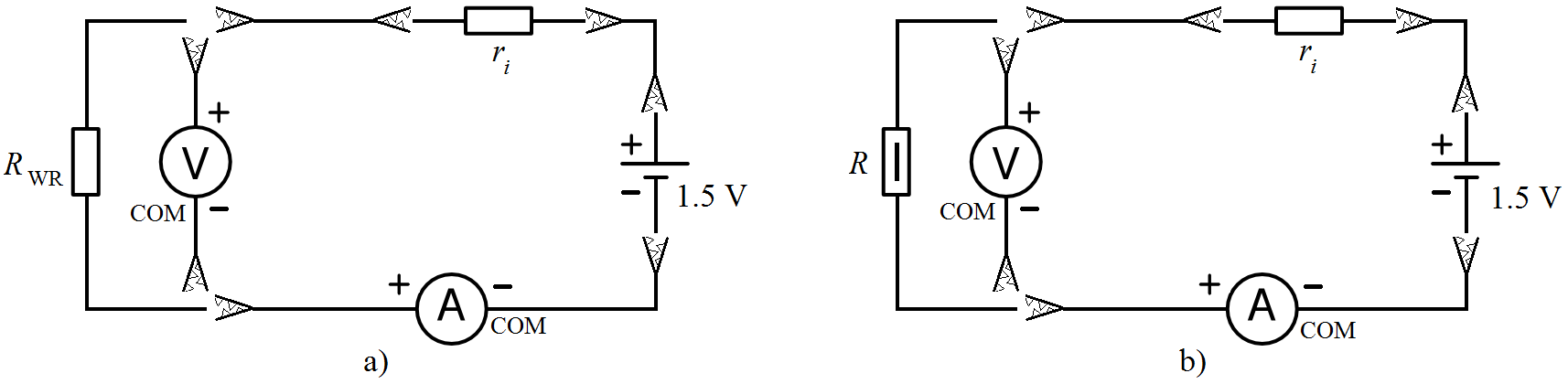}
\caption{Схема на свързване на електричната верига от Фиг.~\ref{BG_resistance-measurement}, чрез наличните съединителни проводници завършващи с накрайници -- ``крокодили''.}
\label{BG_R_neg_R_with_crocodiles}
\end{figure}

Последователно на амперметъра включете батерията от 1.5~V. 
Затваряйте веригата последователно със съпротивления 
от 0~$\Omega$ и по отделно 5-те $r_i.$ 
За всяко от измерванията запишете в таблица с 5 стълба и 6 реда по образеца даден в Табл.~\ref{BG_template}:
1)~номера на резистора $i$,
2)~съпротивлението $r_i$,
3)~тока $I_i$ и
4)~напрежението на волтметъра $U_i$ 
5)~пресметнатото съпротивление на изследвания елемент $U_i/I_i$.

\begin{table}[ht]
\caption{Образец за таблица за обработка на експериментални данни за опитите показани на 
Фиг.~\ref{BG_resistance-measurement}.
Стълбовете означават:
1)~номера на резистора $i$,
2)~съпротивленията на резисторите  $r_i$;
късото съединение $r_0=0$ е нулевия ред на таблицата,
3)~тока $I_i$, който тече през веригата за различни съпротивления $r_i$ последователно свързани с батерия с напрежение $\mathcal{E}$,
4)~напрежението $U_i$ върху белия резистор или ``черната кутия'',
5)~Отношението на напрежението $U_i$ и тока $I_i$.
}
\begin{tabular}{| r | r | r | r | r | r | r |}
\tableline
 i& $r_i \, [\Omega]$ & $I_i \,[\mu \mathrm{A}]$ & $U_i\,[\mathrm{V}]$ & $U_i/I_i\,[\Omega]$ \\
\tableline
0 & 0 &  & & \\
1 &   &  &  &\\
2 &   &  &  &\\
3 &   &  &  &\\
4 &   &  &  &\\
5 &   &  & &\\
\tableline
\end{tabular}
\label{BG_template}
\end{table}

\item  \textbf{С помощта на петте малки резистора и батерийката от $\mathcal{E}=$1.5 V измерете зависимостта между тока и напрежението на ``черната кутия''. (7~точки)}

За целта повторете същите измервания, както в предходното подусловие, като замените белия резистор с ``черната кутия''

\item  \textbf{Начертайте волт-амперната характеристика (ВАХ) на ``черната кутия'' и белия резистор на една обща графика, използвайки данните от таблиците на предишните две подусловия. (7~точки)}

Данните от таблиците представете графично, 
по абсцисата (оста~\textit{x}) -- напрежението $U_i$
по ордината (оста~\textit{y}) -- тока $I_i$.
Ние препоръчваме отначало да направите малка графика в мащаб
$\mathrm{1\,V=1\,cm}$ и
$\mathrm{1\,A=1\,cm},$
която да съдържа координатното начало $U=0$ и $I=0$.
Само ако Ви остане време накрая, може да я прерисувате в по подходящи координати. 
Това е малка част от волт-амперната характеристика (ВАХ) 
на ``черната кутия'' и на белия резистор, и от зависимостта $I(U)$ имаме само 6 точки за всеки от изследваните елементи. 
През точките от графиката прекарайте права линия, която минава най-близо до тях. Означете коя от тези прави за кой изследван елемент се отнася.
\item \textbf{Определете наклона $\Delta U / \Delta I$ на ВАХ начертани в предното подусловие. (9~точки)}

Символа $\Delta$ означава разлика $\Delta U=U_2-U_1$ и $\Delta I=I_2-I_1$. Изберете две точки от начертаните прави линии.
С каква характеристика на изследваните елементи е свързан наклона  $\Delta U / \Delta I$?
Намирате ли нещо необичайно за тази характеристика на ``черната кутия''?

\subsection{Светът не е идеален. Детайлно изследване на ВАХ. Раздел за по-глемите ученици имащи практика в изследване на ВАХ}
\label{BG_the_world_is_not_perfect}
\textit{По-големите ученици, които имат практика в изследване на ВАХ могат да започнат с този раздел~\ref{BG_the_world_is_not_perfect} и да се върнат към началните подусловия от раздел~\ref{BG_black_box_investigation} само, ако им остане време.}

При малки напрежения ВАХ на ``черната кутия'' е част от права линия, но как изглежда ВАХ при по-големи стойности на напреженията ще откриете след малко.

\begin{figure}[h]
\includegraphics[width=8cm]{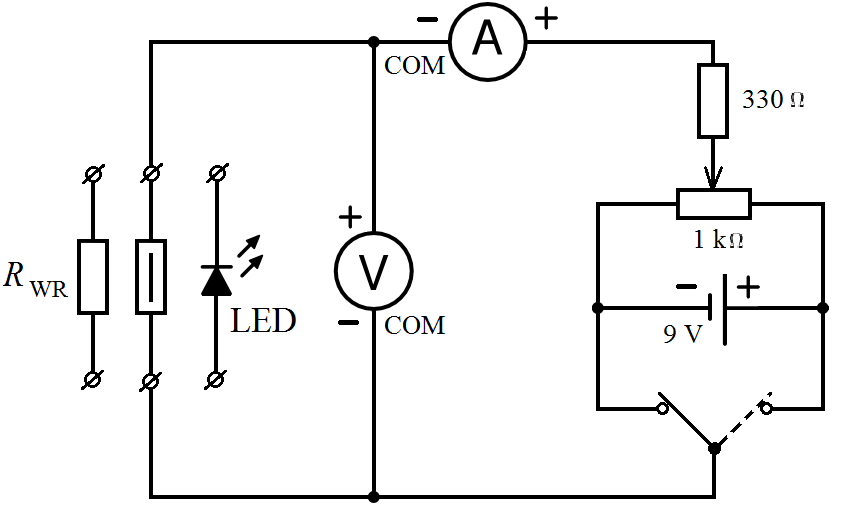}
\caption{Постановка за изследване на ВАХ $I(U)$. Първо се изследва белия резистор  $R_\mathrm{WR}$ (най-отляво), после той се заменя с ``черната кутия'' включена в схемата и накрая към постановката включвате само светодиода~(LED)~(от дясно).  
Волтметърът измерва напрежението $U$ върху изследвания елемент, а амперметърът измерва тока $I$.
Напрежението се създава от батерия с напрежение $\mathcal{E}=9\,\mathrm{V}$. 
Когато въртим с ръка оста на потенциометъра със съпротивление 
$1\,\mathrm{k}\Omega$
напрежението $U$ се изменя 
от 0 до $+\mathcal{E}$.
При превключване на ``крокодила'' от единия краен извод на потенциометъра към другия (както е показно в долния десен ъгъл) напрежението $U$ пробягва интервала от $-\mathcal{E}$ до 0.
Така напрежението се изменя $-\mathcal{E}<U<+\mathcal{E}$.
Съпротивлението от $330\,\Omega$
ограничава тока в схемата и предпазва светодиода от изгаряне.
}
\label{BG_resistance-measurement-I-V-curve}
\end{figure}

\item \textbf{Изследвайте детайлно зависимостта между тока и напрежението на белия резистор. (7~точки)}
 
Включете батерията от 9~V към съединителните контакти
закачени към потенциометъра. 
Това е един източник на напрежение, към който включвате волтметър и измервате обхвата на напрежението при крайните завъртания на потенциометъра от $U_\mathrm{min}$ до 0 и от 0 до $U_\mathrm{max}$ за двете включвания на ``крокодила'' към потенциометъра, както е показано на Фиг.~\ref{BG_resistance-measurement-I-V-curve}.
Символично начертания ключ се реализира чрез превключване на ``крокодил''.
Към белия резистор включвате последователно амперметър 
и ги закачате към източника на напрежение.
Амперметърът измерва тока $I_\mathrm{WR}$ през белия резистор, а волтметърът напрежението $U$. 
Когато въртите оста на потенциометъра напрежението се мени
и така изследвате ВАХ. 
Първо превъртете набързо потенциометъра между крайните му положения и разберете в какви обхвати на уредите трябва да работите.
После изследвате зависимостта $I_\mathrm{WR}(U)$, като двойките величини $(I_\mathrm{WR},U)$ записвате в таблица. За белия резистор е в сила законът на Ом. За построяване на линейната зависимост $I_\mathrm{WR}(U)$ са напълно достатъчни и 5 точки, две от които да бъдат при напрежения $U_\mathrm{min}$ и $U_\mathrm{max}$.
Пак съгласно закона на Ом знаците на напрежението и тока трябва да бъдат еднакви. При различие в знаците потърсете грешка в схемата.

\item \textbf{Изследвайте детайлно зависимостта между тока и напрежението на ``черната кутия''. (5~точки)}

В постановката от предишното подусловие заменете белия резистор с ``черната кутия'' без да правите други промени.
Въртейки оста на потенциометъра променяйте напрежението $U$ между $U_\mathrm{min}$ и $U_\mathrm{max}$ през около 1 $V$ и записвайте в таблица двойките числа $(I, U)$.
Запишете в таблицата и напреженията, при които токът има максимум или минимум.

\item \textbf{Изследвайте детайлно зависимостта между тока и напрежението на светодиода (LED). (3~точки)}

В постановката от предишното подусловие заменете ``черната кутия'' със светодиода отново без да правите други промени.
Вижте в какво крайно положение на потенциометъра диодът свети най-ярко. 
Сега въртете оста на потенциометъра и следете показанията на амперметъра. 
При ток по-малък от 6~mA записвайте в таблица двойките числа $(I_\mathrm{LED}, U)$ през 1~mA.
При токове по-малки от 2~mA=2000~$\mu$A записвайте двойките числа през около 200~$\mu$A, 
докато достигнете ток по-малък от 200~$\mu$A.

\textit{Когато свършите с тази 3-та ВАХ изключете батерията от 9~V от контактите,
те бързо се изтощават.}

\item \textbf{Представете върху една обща графика ВАХ на: белия резистор $I_\mathrm{WR}(U)$, ``черната кутия''  $I(U)$ и светодиода $I_\mathrm{LED}(U)$. (10~точки)}

Първо анализирайте най-малката и най-голямата стойност на тока и напрежението. Тези параметри определят правоъгълника, в който ще бъдат представени ВАХ на трите елемента. 
Ние препоръчваме мащаб по хоризонтала 
$\mathrm{1\,V=1\,cm}$
и по вертикала $\mathrm{1\,mA=1\,cm}$.
И за трите елемента ВАХ $I(U)$ са непрекъснати криви.
Начертайте през точките водеща очите прави (или криви), 
които максимално добре да описват експерименталните данни.

\item \textbf{Защо са важни ВАХ? (5~точки)}

Нарисувайте огледален образ на ВАХ на светодиода, при която $I$ се заменя с $-I$, т.е. превъртаме ВАХ около хоризонталната ос на напрежението.
Върху ВАХ на светодиода нанесете с малко кръгче стойностите $(-I^*,U^*)$ от таблица~\ref{BG_template_4_setups}.
Обърнете внимание, че тази точка е близо до пресечницата на ВАХ на ``черната кутия'' и ВАХ светодиода.
Случайно ли е това?

\item \textbf{Анализ на ВАХ на: белия резистор, ``черната кутия'' и светодиода. (7~точки)}

Ако ВАХ $I(U)$ се състои от отделни прави, 
от техния наклон определете съответстващите съпротивления
$R=\Delta U/\Delta I.$ 
Символът $\Delta$ означава разлика (\textit{difference}). Избирате 2 точки от отсечката и измервате разликата в напрежението $\Delta U= U_2-U_1$ и тока $\Delta I=I_2-I_1.$  
Когато в разглеждания участък кривината е пренебрежима 
вместо $\Delta$ се пише $\mathrm{d}.$ Съпротивление, получено от наклон на ВАХ, се нарича диференциално съпротивление, а реципрочната му стойност диференциална проводимост $\sigma_\mathrm{diff}=\mathrm{d}I/\mathrm{d}U$.

Намерете: a) съпротивлението на белия резистор, b) съпротивлението на централния участък на ``черната кутия'', c) съпротивлението на левия участък на ``черната кутия'', d) съпротивлението на десния участък на ``черната кутия'' и e) съпротивлението на светодиода, когато е отпушен~(свети) в интервала от 2-5 mA.

\item \textbf{Каква особеност на ВАХ на ``черната кутия'', е съществено важна за постоянното светенето или мигането на диода в двете качествени задачи. (4~точки)}

\textit{За ``черната кутия'' е най-важен централният участък на ВАХ, който включва нулевото напрежение. 
На този участък от ВАХ се дължи постоянното светенето на диода и възбуждането на променливи токове в резонансния контур, които наблюдавахте в качествените задачи показани на Фиг.~\ref{BG_LEDNR} и Фиг.~\ref{BG_LCNR}, както и многото възможни технически приложения на устройства от типа на скритото в черната кутия.}

Някакво различие във ВАХ на ``черната кутия'' и белия резистор разкрива каква е причина за светването на диода.
Кое е това различие?

\item \textbf{Опишете качествено как се създава токът, който предизвиква постоянното светенето или мигането на диода в двете качествени задачи. (10~точки)}

\subsection{Измерване на честотата на електричните трептения, създадени от ``черната кутия''}
\label{BG_dynamics}
Нека свържем отново схемата от втората качествена задача показана на 
Фиг.~\ref{BG_LCNR}
когато всички елементи на веригата са свързани успоредно:
``черната кутия'',
$LC$ резонатора,
пиезо-пластинката (зумера),
светодиода.
Ако вместо зумер се използва слушалка за GSM, то към нея трябва да се добави последователно свързано съптотивление по-голямо от 10~k$\Omega$.
\item \textbf{Механично определете честотата на осцилации. (7~точки)}

Електричните измервания са по-лесни и по-точни.
За да измерим честотата на осцилациите без честотомер при дадените условия е необходимо да се прояви малко сръчност.
В аудиторията разполагате със залепваща лента.
Закрепете светодиода към края на еластична линия със залепваща лента. 
Когато притиснете с една ръка линията към края на масата, а с другата възбуждате трептения на линийката, в близкия до масата край, се наблюдават няколко
почти неподвижни светли петна.
Ако електричната честота $f_\mathrm{res}$ е кратна на механичната $f_\mathrm{mech}$, светлите петна са неподвижни.
Преброявате или само оценявате броя на светлите петна $N$ и като броите бързо определяте и честотата на механичните трептения $f_\mathrm{mech}$. 
Постарайте се да клатите линийката с постоянна честота и амплитуда.
Варирайте и дължината на свободния край на линийката за да постигнете неподвижни светли петна.
Пребройте за 10 секунди колко пъти осцилира линийката 
и така оценете механичната честота $f_\mathrm{mech}$.
Честотата на електричните трептения $f_\mathrm{res}$  изразете 
чрез броя на петната и честотата на механичните трептения.
Колко херца за резонансната честота на електричните трептения дава вашата формула $f_\mathrm{res}(N,f_\mathrm{mech})$?

\item \textbf{Изчислете теоретично честотата на осцилации. (2~точки)}

Един алтернативен метод за определяне на тази честота е нейното теоретично оценяване по формулата на $f_\mathrm{res}=1/(2\pi\sqrt{LC}).$ 
На кондензатора е означено (по инженерски) 4.7~$\mu F.$ А на тороидалната индуктивност пише, че има 2 намотки по 100~mH или при последователно свързване индуктивността и става $L$=400~mH.
Как пресметнатата честота се съгласува с експериментално определената?

\item \textbf{Определете по слух честотата на вибрации на зумера. (1~точки)}

Ако имате музикален слух 
и помните честотите на музикалните тонове може да определите
на кой музикален тон съответства честотата на бръмченето на зумера и каква е неговата честота?
Не се очаква голяма точност и 50\% грешка би било удовлетворително добро съгласие на тези 3 метода за определяне на честота без честотомер.

\end{enumerate}%

\section{Теоретична задача, 35~точки}
\label{BG_TheoreticalProblem}

\subsection{Условие}
\begin{figure}[h]
\includegraphics[width=9cm]{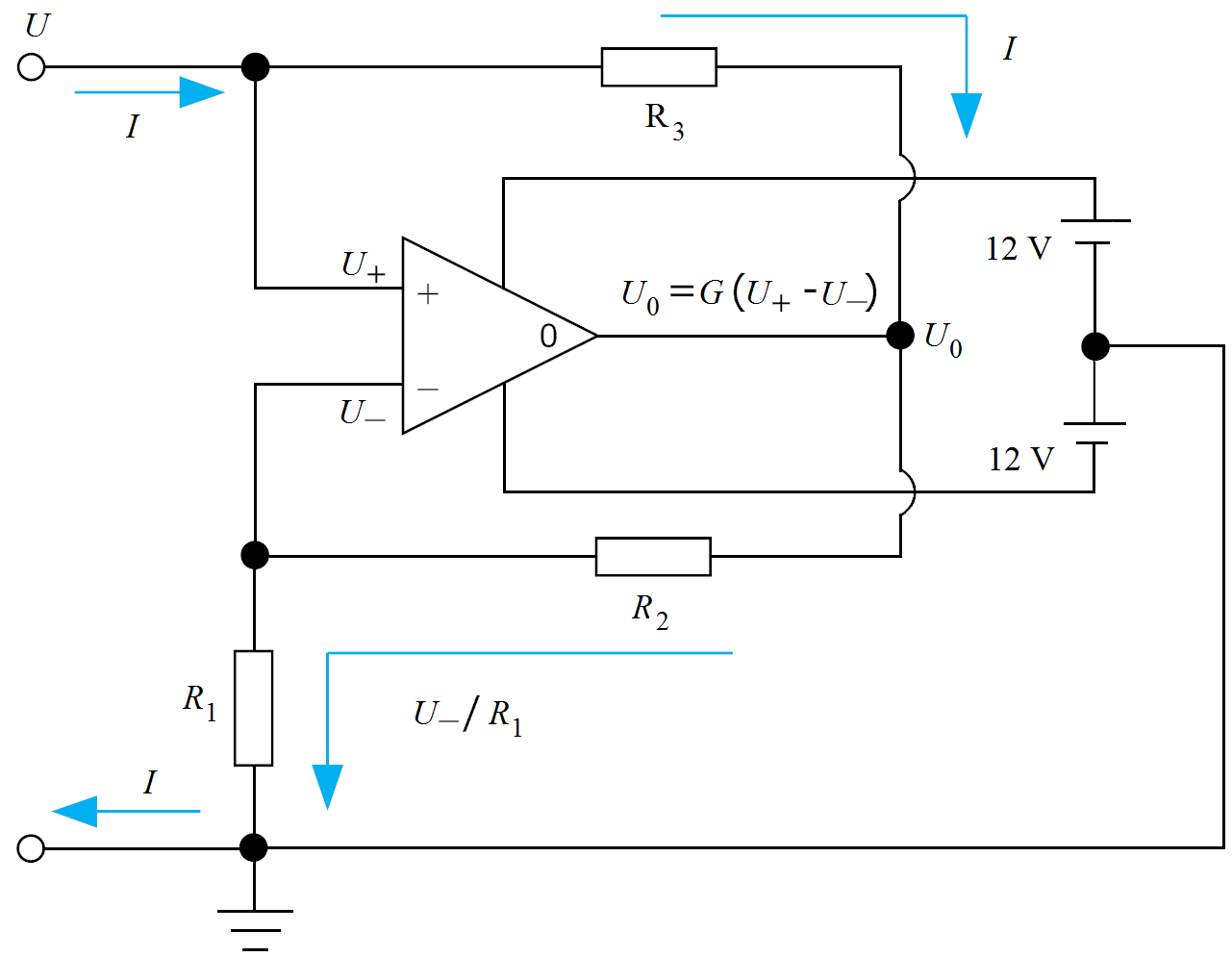}
\caption{Намерете с процентна точност ефективното съпротивление $R=U/I$ на веригата от три резистора $R_1,$ $R_2$ и $R_3$ 
и представения с триъгълник
усилвател по напрежение с коефициент на усилване $G=10^5,$ 
който се захранва от две батерии с напрежение $V_S$. 
Напреженията $U_0$, $U_{-}$, както и токът $I$ са неизвестни.}
Означението $U_0$ е произлиза от $U_\mathrm{output}=(U_+-U_-)G.$
\label{BG_circuit}
\end{figure}

На електричната верига, показана на Фиг.~\ref{BG_circuit}, 
входните токове в точките (+) и (--) са нула, 
а изходния ток в точката (0) е такъв, че
съответните напрежения са свързани със съотношението 
$U_0=(U_+ - U_-)G$,  където
коефициентът на усилване $G=10^5 \gg 1$ е много голямо число. 
Означеният с триъгълник електронен елемент (усилвател) 
се захранва с две батерии и в условието на задачата се предполага, 
че споменатите напрежения са по-малки от
напрежението на захранващите батерии $\mathcal{E}_\mathrm{B}=12\;\mathrm{V}.$ Точката между двете батерии е свързана с проводник с края на резистора $R_1$ и един от входните електроди на
схемата. 
Удобно е напрежението в тази обща точка да се избере
за нула $U_\mathrm{CP}=0$ 
или както техниците казват това е “земя” (\ground). 
Индексът $\mathrm{CP}$ идва от английското Common Point 
и на мултиметрите се означава с COM.
В противен случай, ако $U_\mathrm{CP} \neq 0$, уравнението за
усилването по напрежение приема вида 
$U_0=(U_+ - U_-)G+U_\mathrm{CP}$. 
Токът, който изтича в ``земята'', е нула. 
Пресметнете с 1\% точност (три значещи цифри) отношението на
входното напрежение U и протичащия през веригата ток $I$ и изразете това
ефективно съпротивление $R=U/I=R(R_1,R_2, R_3)$ чрез трите съпротивления на схемата. 
За простота може да предположите, 
че коефициентът на усилване клони към
безкрайност $G \rightarrow \infty$. 
В получения израз заместете 
$R_1=R_2=10\;\mathrm{k} \Omega,\; R_3=1.5 \;\mathrm{k} \Omega$.
Накратко: търси се 
(1) крайна формула за съпротивлението на схемата и 
(2) пресметната по нея числена стойност, като 1\% точност е приемливо приближение.
Какъв е знакът на $R=U/I$ и колко ома е неговия модул. 
Как това съпротивление запалва светодиод?
\subsection{Гатанка, 2~точки} 
Червено и здраво дето лети. Що е то?

\section{Задача за домашно, $\mathbf{137\,\$}$}
\label{BG_Homework}
\textit{След приключване на Олимпиадата, намерете отверка и извадете винтовете от капака на ``черната кутия''.
Извадете батериите или превключете един от ключовете от положение ``On'' на ``Off''.}

За малки напрежения волт-амперната характеристика (ВАХ)
на ``черната кутия'' е права линия с постоянно отношение $U/I$ с размерност съпротивление, също както в закона на Ом. Опитайте да измерите съпротивлението на ``черната кутия'' с омметър. 
Сравнете показанията на омметъра и съпротивлението, определено чрез изследване на ВАХ.

Обяснение защо съпротивлението определено от ВАХ не може да се измери с омметър?
Какво изменение в схемата от Фиг.~\ref{BG_circuit} реализирана в ``черната кутия'' и защо ще направи възможно измерването на съпротивлението с омметър, например с мултицет DT-830B в обхват от 20~k$\Omega$, който Ви беше даден на Олимпиадата?

Първият, който намери отговор на поне на един от двата въпроса и го изпрати от адреса, с който се е регистрирал за Олимпиадата на \texttt{epo@bgphysics.eu} до 07:00 на 1 ноември 2015~г. печели паричната награда от 137~\$. 
Можете да работите в екип, да четете книги, да ползвате \textit{Google} и чрез Интернет да се консултирате с радиоинженери и университетски професори по електроника навсякъде по света. 
Когато в Куманово е полунощ, в Калифорния е късен следобед, 
а Япония започва деня -- винаги по глобуса има работещи колеги.

\newpage
\section{Задачи за понататъшна изследователска работа}
\begin{enumerate}
\item \textbf{За всяко от измерванията в предното подусловие изчислете и допълнете в горните таблици съпротивлението на изследвания елемент, пълното съпротивление на веригата и тяхната разлика}

Допълнете 2-те таблици с още 3 стълба:
5) съпротивлението на изследвания елемент
определено като $R=U/I$, т.е. числото от 4-тия стълб разделено на числото от третия,
6) пълното (тоталното) съпротивление на веригата
дефинирано като $\mathcal{R}_\mathrm{tot}=\mathcal{E}/I$,
т.е. делите напрежението на батерията върху тока от 3-тата колона,
7) пълното съпротивление определено като разлика на тоталното и допълнителния резистор 
$\tilde{R}=\mathcal{R}_\mathrm{tot}-r_i
=\mathcal{E}/I-r_i$.
Така експерименталните данни се обработват в таблици от 7 колони
представени в таблицата за образец \ref{BG_template_ext}.

\begin{table}[ht]
\caption{Образец за таблица за обработка на експериментални данни за експеримента показан на 
Фиг.~\ref{BG_resistance-measurement}.
Стълбовете постепенно означават:
1) номера на резистора $i$,
2) съпротивленията на резисторите  $r_i,$;
късото съединение $r_0=0$ е нулевия ред на таблицата
3) тока $I_i$, който тече през веригата за различни съпротивления $r_i$ последователно свързани с батерия с напрежение 
$\mathcal{E}$.
4) напрежението върху ``черната кутия'' или белия резистор $U_i$, 
5) съпротивлението на ``черната кутия'' 
$R=U/I$ определено като отношението
между тока и напрежението,
6) пълното съпротивление на веригата
определено като отношението на електродвижещото напрежение и тока $\mathcal{R}_\mathrm{tot}=\mathcal{E}/I,$
7) съпротивлението на изследвания елемент
$\tilde{R}=\mathcal{R}_\mathrm{tot}-r$
определено като разлика между пълното съпротивление 
и съпротивлението на на последователно свързания резистор $r$. 
}
\begin{tabular}{| r | r | r | r | r | r | r |}
\tableline
 i& $r_i$ [$\Omega$] & $I$ [$\mu$ A] & $U$ [V] & $U/I\,$[k$\Omega$]& $\mathcal{E}/I\,$[$\Omega$]& $(\mathcal{E}/I - r_i) \,$[$\Omega$]\\
\tableline
0 & 0 &  & & & & \\
1 &   &  & & & & \\
2 &   &  & & & & \\
3 &   &  & & & & \\
4 &   &  & & & & \\
5 &   &  & & & & \\
\tableline
\end{tabular}
\label{BG_template_ext}
\end{table}

\item \textbf{Изчислете средната стойност на съпротивлението на изследвания елемент и пълното съпротивление на веригата. Защо тези стойности се различават?}

Усреднете по колоните от 2-те таблици съпротивленията 
$\left<R\right>$ и $\left<\tilde{R}\right>$, т.е. намерете средното аритметично
на 5-тата и 7-мата колона колко процента е разликата $(\left<R\right>-\left<\tilde{R}\right>)/\left<R\right>$ 
и анализирайте откъде идва различието.

\item \textbf{За ``черната кутия'' и белия резистор представете графично зависимостта на
пълното съпротивление 
$\mathcal{R}_\mathrm{tot}$ (ордината) 
от съпротивленичта на отделните ресистори $r_i$ (обща абсциса).}

\item \textbf{От графиката начертана в предното подусловие определете отново неизвестното съпротивление $R$.} 

Свържете точките с гладка крива и определете отново неизвестното съпротивление $R$ на ``черната кутия'' и белия резистор $R_\mathrm{WR}$. 
Намерихте ли нещо смущаващо, необичайно или интересно? 

Ако отговорът на въпроса от предното подусловие е ``Да'' опитайте се 
да намерите възможни технически приложения на ``черната кутия''.

\item \textbf{Представете графично зависимостта $I(r_i)$ за ``черната кутия'' и за белия резистор. Изведете формула която описва тази зависимост.}

Представете данните на обща графика:
по абсцисата общите съпротивления $r_i$,
а по ордината токовете.
Нарисувайте водещи очите криви които описват двете зависимости $I(r)$.
Опишете дали функциите са намаляващи или разтящи.
Изведете формула която описва тази зависимост.
Отговорете също така на въпроса, 
ако $r_i$ нараства дали тока $I$ намалява или се увеличава 
съгласно получените от Вас данни.

\item \textbf{Опишете качествено зависимостта на пълното съпротивление на веригата от съпротивлението  $r_i$.}

Веригата от последователно свързани 
``черна кутия'' със съпротивление $R$ и резистор $r_i$ има сумарно 
съпротивление $\mathcal{R}_i=R + r_i$. Това е и пълното съпротивление на веригата. Опишете качествено графиката на функцията $\mathcal{R}(r)$. 
Дали това е периодична, намаляваща или растяща функция? Има ли минимуми и максимуми? 
Какъв е наклонът (производната на графиката) положителен или отрицателен.?

\item 
Включвате последователно ``черната кутия'' и амперметър, а волтметъра
успоредно на черната кутия. 
Вместо към източника на напрежение този път включете 
``черната кутия'' към светодиода.
Така както е показано на Фиг.~\ref{BG_LEDNR},
но сега добавяте ампреметър и волтметър.
Ако диода не светва обърнете полярността му.
Ако и при обратна полярност не светва потърсете съдействие от комисията на Олимпиадата. 
Амперметърът измерва общия ток $I^*$, 
който циркулира през ``черната кутия'' и светодиода.
Обърнете внимание сега нямаме външен източник на напрежение.
А волтметърът измерва с точност до знак общото напрежение 
$U^*$ на ``черната кутия'' и светодиода. 
Проблемът със знака възниква пореди успоредно свързаните елементи.
Циркулиращия ток можем да опишем като ток $I$ минаващ през ``черната кутия''
и успоредно на него ток $-I$ минаващ през светодиода, като напрежението през тези два елемента е общо.
Следователно с точност до знак показанията на приборите
са точка от ВАХ за ``черната кутия'' и светодиода
$I(U)=-I(U)_\mathrm{LED}$ или $I(U)=I(-U)_\mathrm{LED}.$ 
Нанесете тази точка $(I^*,\, U^*)$ на ВАХ от общата графика.
Проблемът със знака се решава като прерисувате отразената през ветикалната права ВАХ на светодиода.
Колко mA е тока $I^*$, който циркулира по затворения контур без външно напрежение, сякаш контура е от свръхпроводници точно нулево съпротивление?
А колко волта е общото напрежение $U^*$?
В свръхпроводящ пръстен тока може да циркулира с години, 
можете ли да обясните качествено физиката на светодиода запален от черна кутия с постоянно отношение на напрежение и ток $R=U/I=\mathrm{const},$
също както при резистор. 

\item \textbf{Измерете променливото напрежение върху пулсиращия светодиод. Сравнете това напрежение с $U^*$.}

Нека свържем отново схемата от втората качествена задача показана на 
Фиг.~\ref{BG_LCNR}
когато всички елементи на веригата са свързани успоредно:
``черната кутия'',
$LC$ резонатора,
пиезо-пластинката (зумера),
светодиода,
и волтметъра който сега е включен да измерва променливи токове,
по-точно тяхната средноквадратична амплитуда. 

Качестено видяхме, светодиода пулсира, зумера бръмчи. 
Какво напрежение $\tilde{U}$ показва волтметъра включен да измерва променливи токове.
Как това напрежение $\tilde{U}$ е свързано с постоянното напрежение $U^*$ на светодиода от предишното подусловие?

\item \textbf{Предложете възможни технически приложения на ``черната кутия''} 

\item \textbf{ВАХ на Омметър}

\end{enumerate}
\newpage
\section{Решение}

\subsection{Решение на експерименталната задача}
Тоз раздел е помагало за бърза проверка на работите на учениците.
Без съпровождащ текст са дадени: числата, таблиците и графиките 
които трябва да описват решението на експерименталната задача.
Във скоби са дадени точките съответстващи на всяко от тези подусловия.
Задачата на проверяващите е бързо да намерят и проверят съответния детайл в ръкописа.

\begin{table}[ht]
\caption{Резултати от експеримента показан на 
Фиг.~\ref{BG_resistance-measurement}.
Стълбовете постепенно означават:
номера на резистора $i$,
съпротивленията на резисторите  $r_i,$
напрежението върху ``черната кутия'' $U_i$, 
тока $I_i$, който преминава през нея за различни съпротивления $r_i$ последователно свързани с батерия с напрежение 
$\mathcal{E}=1.634\;\mathrm{V}$, когато не е натоварена.
5-тата колона е съпротивлението на ``черната кутия'' 
$R=U/I$ определено като отношението
между тока и напрежението,
6-тата е пълното (тоталното) съпротивление на веригата
определено като отношението на електродвижещото напрежение и тока $\mathcal{R}_\mathrm{tot}=\mathcal{E}/I,$
7-тата колона е съпротивлението на ``черната кутия'' 
$^{\prime\prime\!}R^{\prime\prime}=\mathcal{R}_\mathrm{tot}-r$
определено като разлика между пълното съпротивление и съпротивлението на на последователно свързания резистор. 
Този втори метод е по-неточен защото 
$^{\prime\prime\!}R^{\prime\prime}$
вътрешното съпротивленние на амперметъра,
около 100~$\Omega$ за използвания обхват,
и вътрешното съпротивление на батерията.
Освен това електродвижещото напрежение на батерията 
слабо зависи от малкия ток който протича през нея 
$\mathcal{E}_i(I_i)$; светът не е идеален.
Анализа на данните от 5-тата колона показва, 
че черната кутия има отрицателно съпротивление 
$R=-1.56\mathrm{k}\mathrm{\Omega}.$
}
\begin{tabular}{| r | r | r | r | r | r | r |}
\tableline
 i& $r_i$ [$\Omega$] & $I$ [$\mu$ A] & $U$ [V] & $U/I\,$[k$\Omega$]& $\mathcal{E}/I\,$[$\Omega$]& $(\mathcal{E}/I - r_i) \,$[$\Omega$]\\
\tableline
0 &    0 & -1123 & 1.76 & -1.57 & -1455 & -1455 \\
1 & 100 & -1209 & 1.89 & -1.56 & -1352 & -1452 \\
2 & 180 & -1288 & 2.02 & -1.57 & -1269 & -1449 \\
3 & 300 & -1429 & 2.24 & -1.57 & -1144 & -1444 \\
4 & 390 & -1554 & 2.43 & -1.56 & -1052 & -1442 \\
5 & 510 & -1764 & 2.76 & -1.56 & -926.3 & -1436 \\
\tableline
\end{tabular}
\label{BG_experimental_data_black_box_6}
\end{table}

\begin{table}[ht]
\caption{Това са резултати точно съответстващи на данните от 
таблица~\ref{BG_experimental_results_black_box}.
Единственното различие е, че при този експеримент
``черната кутия''  е заменена с белия резистор със съпротивление
$1.5\,\mathrm{k}\Omega.$
За резистора зависимостта $U(I)$, т.е. ВАХ се описва много добре от закона на Ом. 
Отклонението на съпротивлението от средната му стойност е по-малко от 1 хилядна.
Анализа на данните от 5-тата колона показва, 
че белият резистор има положително съпротивление съпротивление 
$R_{WR}=+1.56\mathrm{k}\mathrm{\Omega}.$
С процентна точност съпротивлението на белия ресистор и 
модула на съпротивлението на ``черната кутия'' съвпадат
$R_{WR}\approx\overline{R}.$
Това означава, че наклоните на ВАХ им характеристики ще бъдат точно противоположни, както се вижда на Фиг.~\ref{BG_IV6} и Фиг.~\ref{BG_IVdetail}.
}
\begin{tabular}{| r | r | r | r | r | r | r |}
\tableline
 i& $r_i$ [$\Omega$] & $I$ [$\mu$ A] & $U$ [mV] & $U/I\,$[$\Omega$]& $\mathcal{E}/I\,$[$\Omega$]& $(\mathcal{E}/I - r_i) \,$[$\Omega$]\\
\tableline
1 &    0 & 981 & 1528 & 1558 & 1666 & 1666 \\
2 & 100 & 923 & 1439 & 1559 & 1770 & 1670 \\
3 & 180 & 882 & 1375 & 1559 & 1853 & 1673 \\
4 & 300 & 827 & 1288 & 1557 & 1976 & 1676 \\
5 & 390 & 790 & 1231 & 1558 & 2068 & 1678 \\
6 & 510 & 744 & 1160 & 1559 & 2196 & 1686 \\
\tableline
\end{tabular}
\label{BG_experimental_data_white_resistor_6}
\end{table}

\begin{figure}[hp]
\includegraphics[width=15cm]{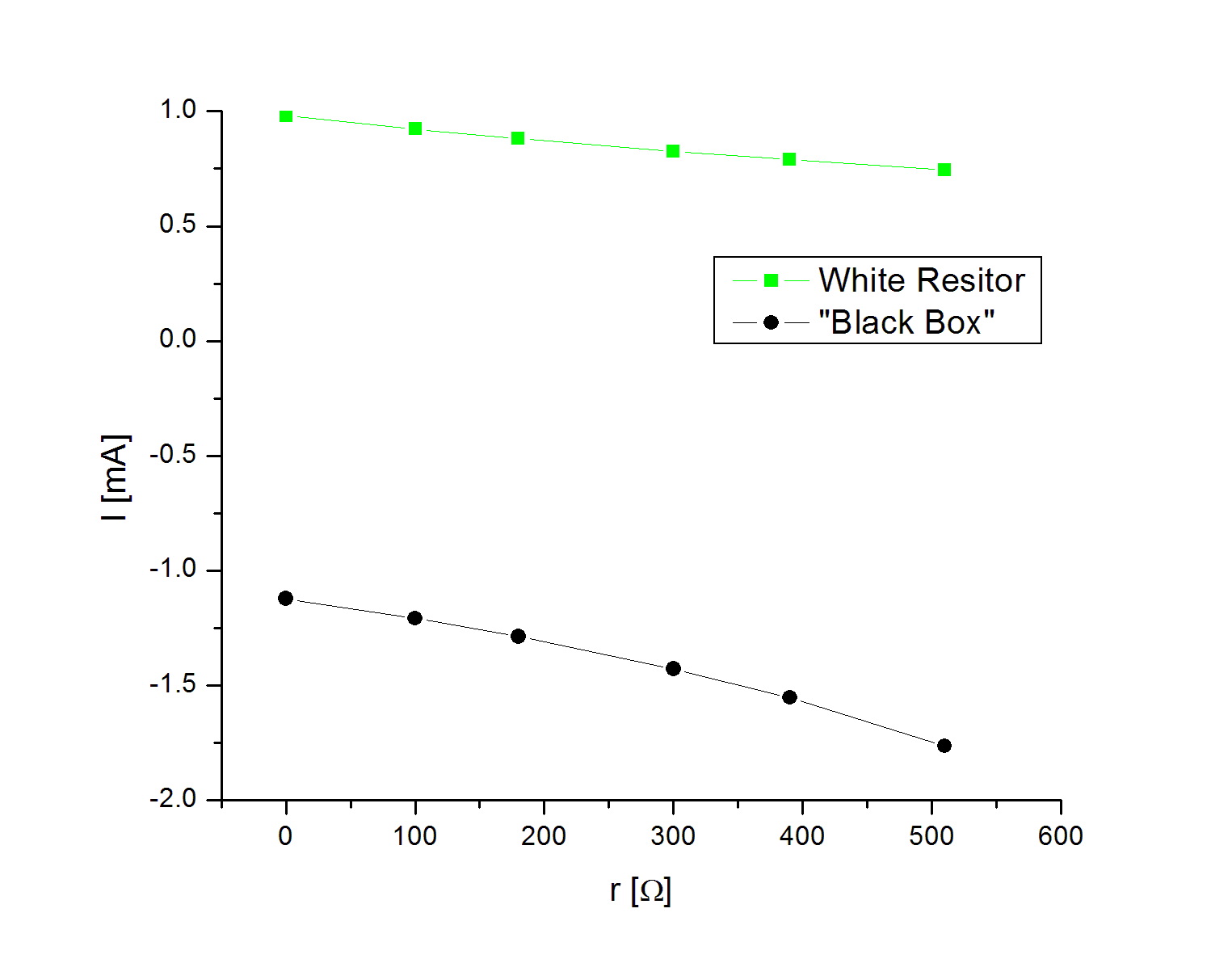}
\caption{
Токът $I$ през контура от Фиг.~\ref{BG_resistance-measurement}
като функция от допълнителните сеъпротивления $r_i.$
Черните точки са данните за ``черната кутия'' от
Табл.\label{BG_experimental_data_black_box_6},
а зелените триъгълници са данните за белия резистор от
Табл.\label{BG_experimental_data_white_resistor_6}.
Черната линия е фитиращата хипербола 
$I=\mathcal{E}/(-\overline{R}+r_i)$,
зелената линия е хиперболата
$I=\mathcal{E}/(R_\mathrm{WR}+r_i)$.
И в двата случая имаме намаляващи функции с отрицателен наклон.
Ако, обаче подразбираме само модул зависимостта 
$|I|=\mathcal{E}/|\overline{R}-r_i|$
разтяща функция в областта в която я изследваме.
}
\label{BG_I-r_plot}
\end{figure}

\begin{figure}[hp]
\includegraphics[width=15cm]{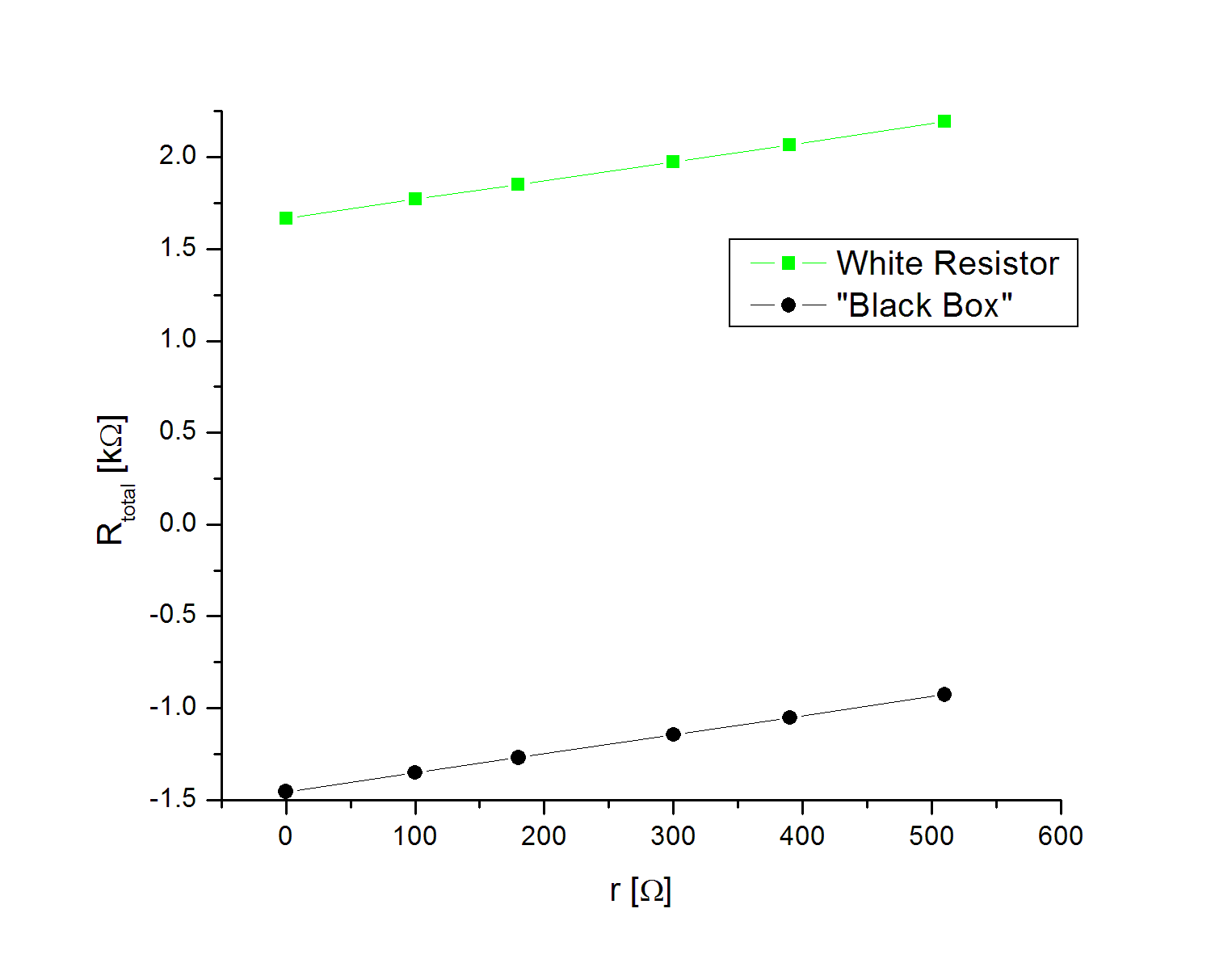}
\caption{
Пълното (тоталното) съпротивление 
$\mathcal{R}_\mathrm{tot}\equiv \mathcal{E}/I$ на контура от 
Фиг.~\ref{BG_resistance-measurement}
като функция от допълнително включваните съпротивления $r_i.$
Както във Фиг.~\ref{BG_I-r_plot}
Черните точки са данните за ``черната кутия'' от
Табл.\label{BG_experimental_data_black_box_6},
а зелените триъгълници са данните за белия резистор от
Табл.\label{BG_experimental_data_white_resistor_6}.
Черната права е линейната регресия  
$\mathcal{R}_\mathrm{tot}=-\overline{R}+r_i$,
а зелената права е  
$\mathcal{R}_\mathrm{tot,\,WR}=R_\mathrm{WR}+r_i$.
И в двата случая имаме разтящи функции с единичен наклон
$\mathrm{d}\mathcal{R}_\mathrm{tot}/\mathrm{d}r=1.$
Ако, обаче подразбираме само модул зависимостта 
$\mathcal{R}_\mathrm{tot}=|\overline{R}-r_i|$,
намаляваща функция за областта на използваните резистори.
Единичния наклон означава просто, че сме проверили, че
при последователно свързване съпротивленията се събират.
Точката с късо съединение $r_i=0$ дава съпротивлението
на ``черната кутия'' което не може да се измери с волтметър 
и съпротивлението на белия резистор което може да се измери с омметър.
}
\label{BG_R_tot_versus_r_i}
\end{figure}

\begin{figure}[hp]
\includegraphics[width=15cm]{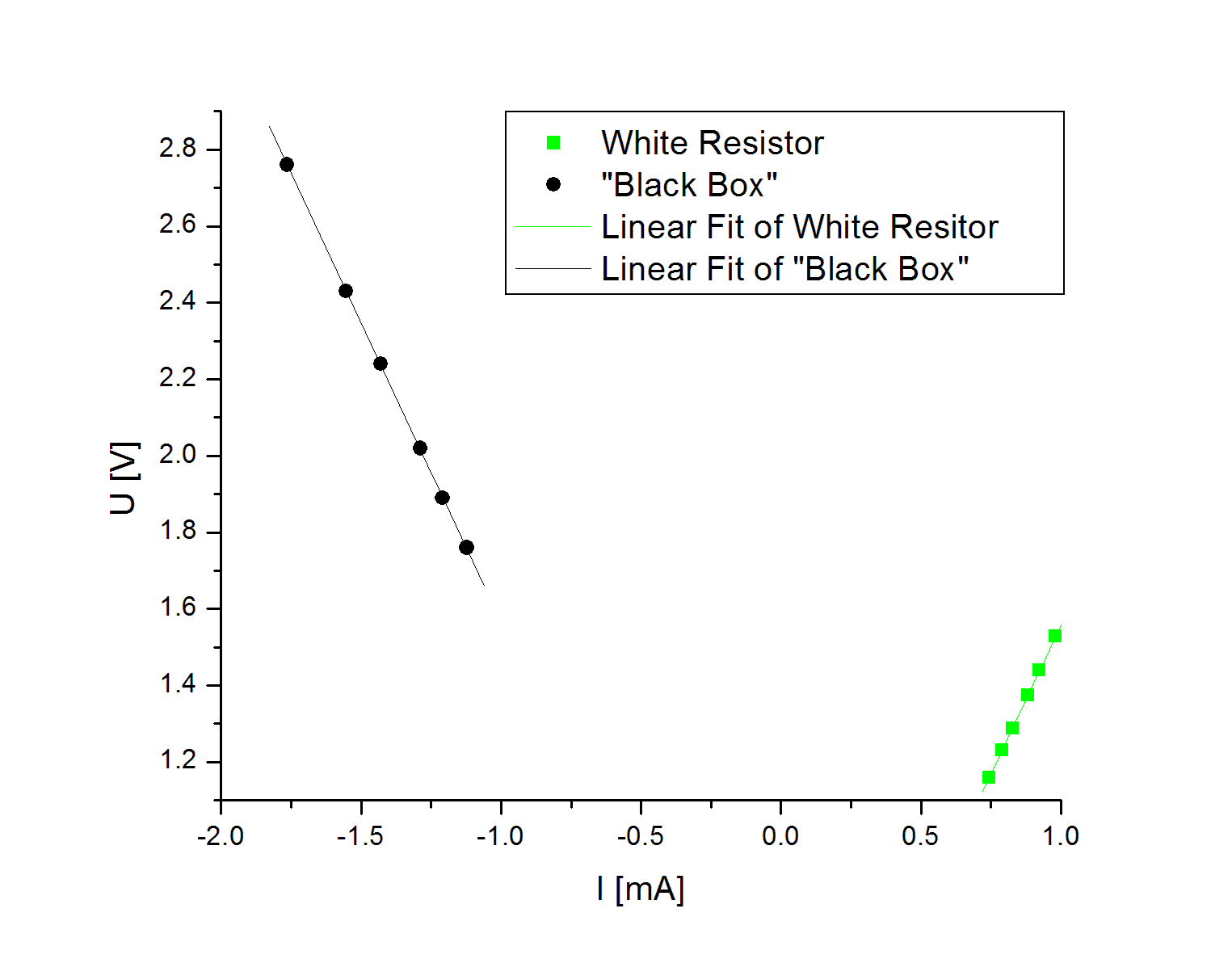}
\caption{Зависимост между ток и напрежение $U(I)$ 
за 6 точки от изследваните ``черната кутия'' и бял резистор. 
По абсцисата е даден тока $I$, 
а по ординатата напрежението
$U.$  
Експерименталната постановка е представена 
на Фиг.~\ref{BG_resistance-measurement},
a данните са взети от таблица~\ref{BG_experimental_data_black_box_6}
и таблица~\ref{BG_experimental_data_white_resistor_6}.
За белия резистор (зелените точки) 
наклона на ВАХ съответства на съпротивление 
$R_\textrm{WR}=\Delta U/\Delta I \approx 1.56\,\mathrm{k}\Omega$,
докато за ``черната кутия'' (черните точки) 
наклона е отрицателен и дава съпротивление
$R=\Delta U/\Delta I =-\overline{R}$, където параметърът
$\overline{R}\approx 1.56\,\mathrm{k}\Omega$ е положителен.
Символа $\Delta$ осначава разлика,
а номерата на точките описват 
номерата на резисторите от таблиците. 
Правите съответстващи на закона на Ом
са поличени чрез линейна регресия на експерименталните данни.
Отрицателният наклон на ВАХ на ``черната кутия''
може формално да се опише като резистор с отрицателно 
Омово съпротивление.
Съпротивлението описва триене и топлинна дисипация на енергията. 
Много интересни ефекти могат да се наблюдават в системи с отрицателно триене; представете си как биха ``затихвали'' трептенията на махало с отрицателно триене.
}
\label{BG_IV6}
\end{figure}

\begin{figure}[hp]
\includegraphics[width=15cm]{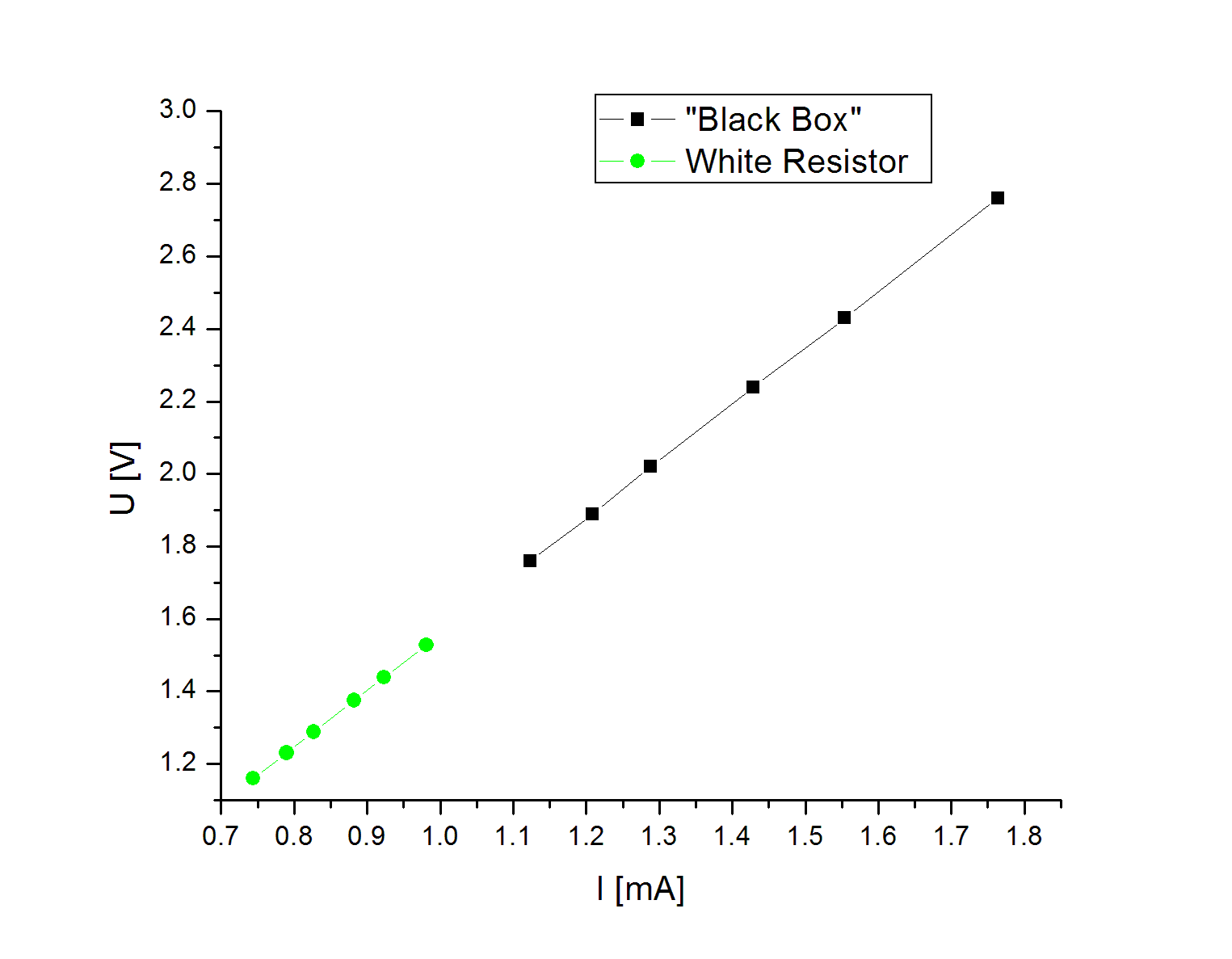}
\caption{Това са данните от Фиг.~\ref{BG_IV6}
които са насложени като на тока се отчита по модул.
Имаме точки от ВАХ $U(|I|)$.
Вижда се при това наслагване,
че черните точки и зелените триъгълници се описват с
една и съща права линия $U\approx (1.56\,\mathrm{k}\Omega)\times|I|.$
Това означава, че модула на съпротивлението на ``черната кутия'' 
$\overline{R}$
е приблизително равен на съпротивлението на белия резистор
$R_\mathrm{WR}$.
}
\label{BG_R_abs_R_New}
\end{figure}

\begin{figure}[hp]
\includegraphics[width=15cm]{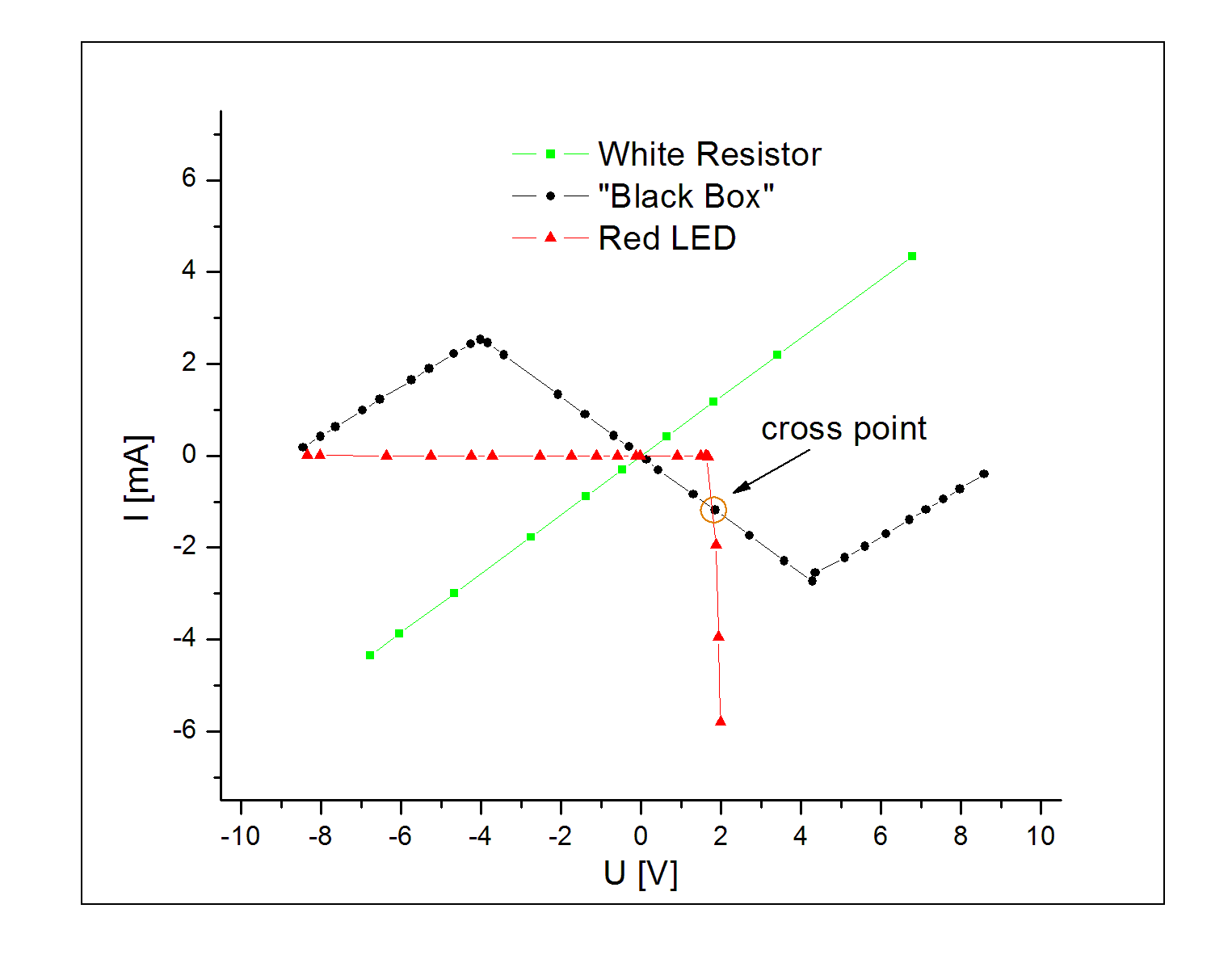}
\caption{Детайлна ВАХ $I(U)$ на 
``черната кутия'' (черните точки) 
от таблица~\ref{BG_experimental_results_black_box},
белия резистор (зелените точки) 
от таблица~\ref{BG_experimental_results_white_resistor}
и червения светодиод (червените точки) от таблица~\ref{BG_experimental_results_red_LED}.
По за разлика от Фиг.~\ref{BG_IV6} 
сега тока $I$ е даден по ординатата (верткалната ос) 
като функция от създалото го напрежение $U$
представено по абсцисата (хоризонталната ос).
Експерименталната постановка е представена 
на Фиг.~\ref{BG_resistance-measurement-I-V-curve}.
Най-проста е ВАХ на резистора
зелената права линия
$I_\mathrm{WR}=U/R_\mathrm{WR}$
точно съответства на закона на Ом
с положително съпротивление от 
$R_\mathrm{WR}\approx 1.56\,\mathrm{k}\mathrm{\Omega}$.
``Черната кутия'' има непрекъснатата ВАХ има $N$-образна форма
с два участъка отляво и отдясно с положителен наклон
$\mathrm{d}I/\mathrm{d}U=\dots\,\mathrm{k}\mathrm{\Omega}$;
символа $\mathrm{d}$ означава разлика когато кривината на разглеждания учатък е пренебрежима.
И най-важното ВАХ има един отрицателен сегмент по средата 
съответстващ на закона на Ом с отрицателно съпротивление
$\mathrm{d}I/\mathrm{d}U=-1/\overline{R},$ където
$\overline{R}\approx 1.56\,\mathrm{k}\mathrm{\Omega}$.
При този мащаб ВАХ на светодиода се описва приближено с една хоризонтална права с нулев ток при $U>U_c=\,\dots\mathrm{V}$}
и една стръмна почти права линия с много малко диференциално съпротивление
$\mathrm{d}I_\mathrm{LED}/\mathrm{d}U=1/r_\mathrm{LED},$ 
където 
$r_\mathrm{LED}=\dots\,\mathrm{\Omega}$.
Уравнението $I(U^*)=I_\mathrm{LED}(-U^*)=I^*$
описва работната точка $(U^*,\, I^*)=(\dots\mathrm{V},\,\dots\mathrm{I})$
при която ``черната кутия'' запалва светодиода с прав ток.
Тази работна точка е нанесена върху червената линия с оранжево кръгче.
Напрежението в тази работна точка е почти равно на напрежението на отпушване и запалване на светодиода $U^*\approx U_c$ и правия ток който тече през сжетодиода и черната кутия от постановката на Фиг.~\ref{BG_LEDNR} е $I^*\approx U_c/\overline{R}.$
\label{BG_IVdetail}
\end{figure}

\begin{table}[ht]
\caption{Напрежението подадено върху ``черната кутия'' $U$ и тока $I$ който преминава през нея за различни положения на плъзгача на потенциометъра
от експерименталната постановка начертана на 
Фиг.~\ref{BG_resistance-measurement-I-V-curve}.
}
\begin{tabular}{| r | r | r |}
\tableline
 i& $U$ [V] & $I$ [mA] \\
\tableline
1 & -8.45   & 0.18  \\
2 & -8.01  & 0.42  \\
3 & -7.64 & 0.63  \\
4 & -6.96  & 0.99  \\
5 & -6.53 & 1.23  \\
6 & -5.74  & 1.65  \\
7 & -5.30 & 1.90  \\
8 & -4.68 & 2.22  \\
9 & -4.26 & 2.43  \\
10 & -4.01  & 2.53  \\
11 &  -3.83 & 2.46  \\
12 &  -3.43 & 2.20  \\
13 &  -2.07 & 1.33  \\
14 &  -1.40 & 0.90  \\
15 &  -0.68 & 0.44  \\
16 &  -0.30 & 0.20  \\
17 &  0.14 & -0.08  \\
18 &  0.43 & -0.31  \\
19 &  1.31 & -0.84  \\
20 &  1.86 & -1.18  \\
21 &  2.71 & -1.73  \\
22 &  3.58 & -2.29  \\
23 &  4.29 & -2.73  \\
24 &  4.36 & -2.55  \\
25 &  5.10 & -2.22  \\
26 & 5.60  & -1.98  \\
27 & 6.13 & -1.70  \\
28 & 6.71 & -1.39  \\
29 & 7.13 & -1.17  \\
30 & 7.56 & -0.94  \\
31 & 7.98 & -0.72  \\
32 & 8.58 & -0.40 \\
\tableline
\end{tabular}
\label{BG_experimental_results_black_box}
\end{table}

\begin{table}[ht]
\caption{Напрежението $U$ върху белия резистор от 1.56 k$\Omega$ и тока $I$ за различни положения на плъзгача на потенциометъра
от постановката представена на Фиг.~\ref{BG_resistance-measurement-I-V-curve},
където ``черната кутия'' е заменена с белия резистор.}
\begin{tabular}{| r | r | r |}
\tableline
 i& $U$ [V] & $I$ [mA] \\
\tableline
1 & -6.77   & -4.35  \\
2 & -6.05 &  -3.88 \\
3 & -4.66 & -2.99  \\
4 & -2.76  & -1.77  \\
5 & -1.38 & -0.88  \\
6 & -0.47 & -0.30 \\
7 & 0.65 & 0.42 \\
8 & 1.82 & 1.17 \\
9 &  3.42 & 2.20 \\
10 &  6.78 & 4.35 \\
\tableline
\end{tabular}
\label{BG_experimental_results_white_resistor}
\end{table}

\begin{table}[ht]
\caption{Напрежението $U$ върху червения светодиод, тока $I$ който преминава през диода за различни завъртания на оста на потенциометъра.
При измерването 
схематично показано на 
Фиг.~\ref{BG_resistance-measurement-I-V-curve}
``черната кутия'' е заменена с червения светодиод.}
\begin{tabular}{| r | r | r |}
\tableline
 i& $U$ [V] & $I [\mathrm{A}]\qquad$ \\
\tableline
1 & -2.16 & -16.83 $\times\,10^-3$  \\
2 & -2.12 & -14.16 $\times\,10^-3$  \\
3 & -2.09 & -12.31 $\times\,10^-3$ \\
4 & -2.08 & -10.20 $\times\,10^-3$\\
5 & -2.02 & -7.90 $\times\,10^-3$ \\
6 & -1.99 & -5.80 $\times\,10^-3$ \\
7 & -1.94 & -3.95 $\times\,10^-3$ \\
8 & -1.88 & -1.94 $\times\,10^-3$ \\
9 & -1.63 & -0.01 $\times\,10^-3$ \\
10 & -1.678 & -25.8 $\times\,10^-6$  \\
11 & -1.620 & -9.6  $\times\,10^-6$  \\
12 & -1.492 & -2.2  $\times\,10^-6$ \\
13 & -0.914 & -0.9  $\times\,10^-6$ \\
14 &  0.020 & 0.1  $\times\,10^-6$ \\
15 &  0.132 & 0.2  $\times\,10^-6$ \\
16 &  0.589 & 0.6  $\times\,10^-6$ \\
17 &  1.105 & 1.1  $\times\,10^-6$ \\
18 &  1.743 & 1.7  $\times\,10^-6$ \\
19 &  2.52 & 2.4  $\times\,10^-6$ \\
20 &  3.72 & 3.6  $\times\,10^-6$ \\
21 &  4.24 & 4.1  $\times\,10^-6$ \\
22 &  5.25 & 5.1  $\times\,10^-6$ \\
23 &  6.37 & 6.1  $\times\,10^-6$ \\
24 &  8.03 & 7.7  $\times\,10^-6$ \\
25 &  8.34 & 8.8  $\times\,10^-6$ \\
\tableline
\end{tabular}
\label{BG_experimental_results_red_LED}
\end{table}

\begin{figure}[hp]
\includegraphics[width=10cm]{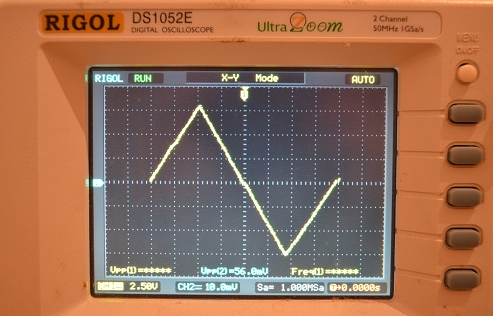}
\caption{N-образна ВАХ на ``черната кутия'' фотографирана от екрана на осцилоскоп. 
По хоризонтала е подадено напрежението $U$ създадено от трионообразен сигнал на генератор, 
а по вертикала се на осцилоскопа се подава напрежение 
пропорционално на тока $I$ през ``черната кутия''.}
\label{BG_N-shape_oscilloscope}
\end{figure}

\begin{figure}[hp]
\includegraphics[width=10cm]{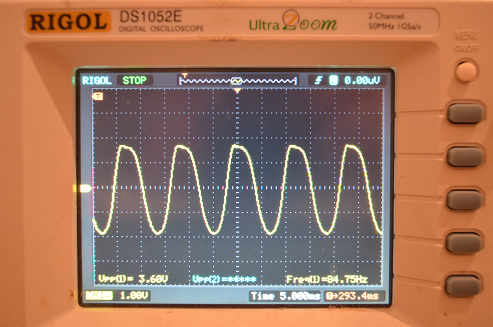}
\caption{Осцилациите на напрежението $U(t)$ на резонансния $LC$ контур 
от постановката показана на Фиг.\label{BG_LCNR}.  
Напрежението като финкция от времето $U(t)$ има почти синусоидална форма.
Само върха на синусоидата е пречупен от запалването на светодиода.}
\label{BG_sin_oscillator_oscilloscope}
\end{figure}

\begin{figure}[hp]
\includegraphics[width=10cm]{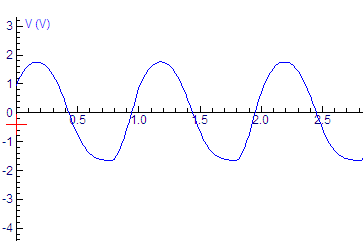}
\caption{Подобно на Фиг.~\ref{BG_sin_oscillator_oscilloscope}
oсцилации на напрежението $U(t)$ на резонансния $LC$ контур 
измерени чрез училищната система Coach-5. 
Синусоидата е леко деформирана от светодиода, който 
фиксира амплитудата на осцилациите; светът не е идеален.
За ниски честоти училищната система работи 
също толкова добре както електрониката за професионалисти.
Това е едно голямо постижение за учебната техника.
Възниква социалният въпрос: 
Как да се стумулира в образованието по физика
използването на наличните в училищата системи Coach-5 и Nova
и други системи за компютъризиран експеримент?}
\label{BG_sin_oscillator_oscilloscope_coach5}
\end{figure}

\begin{figure}[hp]
\includegraphics[width=10cm]{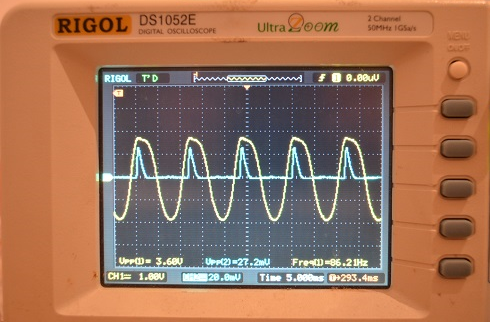}
\caption{Синусоидалната форма на напрежението $U(t)$ върху успоредно свързаните кондензатор и индуктивност е показано заедно с тока през светодиода $I_\mathrm{LED}(t)$ който е пропорционален на интензивността на излъчената светлина. Осцилациите в резонансния контур се възбуждат от отрицателното съпротивление, а амплитудата им се ограничава от нелинейнейната електрическа проводимост на диода.}
\label{BG_oscillations_and_diode_current_oscilloscope}
\end{figure}

\begin{figure}[hp]
\includegraphics[width=10cm]{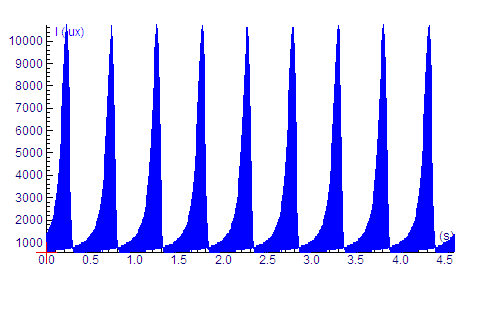}
\caption{Пулсациите на светлината от светодиода като функция от времето
получени със сензора за светлина на системата Coach-5.
В добро приближение интенсивността на излъчената светлина е пропрорционална тока през светодиода, затова тази фигура и 
Фиг.~\ref{BG_oscillations_and_diode_current_oscilloscope} се различават само по мащаба на осите си.
}
\label{BG_LED_pulsations_versus_t_Coach5_photo_sensor}
\end{figure}

\begin{figure}[hp]
\includegraphics[width=10cm]{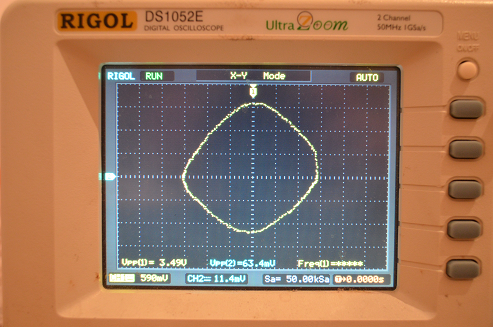}
\caption{Фазов портрет на осцилациите на $LC$ резонатора възбудени чрез отрицателното съпротивление на ``черната кутия.'' 
Верикалната е развивка е пропорционална на напрежението,
а по хоризонтала тока. 
Така получаваме портрета на осцилациите в I-V разнината.
Ако мислено си представим времето начертано перпендикулярно на екрана
осцилациите се описват от винтова линия, а в равнината I-V се вижда фазата на това въртене. Когато модула на отрицателното съпротивление 
$\overline{R}$ е по-голям,
формата на осцилациите е по близка до синусовата,
фазовия портрет на осцилациите е по-близък до окръжност,
но интензивността на светенето на светодиода намалява.
В този случай в техниката се казва, 
че контура има меко възбуждане.
}
\label{BG_phase_portrait}
\end{figure}

\newpage
\subsection{Решение на теоретичната задача}
\subsubsection{Извод формулата за съпротивлението}
От уравнението за изходното напрежение $U_0=(U_+-U_-)G$ 
(виж Фиг.~\ref{BG_circuit})
изразяваме разликата на входните напрежения 
\begin{equation}
U - U_-= U_{0}/G\approx 0, 
\end{equation}
като отчитаме, 
че входът на схемата е свързан с точка (+) и $U_{+}=U.$ 
Тъй като
$1/G=10^{-5} \ll 1/100$ 
в рамките на допустимата точност от 1\% получаваме, 
че двете входни за усилвателя напрежения 
в точките (--) и (+) са равни 
$U_{-}=U.$  
Приближеното уравнение $U_{+}\approx U_{-}$ 
изведено при условието $|1/G|\ll1$ се нарича златно правило на операционните усилватели.

От точката (--) към “земята” се спуска ток $U_{-}/R_1$ различен от входния.
Този ток тръгва от
точка с напрежение $U_{0}$ 
и преминава преди това през съпротивлението $R_2$.
Съгласно закона на Ом, 
напрежението в точка (0) ще получим като умножим този ток
с последователно свързаните съпротивления $R_1+R_2$ 
и така получаваме 
\begin{equation}
U_0=(R_1+R_2)\frac{U_{-}}{R_1}=\left(1+\frac{R_2}{R_1}\right)U.
\label{BG_U_0}
\end{equation}
Горната формула може да се интерпретира и 
като резултат за делител на напрежение:
напрежението $U_{-}$ се разделя от общото напрежение $U_{0}$ 
в пропорция
$R_1/(R_1+R_2)$ и така 
имаме $U_{-}=U_{0}R_1/(R_1+R_2).$

Нека сега проследим как входният ток 
$I$ се спуска през съпротивление $R_3$ от
входното напрежение $U$ до напрежение $U_0$  
под действие на разликата на потенциалите $U-U_0$. 
Във формулата за тока $I=(U-U_0)/R_3$ 
заместваме получения израз за $U_0$
от уравнение \eqref{U_0} и получаваме
\begin{equation}
I=\frac{U-U_0}{R_3}
 =\left[1-\left(1+\frac{R_2}{R_1}\right)\right]\frac{U}{R_3}
=-\frac{R_2}{R_1 R_3}U =\frac{U}{R},
\end{equation}
откъдето за входното съпротивление на веригата извеждаме 
\begin{equation}
R=\frac{U}{I}=-\frac{R_1R_3}{R_2}=-\overline{R},
\qquad \overline{R}\equiv R_1R_3/R_2>0.
\end{equation}
За удобство отрицателното съпротивление $R$. се параметризира с положителния  параметър $\overline{R}$.

Заместването в горната формула на 
$R_1=R_2=10\,\mathrm{k}\Omega$ 
и 
$R_3=1.5\,\mathrm{k}\Omega.$ 
дава на края
$R=-1.5\,\mathrm{k}\Omega.$ 
Знакът показва, че анализираната схема има формално отрицателно съпротивление, например при постоянно подадено напрежение токът има противоположна посока на тока при обикновенните резистори. 

Ние изведохме този резултат предполагайки,
че напрежението на изхода на операционния усилвател 
$|U_0|$ е по-малко от захранващото напрежение $\mathcal{E}_s.$
Тогава напрежението на входа $U$ трябва да бъде по модул по-малко от максималното напрежение
\begin{equation}
U_\mathrm{max}=\frac{R_1}{R_1+R_2}\mathcal{E}_s.
\end{equation}
При избраните параметри на задачата $U_\mathrm{max}/\mathcal{E}_s=\frac12.$
Извън тази област на малки напрежения $|U|<U_\mathrm{max}$ 
ВАХ има положителeн наклон или както още се казва положително диференциално съпротивление равно на $R_3$.

\subsubsection{Как отрицателното съпротивление запалва светодиода}

Ще разкажем накратко как отрицателното съпротивление генерира осцилации в успореден резонансен контур с резонансна честота
\begin{equation}
f_\mathrm{res}=\frac1{2\pi \sqrt{LC}}
\end{equation}
и голямо съпротивление при резонанса 
\begin{equation}
R_\mathrm{res}=\frac{L/C}{r},
\end{equation}
където $r$ е малкото съпротивление на индуктивността;
детайлната теория ще бъде дадеда в отделна секция.

Когато отрицателната проводимост $1/R$ е по голяма по модул от 
проводимостта резонансния контур $1/R_\mathrm{res}$
осцилациите в контура не затихват а се усилват.
Чрез съпротивленията и параметрите на задачата
критерия за възникване на осцилациите може да се запише 
като
\begin{equation}
\label{BG_criterion_oscillations}
\overline{R}=\frac{R_1R_3}{R_2}<R_\mathrm{res}=\frac{L/C}{r}.
\end{equation}
Амплитудата на осцилациите $U_m$ нараства докато малко превиши
напрежението на отпушване и запалване на светодиода $U_c$.
Средноквадратичното напрежение на установилите се осцилаци е
$\tilde{U}=U_c/\sqrt{2}.$
Тогава мощността която се консумира от отрицателното съпротивление 
е 
\begin{equation}
P=\tilde{U}^2/\overline{R}\approx \frac{U_c^2}{2\overline{R}}.
\end{equation}
При голям Q-фактор на резонансния контур
\begin{equation}
\mathcal{Q}\equiv\frac{\sqrt{L/C}}r
=\frac{R_\mathrm{res}}{\sqrt{L/C}}
\gg1
\end{equation}
тази мощност дисипира върху мигащия светодиод
и чрез параметрите на задачата средната мощност която консумира 
светодиода може да се препише като
\begin{equation}
P_\mathrm{LED}=\tilde{U}^2/\overline{R}
\approx\frac{U_c^2}{2\overline{R}}=\frac{U_c^2R_2}{2R_1R_3}.
\end{equation}

При работа светодиода се нагрява, но
с голям коефицент на полезно действие $\eta$ 
тази мощност се трансформира в светлина.
Може да пресметнем средния брой фотони 
излъчени за единица време 
$\dot{\mathcal{N}}= P_\mathrm{LED}/\eta h \nu_\gamma,$ 
където е константата на Планк, 
а $\nu_\gamma$ е честотата на светлината. 
Енергията на фотона е близка до енергията на спиращото напрежение 
$h \nu_\gamma\approx q_eU_c,$ където $q_e$ е заряда на електрона.

Излъчването не е равномерно, а е концентрирано в моментите когато променливото напрежението запалва светодиода. 
Честотата на светлинните импулси e разбира сe резонансната честота
$f_\mathrm{res}$ и ако светодиода се движи със скорост $v$, 
разстоянието между светлите петна е $l=v/f_\mathrm{res}$;
например при $v=1\,\mathrm{m/s},$ 
$f_\mathrm{res}=100\,\mathrm{Hz},$ 
$l=1\,\mathrm{cm}$, което би дало една оценка за честотата.
Ако светодиода вибрира механично с известна честота $f_\mathrm{mech}$ 
при целочислено отношение $f_\mathrm{res}/f_\mathrm{mech}$ се
виждат стоящи цветни петна.

Друг начин да определим честотата е да я чуем със слушалки. 
В този случай трябва да знаете честотите на музикалните тоновете.

Kритерия за електростатична неусточивост, когато към отрицателното съпротивление е включен външен резистор $R_\mathrm{ext}$
е същия както за възбувдане на осцилациите Уравнение~(\ref{BG_criterion_oscillations}).
Когато вънщното съпротивление е по-голямо от модула на отрицателното съпротивление 
\begin{equation}
\label{BG_criterion_stability}
\overline{R}<R_\mathrm{ext}
\end{equation}
токовете във веригата се усилват,
докато напрежението на изхода на операционния усилwател достигне захранващото напрежение $U_0=\mathcal{E}_s$ или светодиода се запали $U^*=U_c$.
Когато светодиода свети с постоянен ток напрежението му е малко по-голямо
от напрежението на запалване $U_c$ мощността $U_c^2/\overline{R}$
e почти два пъти по голямя от мощността при пулсиращия режим.
По прецизен анализ дава 
\begin{equation}
\frac{U^*-U_c}{U_c}=1/(\overline{R}/r_{_\mathrm{LED}}-1)
\approx r_{_\mathrm{LED}}/\overline{R},
\end{equation}
където 
$r_{_\mathrm{LED}}$
е диференциалното съпротивление на светещия диод, което параметризира
наклона на почти линейната му ВАХ при големи токове.

Когато се опитваме да измерим отрицателното съпротивление с волтметър 
той показва значително напрежение.
Това напрежение $U_\mathrm{max}\approx\frac12\mathcal{E}_s$ остава 
и при свободни електроди на черната кутия 
и затова батериите в нея трябва да се изключват.

Чрез проведения теоретичен анализ или чрез експериментално изследвване
ние получихме, че ``черната кутия''
се държи като отрицателно съпротивление от около $-1.5$~k$\Omega$. 
Статии за отрицателно съпротивление и неговите приложения\cite{NegativeResistanceWiki}
могат лесно да се намерят чрез Интернет,  
ако потърсите в \textit{Google}
\textit{Negative resistance} или
\textit{Negative Impedance Converter(NIC)}.
Отворете учебник по електроника и намерете описанието как се генерират 
хармонични електрични осцилации.
В чертежите лесно ще намерите къде се намира отрицателното съпротивление.
От книгите по физика бихме препоръчали монографията на Пипард,
\cite{Pipard:07} (1978 г.)
а за историческа справка вижте например
оргиналната статия Линвил\cite{Linvill:53} (1953 г.)
и патента на Дебу\cite{Deboo:68} (1968 г.).

В заключение искаме да отбележим, че теоретичната задача е просто теорията на предложените  експерименти. 
И както анализа на информацията от черните кутии 
помага за повишаване на надеждността на авиацията,
така и проведеното решение ни помага да разберем един много общ
принцип за генериране на трептения 
чрез отрицателно триене.


%
%
%

\newpage
\subsection{Упътване за консултантите към задачата за домашно. Електроника за напреднали}
Тук ние ще разгледаме как това съпротивление може да бъде измерено с омметър, какви са условията за устойчивост и приложимост 
на такива отрицателни съпротивления в електронни схеми.
Тази секция е написана не за учениците а за консултантите и колеги които подготвят подобни задачи, затова обясненията са на студентско ниво.
\subsubsection{Използвана математика. Операционно смятане}
 Очакваме читателя на тази секция да знае, например производната и степенния ред на експоненциалната функция
\begin{equation}
\label{BG_exp}
\mathrm{d}_t\mathrm{e}^{st}=s\mathrm{e}^{st},
\qquad
\mathrm{d}_t\equiv\frac{\mathrm{d}}{\mathrm{d} t},
\qquad
\exp(x)=\mathrm{e}^{x}=1+x+\frac12x^2+\frac1{3!}x^3
+\dots+\frac1{n!}x^n+\dots,
\end{equation}
формулата за реда на Тейлор-Маклорен
\begin{eqnarray}
\label{BG_tail}
\left.\exp(\tau\mathrm{d}_t)f(t)\right|_0
&=&\left.\left(1+\tau\mathrm{d}_t+\frac{\tau^2}2\mathrm{d}_t^2
+\frac{\tau^3}{3!}\mathrm{d}_t^3
+\dots+\frac{\tau^n}{n!}\mathrm{d}_t^n+\dots\right)f(t)\right\vert_{t=0}
\nonumber\\
&=&f(0)+\tau f^{\prime}(0)+\frac{\tau^2}2 f^{\prime\prime}(0)
+\frac{\tau^3}{3!} f^{\prime\prime\prime}(0)
+\dots+\frac{\tau^n}{n!} f^{(n)}(0)+\dots\nonumber\\
&=&f(\tau)
\end{eqnarray}

За формула~(\ref{BG_exp}) се казва, че експонентата е собстствена функция на оператора на диференциране и тя е приложима и ако собствената стойност $s\equiv\mathrm{j}\omega$ e чисто имагинерна. 
За аналитични функции от оператора на диференциране спрямо времето имаме 
\begin{equation}
f(\mathrm{d}_t)\mathrm{e}^{st}
=f(\mathrm{d}_t)\mathrm{e}^{\mathrm{j}\omega t}
=f(\mathrm{j}\omega)\mathrm{e}^{j\omega t}
=f(\mathrm{j}\omega)\mathrm{e}^{st},
\end{equation}
например
\begin{equation}
L\mathrm{d}_t\mathrm{e}^{st}=Ls \mathrm{e}^{st}
=L\mathrm{j}\omega\mathrm{e}^{\mathrm{j}\omega t},
\qquad
\frac1{C\mathrm{d}_t}\mathrm{e}^{st}=\frac1{Cs}\mathrm{e}^{st}
=\frac1{C\mathrm{j}\omega}\mathrm{e}^{\mathrm{j}\omega t}.
\end{equation}
В контекста на приложение в теорията на електричнитет вериги
тези операциите на диференциране  $L\mathrm{d}_t$ и интегриране $1/C\mathrm{d}_t$ приложени към експоненти от времето дават импедансите на индуктивност $Z_L=\mathrm{j}\omega L$ и капацитет $Z_C=1/C\mathrm{j}\omega$.
Така ние преговорихме освовните формули на електритехниката за връзката между тока $I$ напреженията за индуктивност, омово съпротивление и кондензатор
\begin{equation}
U_L=L\mathrm{d}_tI,\qquad
U_R=RI,\qquad
\mathrm{d}_tU_C=CI.
\end{equation}

\subsubsection{Стандартен модел на операционен усилвател}
Сега можем да анализираме стандартния модел на операционен усилвател,
който основно се ползва в електрониката.\cite{eCircuit_Center,Allen:11}
Входната разлика на напрежение $U_+-U_-$ се подава на един генератор за ток който създава $I_a=(U_+-U_-)/r_a;$ параметарът $r_a^{-1}$ има размерност проводимост.
Токът $I_a$ се подавa на успоредно свързани кондензатор и резистор със
сумарен импеданс
\begin{equation}
Z_a=\frac{R_aZ_a}{R_a+Z_a}
=\frac{R_a\frac1{\mathrm{j}\omega C_a}}{R_a+\frac1{\mathrm{j}\omega C_a}}
=\frac{R_a}{1+\mathrm{j}R_aC_a\omega}
=\frac{R_a}{1+\tau_a s},
\qquad \tau_a\equiv R_aC_a,
\qquad \hat{Z}_a\equiv\frac{R_a}{1+\tau_a \mathrm{d}_t}
\end{equation}
и върху импеданса се създава напрежение $U_a=Z_aI_a.$ 
Това напрежение управлява генератора на изходното напрежение напрежение на операционния усилвател $U_0=U_a$ и заместването дава
крайния общ резултат за функцията на предаване на операционния усилвател
изразена в операторна форма
\begin{eqnarray}
\label{BG_GeneralOpAmpEq}
&& U_+(t)-U_-(t)=\left(\frac{1}{G_0}
+\tau_0\frac{\mathrm{d}}{\mathrm{d}t}\right) U_0(t),\qquad
U_0(t)=G({\mathrm{d}_t})\left[U_+(t)-U_-(t)\right],\\&&
\hat{G}^{-1}=\frac{1}{G_0}+\tau_0\mathrm{d}_t,
\quad G({\mathrm{d}_t})\equiv\frac{G_0}{1+\tau_a\mathrm{d}_t}.
\quad G_0\equiv \frac{R_a}{r_a},
\quad \tau_a\equiv R_a C_a,
\quad \tau_0\equiv r_a C_a=\frac{\tau_a}{G_0}
\end{eqnarray}
или като функция от честотата  
$G(\omega)=G_0/(1+\mathrm{j}\omega\tau_a).$

При по детайлен анализ на операционните усилватели се получава
по-прецизна формула за функцията на предаване
\begin{eqnarray}
\label{BG_DetailedOpenOpAmpGain}
\tilde{G}({\mathrm{d}_t})\equiv\frac{G_0}{1+\tau_a\mathrm{d}_t}
\frac{P_m({\mathrm{d}_t})}{P_n({\mathrm{d}_t})},\quad
P_m(s)=1+a_1s+a_2s^2+\dots+a_ms^m,\quad
P_n(s)=1+b_1s+b_2s^2+\dots+b_ns^n,
\end{eqnarray}
като коефицентите на Паде (Pad\'e) апроксимантата се определят от хардуерната реализация на операционния усилвател, трансистори, кондензатори и кондензатори. 
Тези апроксиманти са имплицитно заложени в SPICE модела на операционните усилватели.
Типични стоности на параметрите на операционните усилватели са
$G_0\sim 10^6$ $\tau_a^{-1}\sim 100\,\mathrm{rad/sec}.$ 
За използания в нашата схема операционен усилвател TLO71\cite{TL071} SPICE модела има параметри $R_a=\dots,$ $r_a=\dots,$ $C_a=\dots,$ което дава \dots

За достатъчно големи честоти $\omega R_a C_a\gg1$ 
и бързи изменения на напрежението от уравнение~(\ref{BG_GeneralOpAmpEq}) получаваме най-важното уравнение за операционните усилватели 
\begin{equation}
\label{BG_MasterOpAmpEquation}
\tau_0\mathrm{d}_t U_0 \approx U_+-U_-,
\end{equation}
което описва практически всички високочестотни приложения.

За променливи токове най-често се работи с имагинерни експоненти,
токове и напрежения $\propto \mathrm{e}^{j\omega t}$ 
В този случай заместваме $s=j\omega$ и
главното уравнение на операционните усилватели приема вида
\begin{equation}
G(\omega)=\frac1{j\omega\tau_0}=\frac1{s\tau_0},\qquad |G|\gg1.
\end{equation}
Това уравнение описва работата на операционните усилватели във високочестотни устройства, усилватели или активни филтри например.
В спецификациите (описанията) на операционните усилватели (OpAmp)
често се дава и кросовер честотата $f_\mathrm{crossover}=1/(2\pi\tau_0)$
и $\omega_0\equiv1/\tau_0$.

Като просто следствие на главната формула за усилването на операционните усилватели (\ref{BG_MasterOpAmpEquation}) се получават известните формули за усилването $\Upsilon=U_\mathrm{out}/U_\mathrm{in}$ на инвертиращия и неинвертиращия усилвател\cite{ADA4817,Kobayashi:90}
\begin{eqnarray}&&
\Upsilon_\mathrm{inv}(s)=
-\frac{R_f}{(R_f+r_g)\tau_0s+r_g}
=-\frac{R_f/r}{1+(1+R_f/r)(1/G)},
\\&&
\Upsilon_\mathrm{non}(s)=\frac{R_f+r_g}{(R_f+r_g)\tau_0s+r_g}
=\frac{1}{r/(R_f+r)+(1/G)},\\&&
1/G=1/G_0+\tau_0s\approx\tau_0s=\mathrm{j}\omega\tau_0,
\qquad U(t)\propto \mathrm{e}^{st},
\end{eqnarray}
където $R_f$ е съпротивлението на обратната връзка, а $r_g$ е малкото входно съпротивление определящо усилването. 
Числото $s$ е собствена стойност на оператора за диференциране по време
$\mathrm{D}=\hat{s}=\mathrm{d}_t.$
Oснователят на операционното смятане\cite{OpCalc} Оливър Хевисайд
използва означението D.
В операторен вид стандартното уравнение за усилването на операционните усилватели (\ref{BG_MasterOpAmpEquation}) дава просто една интегрираща верига
\begin{equation}
\hat{G}(\hat{s})=\omega_0\hat{s}^{-1},\qquad
U_0(t)=\hat{G}U_{\pm}(t),\qquad U_{\pm}\equiv U_+-U_-.
\end{equation}
Стандартно в смисъл, че резултатите от прилагането на тази формула към теорията на електрични вериги със операционни усилватели се дават в инженерни приложения без да се цитира извод в книги написани на езика 
на физиката, която е основа на всички инженерни науки. 
Ние също не можахме да намерим в интернет подходящ учебник по физика или електроника измежду ръководствата за софтуер и хардуер, 
но и фирмите производители на операционни усилватели са в същото положение и в специфицациите им не се посочват монографии или учебници, където са дадени основните уравнения на операционните усилватели.

\subsubsection{Отрицателно съпротивление и критерия за електростатична устойчивост}

Нека повторим извода на отрицателното съпротивление като отчетем и честотната му зависимост чрез честотната зависимост на операционния усилвател.
От ``чертежа на Фиг~\ref{BG_circuit} се вижда, че:'' 
\begin{equation}
U_+=U,\quad 
U_{+}-U_{-}=U_0G^{-1},\quad
\frac{U_{-}}{R_1}=\frac{U_0}{R_1+R_2},\quad
U_0\in(-\mathcal{E}_s,\,\mathcal{E}_s).
\end{equation}
Eлементарно заместване тази система дава
\begin{equation}
U_0=\frac{U}{\frac{R_1}{R_1+R_2}+G^{-1}}
\end{equation}
и за входния ток получаваме
\begin{equation}
I=\frac{U-U_0}{R_3}
=\frac{U}{R_3}\left[1-\frac1{\frac{R_1}{R_1+R_2}+G^{-1}}\right]
=\frac{U}{Z},
\end{equation}
където
\begin{equation}
Z=\frac{R_3}{1-\frac1{\frac{R_1}{R_1+R_2}+G^{-1}}}
\end{equation}
е търсения честотно зависим импеданс на отрицателното съпротивление.
За реципрочния усилване на операционния усилвател 
общото уравнение~(\ref{BG_GeneralOpAmpEq}) дава в операторна форма
\begin{equation}
\hat G^{-1}=\frac{1}{G_0}
+\tau_0\frac{\mathrm{d}}{\mathrm{d}t}
\end{equation}
или за експоненциално зависещи от времето напрежения 
$U\propto\mathrm{e}^{st}$
\begin{equation}
\hat G^{-1}=G_0^{-1}+\tau_0s
=G_0^{-1}+\mathrm{j}\omega\tau_0.
\end{equation}
За реални честоти $s=\mathrm{j}\omega.$
Така получаваме честотната зависимост на отрицателния импеданс 
конветор \textit{(Negative-Impedance Converter, NIC)}
\begin{equation}
Z=\frac{R_3}{1-\frac1{\frac{R_1}{R_1+R_2}+G_0^{-1}+\tau_0s}}
=\frac{R_3}{1-\frac1{\frac{R_1}{R_1+R_2}+G_0^{-1}+\mathrm{j}\omega\tau_0}}.
\end{equation}

Когато отрицателното съпротивление е свързано 
с външно съпротивление $R_\mathrm{ext}$ 
сумарната проводимост е нула
\begin{equation}
\label{BG_EigenFrequenz}
1/Z+1/R_\mathrm{ext}=0,
\end{equation}
а в системата имаме собствено напрежение $U.$
Това уравнение за собственните честоти има явно решение
\begin{equation}
\tau_0s+G_0^{-1}=\frac1{1+\frac{R_3}{R_\mathrm{ext}}}
-\frac1{1+\frac{R_2}{R_1}}.
\end{equation}
Това уравнение е приложимо за произволна честотна зависимост на импедансите, но за реални съпротивления и пренебрежимо $G_0^{-1}$ инкремента $s$ е положителен, ако $R_3/R_\mathrm{ext}<R_2/R_1$ или
\begin{equation}
\label{BG_instab}
R_\mathrm{ext}>\overline{R}\equiv{R_1R_3}{R_2}.
\end{equation}
С други думи, ако към отрицателното съпротивление бъде включено външно съпротивление по голямо от модула на отрицателното съпротивление при нулева честота системата е неустойчива и напрежението нараства докато не бъде ограничено от нелинейни ефекти.
В нашия пример $\overline{R}=1.5\,\mathrm{k}\Omega$,
а волтметрите имат поне $R_\mathrm{V}=1\,\mathrm{M}\Omega$ 
вътрешно съпротивление и
системата е електростатично неустойчива $R_\mathrm{V}>\overline{R}$. 
Затова омовете на ``черната кутия'' немогат да се измерят с омметър.
Ако включим към ``черната кутия'' волтметър той ще покаже значително напрежение, но ако превключим мултиметъра като амперметър с вътрешно съпротивление от около $R_\mathrm{A}=100\,\Omega$ за най чувствителния обхват, то $R_\mathrm{A}<\overline{R},$ веригата е електростатично устойчива и ампреметърът показва пренебрежим ток. Потърсете в Интернет например
\textit{stability, negative impedance converter} и измежду спама може да намерите нещо, ново, интересно и полезно.

Ако искаме все пак да използваме омметър за измерване на отрицателното съпротивлението на черната кутия в нея трябва да има монтиран сдвоен ключ, който разменя (+) и (-) входа на операционния усилвател.

При инвертирани входове се инвертира и критерия за електростатична неустойчивост. Инвертираното отрицателно съпротивление е устойчиво при големи външни съпротивления и обратно губи устойчивост и може да генерира осцилации в последователен $LC$ резонатор с малко съпротивление $r$ в резонанса обусловено от съпротивлението на индуктивността.

За сложни електронни схеми 
с отрицателен коннвертор на импеданса
уравнението за аналитично продължените импеданси
$R_\mathrm{ext}=-Z$
дава собственните честоти на системата и апроксимира
секулярното уравнвние на електронната схема изведено от $SPICE$-модела.

Нека сега разгледаме възбуждането на резонансни трептения във високочестотен успореден резонансен контур със отрицателно съпротивление.

\subsubsection{Осцилации в успоредния резонансен контур}

Нека припомним основните понятия при анализа на резонансен трептящ кръг 
с капацитет $C$ индуктивност $L$ 
и собственна честота на резонансни трептения 
$\omega_0=1/\sqrt{LC}.$ 
Ако амплитудата на напрежението върху кондензатора е $U_m$
амплитудата на заряда който се натрупва върху кондензатора е 
$Q_m=CU_m,$
a амплитудата на тока през индуктивността е 
\begin{equation}
I_m=\omega_0Q_m=\omega_0 C U_m=\frac{U_m}{\sqrt{L/C}}.
\end{equation}
Ако отчетем малкото съпротивление на индуктивността $r\ll\sqrt{L/C}$
средната мощност която дисипира върху индуктивността е
\begin{equation}
P_r=\frac12rI_m^2= \frac{U_m^2}{2R_\mathrm{res}}
,\qquad R_\mathrm{res}\equiv \frac{L/C}{r}.
\end{equation}
Тази мощност която дисипира като топлина отделена върху индуктивността
трябва да се компенсира от мощността която се получава в трептящия кръг,
ако успоредно на кондензатора е включено отрицателно съпротивление
\begin{equation}
P_{\overline{R}}=\frac12U_m^2/\overline{R}.
\end{equation}
След съкращаването на амплитудата на напрежението,
условието за възникване на осцилации в трептящия контур 
$P_{\overline{R}}>P_r$ се чете като
\begin{equation}
1/\overline{R}>\frac1{R_\mathrm{res}}=\frac1{r\mathcal{Q}^2},
\qquad \mathcal{Q}\equiv\frac{\sqrt{L/C}}r
=\mathcal{Q}_\parallel\equiv\frac{R_\mathrm{res}}{\sqrt{L/C}}
\gg1
\end{equation}
и има простата интерпретация:
успоредно свъзаната отрицателната проводимост $1/\overline{R}$  
трябва да превишава на малката проводимост на успоредния
резонансен контур $1/R_\mathrm{res}$.
При резонансната честота $\omega=\omega_0$ 
импеданса на успоредно свързаните кондензатор и индуктивност 
е много голям
\begin{equation}
Z_\parallel(\omega_0)
=\frac{Z_CZ_L}{Z_C+Z_L}
=\left(\mathrm{j}\omega_0C+\frac1{\mathrm{j}\omega_0L+r}\right)^{-1}
=(1-\mathrm{j}\mathcal{Q}^{-1})r\mathcal{Q}^2
\approx r\mathcal{Q}^2=R_\mathrm{res}\gg r.
\end{equation}
Малката стойност на съпротивлението на индуктивността $r$
осигурява както висока стойност на фактора на качество $\mathcal{Q}$
и голямия импеданс на резонансния контур
така и критерия за електростатична стабилност 
неравенство (\ref{BG_instab}).
При свободни трептения на резонансен контур енергията намалява 
процентно с $1/\mathcal{Q}$ на единица фаза и за контури с високо качество 
$\mathcal{Q}^{-1}\ll1$ е малкия параметър на теорията.

Сумарно за възбуждане на електрични трептения в нашата схема 
трябва да са изпълнени неравенствата
\begin{equation}
r<\overline{R}<R_\mathrm{res}
\end{equation}
или ако изразим величините с параметрите на схемата
\begin{equation}
r<\frac{R_1R_3}{R_2}<\frac{L}{rC}.
\end{equation}
Съпротивленията $R_1$, $R_2$, $R_3$,
както операционния усилвател и захранващите батерии
се монтират върху дънната платка (\textit{Printable Circuit Board, PCB})
показана на Фиг.~PCB. 


Амплитудата на осцилациите се ограничава от диод успоредно свързан на кондензатора и ако това е светодиод (LED)
при ниски честоти от порядъка на звуковите 
могат да се видят пулсациите на светлината. 
При механични вибрации на светодиода $f_\mathrm{mech}$ 
се наблюдава стробоскопичен ефект 
и преброяването на светлите петна $N$ дава груба оценка за честотата на
\begin{equation}
f_0=\frac{1}{2\pi\sqrt{LC}}=2(N-1)f_\mathrm{mech}\ll f_\mathrm{crossover}.
\end{equation} 
Нека разгледаме числен пример: 
$C=4.7\,\mu\mathrm{F},$
$L=400\,\mathrm{mH}$, $r=3.62\,\Omega,$
$R_1=R_2=R_3=1.5\,\mathrm{k}\Omega$.
При тези параметри 
$f_0=116\,\mathrm{Hz}$,
$\sqrt{L/C}=292\,\Omega;$ 
oсцилации с такава честота могат да се чуят със слушалка
$L/r=110\,\mathrm{ms},$
$\mathcal{Q}=80.5,$
$R_\mathrm{res}=23.5\,\mathrm{k}\Omega,$
$\overline{R}=1.5\,\mathrm{k}\Omega,$
$\overline{Q}\equiv\overline{R}/\sqrt{L/C}=5.1\gg1,$ 
$\mathcal{Q}/\overline{Q}=R_\mathrm{res}/\overline{R}=15.6.$
Дисипативния Q-фактор $\mathcal{Q}$
описва отношението на пълната енергия на трептящия контур и
дисипираната енергия когато фазата се завърти на 1 радиан.
Аналогично инкременталния Q-фактор $\overline{Q}$
на успоредния резонансен контур описва отношението 
на енергията в контура и нарастването на енергията 
внесена от отрицателното съпротивление, 
когато фазата нарастне с 1 радиан.
Когато имаме голям инкрементален Q-фактор $\overline{Q}\gg1$
се казва, че имаме ``меко възбуждане''. 
При това меко възбуждане хармоничните електични осцилации 
породени първоначално от топлинните шумове на резисторите
се усилват бавно 
докато амплитудата им не се ограничи от някакъв нелинеен елемент
във веригата, например светодиод който ще разгледаме в следващата секция.

Нека сега повторим отговора на задачата: 
отрицателното съпротивление по модул 
$\overline{R}=10\,\mathrm{k}\Omega$ 
не може да се измери с волтметър
защото вътрешното съпротивление на Омметъра
е много голямо $R_\Omega>100\,\mathrm{k}\Omega$
и се нарушава условието за електростатична устойчивост
неравенство~(\ref{BG_StabilityCriterion}).
Но ако се направи минимално изменение 
в схемата на отрицателното съпротивление от Фиг.~\ref{BG_circuit},
ако само се разменят (+) и (-) входа на операционния усилвател
посоката на неравенството се обръща 
и такова отрицателно съпротивление може вече да се измери с волтметър 
защото външната за отрицателното съпротивление верига е с голямо съпротивление, но пък тогава за малки външни съпротивления веригата
става електростатично нестабилна.
Нещо повече така модифицираната схема може да се използва за генериране на осцилации в последователни резонансни контури когато критерия за възбуждане на осцилации е просто $\overline{R}>r$.
Така, 
ако към схемата се добави сдвоен превключвател на (+) и (-) входовете,
схемата ще може да генерира трептения във всякакви резонансни контури,
в кварцови резонатори например,
а отрицателното съпротивление за тест ще може да се измерва с мултцет.

\subsubsection{Как амплитудата на осцилациите се фиксира от светодиода}

В тази секция ще пресметнем средната мощност която  на светодиода
$P_{_\mathrm{LED}}$ включен успоредно на кондензатора.
ВАХ на светодиода апроксимираме с прави линии
\begin{equation}
I=\chi(U-U_c)/r_{_\mathrm{LED}},\qquad 
\chi(x)\equiv x\theta(x)=\left\{
\begin{array}{ll}
0,&\quad x\le 0\\
x,& \quad x>0
\end{array}
\right.
,\qquad 
\theta(x)\equiv\left\{
\begin{array}{ll}
0,&\quad x <0\\
1,& \quad x>0
\end{array}
\right.
,\qquad\delta(x)=\frac{\mathrm{d}}{\mathrm{d}x}\theta(x),
\end{equation}
Тогава мощността може да се изрази с помощта на въведената за случая функция $\chi$
\begin{equation}
P_{_\mathrm{LED}}(U)=\chi(U-U_c)U/r_{_\mathrm{LED}},
\end{equation}
Амплитудата на тока $U_m$ която малко надхвърла напрежението на запалване на светодиода $U_c$ параметризираме с малкия параметър $\varepsilon$
\begin{equation}
U(t)=U_m\cos(\omega t), \qquad U_m\approx U_c,\qquad
\varepsilon\equiv\frac{U_m-U_c}{U_c}\ll1,\qquad U_m
=(1+\varepsilon)U_c.
\end{equation}
Нека сега намерим усреднената по периода на осцилациите $2\pi/\omega$ 
мощност
\begin{equation}
P_{_\mathrm{LED}}=\left< I(t)U(t)\right>
\approx \frac{U_c^2}{r_{_\mathrm{LED}}}F(\varepsilon),\qquad
F(\varepsilon)\equiv\int_{-\pi}^{\pi}\chi(\cos(\omega t)-1/(1+\varepsilon))
\frac{\mathrm{d}t}{2\pi/\omega}.
\end{equation}
С помощта на честотата $\omega$ и времето $t$
образуваме безразмерната променлива
$x=\omega t,$ 
отчитам е, че косинуса е четна функция
$\cos(-x)=\cos x,$
и за малки стойности на аргумента 
$|x|\ll1$  
взимаме само първите  два члена на тейлоровия ред
$\cos x\approx 1-\frac12 x^2$
Използваме още приближената формула
$1/(1+\varepsilon)\approx 1-\varepsilon.$
При така отчетените приближения за $\varepsilon\ll1$ извеждаме
\begin{equation}
F(\epsilon)\approx \int_{-\pi}^{\pi}
\chi(\varepsilon-\frac12 x^2)
\frac{\mathrm{d}x}{2\pi}
=\frac{2}{2\pi}\int_0^{\sqrt{\varepsilon/2}}\frac12 x^2\mathrm{d}x
=\frac{2}{2\pi}
\left. \frac{x^3}3 \right|_0^{\sqrt{\varepsilon/2}}
=\frac12\frac{\varepsilon^{3/2}}{3\times2^{3/2}\pi }
\end{equation}
и мощността може да се изрази с помощта на въведеното ефективно съпротивление на светодиода
\begin{equation}
\label{BG_Reff}
P_{_\mathrm{LED}}=\frac{U_m^2}{2 R_\mathrm{eff}},\qquad
\frac1{R_\mathrm{eff}}=\frac{\varepsilon^{3/2}}{3\times2^{3/2}\,\pi\, r_{_\mathrm{LED}}},
\end{equation}
Мощността която се черпи от отрицателното съпротивление
дисипира върху индултивността и светодиода
\begin{equation}
P_{\overline{R}}=P_r+P_{_\mathrm{LED}}.
\end{equation}
Този енергиен баланс може да се интерпретира и като нулева сумарна проводимост $\sigma_\mathrm{tot}$ на успоредно свързаните елементи на веригата: резонансен контур, отрицателно съпротивление и светодиод
\begin{equation}
\sigma_\mathrm{tot}=\sigma_\mathrm{res}+\sigma_{\overline{R}}
+\sigma_{_\mathrm{LED}}=0
,\qquad\sigma_\mathrm{res}=1/R_\mathrm{res}
,\qquad\sigma_{\overline{R}}=-1/\overline{R}
,\qquad\sigma_{_\mathrm{LED}}=1/R_\mathrm{eff}
,\qquad Z_\parallel=1/\sigma_\mathrm{tot}=\infty.
\end{equation}
Импеданса на успоредно свързаните елементи е безкраен.

При висококачесвени резонансни контури съпротивлениет в резонанса е много по голямо от модула на отрицателното съпротивление
$\overline{R}\ll R_\mathrm{res}.$
Баланса на енергията е практически между отрицателното съпротивление и светодиода които трябва да имат еднакви усреднени проводимости
$1/\overline{R}=1/R_\mathrm{eff}.$ 
Заместването туk на израза~(\ref{BG_Reff}) получен малко по-горe
дава възможност да определим амплитудата на осцилациите параметризирана
чрез $\varepsilon$ или отношението $\overline{\delta}$ между диференциалното съпротивление на светещия диод и отрицателното съпротивление
\begin{equation}
\overline\delta\equiv\frac{r_{_\mathrm{LED}}}{\overline{R}}=\frac{\varepsilon^{3/2}}{3\times2^{3/2}\,\pi\,}\ll1
,\qquad 
P_{_\mathrm{LED}}=\frac{U_c^2}{2 \overline{R}}
=\frac{U_c^2R_2}{2 R_1R_3}
= h \nu_\gamma \dot N_\gamma/\eta_{_\mathrm{LED}}
\end{equation}
Така чрез ефективното съпротивление $R_\mathrm{eff}=\overline{R}$
изразихме и средната мощност на светодиода и оценихме и броя на фотоните $\dot N_\gamma$ излъчени за единица време. Тук $\nu_\gamma$ е честотата на фотона, $h$ е константата на Планк, а $\eta_{_\mathrm{LED}}$ е коефицентга на полезно действие на светодиода.

Нека отбележим и че зависимостта между малките параметри е неаналитична
\begin{equation}
\label{BG_var_epsilon-delta}
\varepsilon\equiv\frac{U_m-U_c}{U_c}
=2(3\pi)^{2/3}\left(\overline\delta\equiv\frac{r_{_\mathrm{LED}}}{\overline{R}}\right)^{2/3},
\end{equation}
такива степени се срещат с закона на Кеплер за връзката между периода и голямата полуос на елипсата.
При исползваните ниски честоти
среднокрадратичното напрежение може да се измери с мултиметър
$\tilde{U}\equiv U_{_\mathrm{RMS}}=U_c/\sqrt{2}.$ 

Нека накрая пресметнем ``работната точка'' на постоянния ток циркулиращ през светодиода и ``черната кутия.''

\subsubsection{Как черната кутия създава постоянен ток в светодиода}

За да намерим модулите на постоянния ток $I^*$ и съответстващото му напрежение $U^*$
ще търсим пресечната точка на ВАХ на отрицателното съпротивление и светодиода $I_\mathrm{LED}(U^*)=I_\mathrm{NIC}(U^*)=I^*.$
Така получаваме
\begin{equation}
I^*=U^*/\overline{R}=(U^*-U_c)/r_\mathrm{LED},
\end{equation}
което има решение
\begin{equation}
\label{BG_epsilon-delta}
\epsilon\equiv\frac{U^*-U_c}{U_c}
=1/(\overline{R}/r_{_\mathrm{LED}}-1)
\approx\left(\overline\delta\equiv\frac{r_{_\mathrm{LED}}}{\overline{R}}\right)
\end{equation}
За разлика от амплитудата на осцилациите уравнение~(\ref{BG_var_epsilon-delta})
сега епсилон-делта зависимостта е аналитична.

Докато за променливия ток $U_c$ е амплитуда на тока за правия ток $U_c$ е постоянната стойност на тока. 
Затова средноквадратичното напрежение на променливия ток е  
$\tilde{U}=U_c/\sqrt{2}$ и в пулсиращ режим средната излъчвателна мощност е приблизително 2 пъти по малка от светенето при постоянен ток без резонансен контур.

\section{Translation into Macedonian: Услов на задачата}

\begin{figure}[h]
\includegraphics[width=8.8cm]{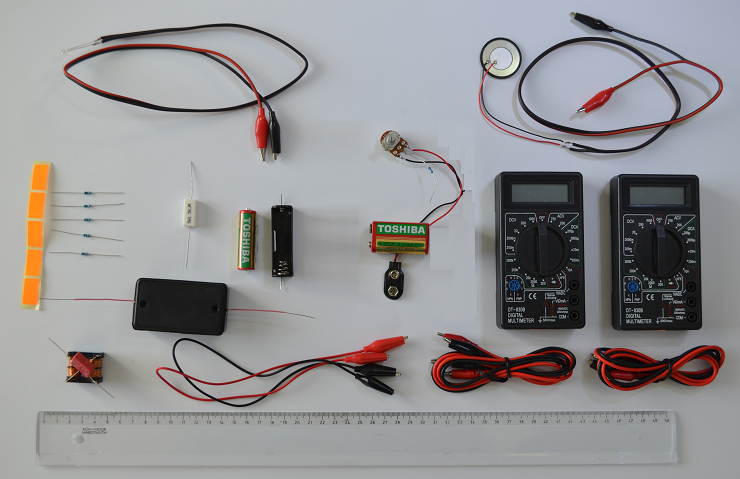}
\caption{
Опис на експерименалниот сет. 
Вие носите еден мултиметар заедно со приклучни кабли и ќе ви биде даден уште еден од организаторот за привремено користење. 
И така, сега треба да имате два мултиметри, 
една батерија од 1.5~V~(AA), 
1~батериjа од 9~V, 
1~бел керамички отпорник со отпор од 1.5~k$\Omega$, 
5 други отпорници, 
жолти етикети, 
1~пластичен линеар од 50~cm, 
еден кондензатор со пиезо-плочката кој игра улога на минијатурен звучник, 
ЛЕД диода сврзана со долги жици,
4 приклучни кабли за мултиметрите (црвени и црни), потенциометар поврзан со контакти за батерии од 9~V,
паралелно поврзани кондензатор и калем (претставувајќи резонантен осцилаторен $LC$ круг), и најважниот дел една црна кутија. 
Проверете дали ги имате сите елементи прикажани на сликата.}
\label{MK_setup}
\end{figure}

Со помош на дадениот комплет од експерименални уреди (сет од уреди) прикажана на Сл.~\ref{MK_setup}, испитајте ја зависноста помеѓу струјата и напонот $I(U)$ или уште како се нарекува волтамперска карактеристика (ВАК) на ``црната кутија'', која не смее да се отвори за време на Олимпијадата.

\subsubsection{Оnис на задачите на Олимnиадата}

Следуваат две квалитативни задачи опишани во делот~\ref{MK_quatity_tasks}, кои имаат за цел да проверите дали уредот скриен во ``црната кутија'' работи. 
Во поглавјето~\ref{MK_black_box_investigation} детално се опишани експериментите кои треба да ги извршите со дадениот комплет 
за да ја истражувате волт-амперската карактеристика (ВАК) на ``црната кутија''. 
Потоа во поглавјето~\ref{MK_the_world_is_not_perfect} ќе направите едно подетално истражување и ќе видите дека физичките својства 
кои сте ги откриле до сега за овој објект се валидни во одредени граници.
После статичката анализа следува поглавјето~\ref{MK_dynamics} во кое ќе ги истражувате динамичките својства на ``црната кутија'' со механички обиди. 
Има и чисто теориски дел, поглавје~\ref{MK_TheoreticalProblem} поврзан со теорискиот опис на предложените експерименти.
3а учениците кои не се многу сигурни во експериментот, но подобро се справуваат со математиката им предложуваме да се концентрираат на теорискиот дел од задачата. 
А за неуморните има и задача за домашно опишана во поглавјето~\ref{MK_Homework} cо парична премија од 137~\$; краен рок за предавање утре наутро 7:00.

По завршувањето на Олимпијадата комплетот ви останува како подарок за кабинетот по физика, 
Ве молиме дадете го на својот наставник за да го покажат експериментот пред учениците во училиштето. 
На организаторите треба да го вратите само дадениот мултиметар. 
Ви посакуваме добро расположение, интересно истражување, забава и успех.

\section{Две квалитативни задачи}

\label{MK_quatity_tasks}
\subsection{Палење на лед диода со ``црната кутија''. Ако диодата не свети обратете се кај наставниците}
Во дадениот комплет прикажан на Сл.~\ref{MK_setup} имате една ``црна кутија'' и една лед диода сврзана на краиштата со подолг кабел со 2 штипки (крокдилки).
Поврзете ја лед-диодата кон ``црната кутија'' и ако не светне сменете ги местата на штипките ``крокодилки''. 
При една комбинација на поврзаност на поларитетот на диодата таа се запалува т.е светнува.
При понатамошните квантитативни задачи ќе треба да се измери струјата и напонот во LED диодите и да се објасни како овие величини се поврзани со волт-амперските карактеристики на елементите.

\begin{figure}[h]
\includegraphics[width=3.3cm]{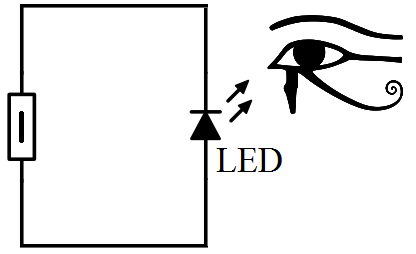}
\caption{
При една комбинација од двата можни поларитети на лед диодата таа се пали, т.е светнува кога е поврзана со електродите на ``црната кутија''. 
Кои се карактеристиките на таа кутија и што има внатре? - тоа е фактички задачата на Олиминадата.}
\label{MK_LEDNR}
\end{figure}

\subsection{Побудување (генерирање) на електрични осцилации со ``црната кутија''. Ако лед диодата не трепка или пиезо-плочката не брмчи контактирајте некој од наставниците}
\begin{figure}[h]
\includegraphics[width=8.8cm]{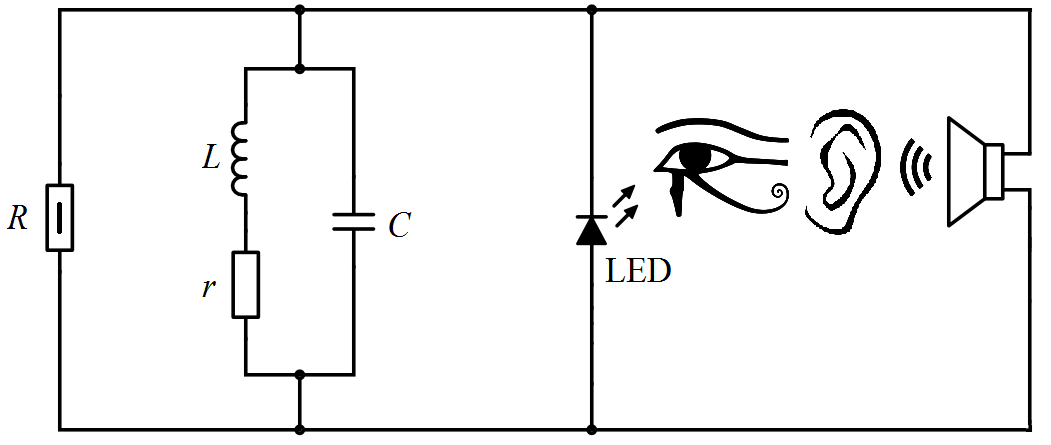}
\caption{
Побудување на електричин осцилации во $LC$ резонантен осцилаторен круг со помош на ``црната кутија''. 
Ако ја придвижиме LED диодата во овој случај забележуваме со окото дека светлината пулсира, а со увото слушаме брмчење на пиезо-плочката предизвикано од наизменичен напон.}
\label{MK_LCNR}
\end{figure}

Паралелно на ``црната кутија'' и лед диодата вклучувате $LC$ контура составена од паралелно поврзани кондензатор со капацитет $С$ и калем со индуктивност $L$ и внатрешен отпор $r$. 
Калемот е намотан на феритно јадро. 
Придвижете ја лед диодата и ќе видите дека сега светлината пулсира. Ако пак ја вклучите паралелно и пиезо плочката, како што е прикажано на Сл.~\ref{MK_LCNR}, ќе слушнете и слабо зуење. 
Проверете дали LED диодата трепка и пиезо-плочката брмчи. 
Наредните задачи се состојат од детално квантитативно истражување и теориско објаснување на светењето и брмчењето.

\begin{figure}[h]
\includegraphics[width=8.8cm]{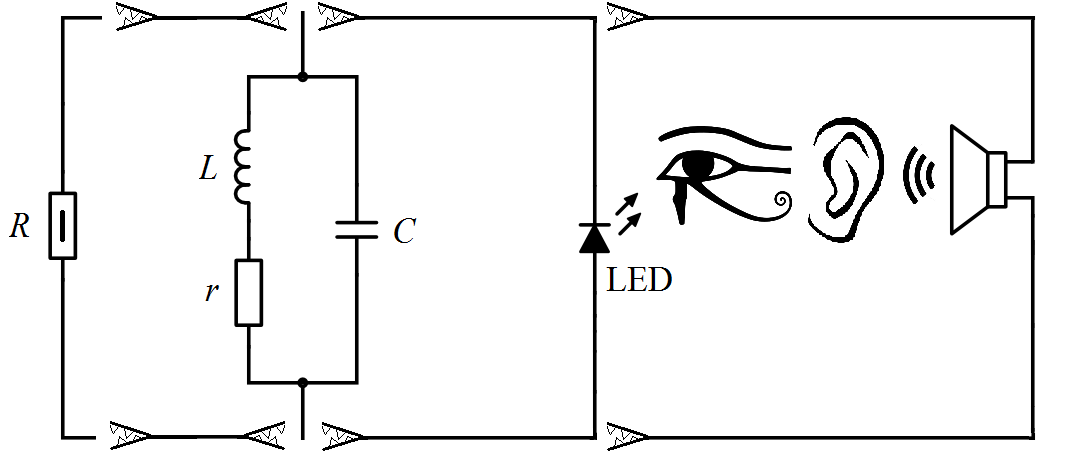}
\caption{
Шема на поврзување на електричното коло од слика~\ref{MK_LCNR}, со помош на приклучните жици, проводинци 
 кои завршуваат со штипки (``крокодилчња').}
\label{MK_LCNR-with-crocodiles}
\end{figure}

При која од двете квалитативни задачи диодата свети -- појако во случајот кога лед диодата е поврзана сама на црната кутија или кога е приклучен и резонаторскот осцилаторен круг т.е при постојаното светење или при пулсирачко светење?

\pagebreak
\section{Експериментална задача, 100~поени}
\subsection{Истражување на статичкото однесување на ``црната кутија''. Подрачје за помалите ученици}
\label{MK_black_box_investigation}

\textit{Задачите од ова подрачје ~\ref{MK_black_box_investigation}, се однесуваат за ученици од пониските класови (ученици со помала возраст). 
Задачите во нив (1--8) се поедноставни и носат помалку поени. Се разбира, ако ви остане време продолжете и со останатите задачи.}

\begin{enumerate}%
\item \textbf{Мерење на напонот и струјата која тече низ ``црната кутија'' поврзана со лед диода и без неа. (7~поени)}

Поврзете ги шемите според Сл.~\ref{MK_Hypothesis_reject} и измерете го напонот $U$ на ``црната кутија'' и струјата $I$, кој тече низ неа. 
Ако лед диодата од првата или последната шема не свети поврзете ја обратно т.е сменете и го поларитетот. 
Резултатите од мерењата внесете ги во табела иста како Табелата~\ref{MK_template_4_setups}, како што е прикажанa подолу. 

На што се должи малата разлика помеѓу напоните$U_a$ и $U^*$?
\begin{figure}[h]
\includegraphics[width=16.2cm]{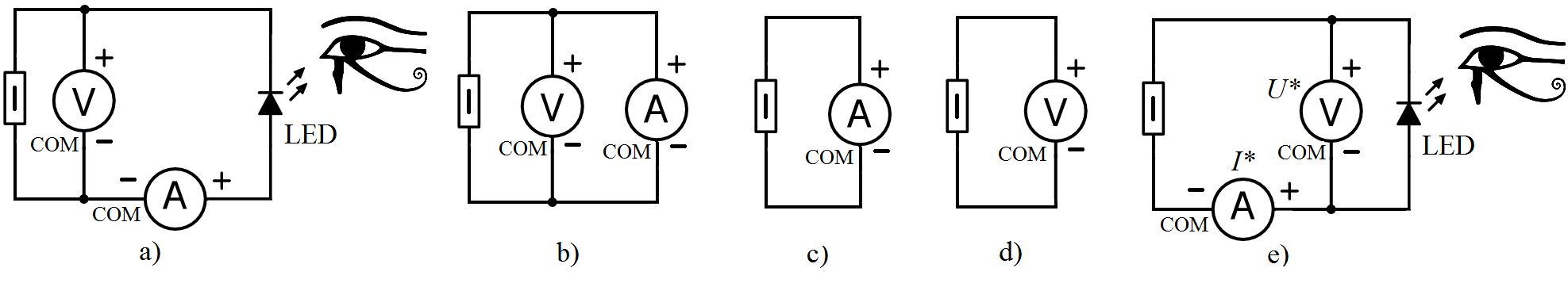}
\caption{
Пет начини за истражување на ``црната кутија'': (a) Кон постановката од Сл.~\ref{MK_LEDNR} се додаваат уште амперметар и волтметар. 
На нив може да се прочитат вредностите $U_a$ и $I_a$. (b) Ја заменуваме лед диодата со проводинк и ги забележуваме покажувањата на уредите $U_b$ и $I_b$. 
(c) Го отсрануваме волтметарот и ja прочитуваме јачината што ја покажува амперметарот $I_c$. 
(d) Заменете го амперметарот со волтметар и запишете ја вредноста на напонот $U_d$. 
(e) Измерете ги напоните $U^*$ и $I^*$ според последната шема. 
Запишете ги мерењата во табела како во примерот модел таблица~\ref{MK_template_4_setups}.
}
\label{MK_Hypothesis_reject}
\end{figure}

\begin{table}[h]
\caption{Модел за табела за обработка на експерименталин податоци за експериментот прикажан на Сл.~\ref{MK_Hypothesis_reject}.}
\begin{tabular}{| c| c | c | }
\tableline
$\#$ & $I$ [$\mu A$] & $U$ [$V$ ] \\
\tableline
a) & $I_a=\qquad \qquad$ & $U_a=\qquad \qquad$ \\
b) & $I_b=\qquad \qquad$ & $U_b=\qquad \qquad$ \\
c) & $I_c=\qquad \qquad$ & \\
d) & & $U_d=\qquad \qquad$ \\
e) & $I^*=\qquad \qquad$ & $U^*=\qquad \qquad$ \\
\tableline
\end{tabular}
\label{MK_template_4_setups}
\end{table}

\item \textbf{Измерете го напонот на батеријата од комплетот со $\mathcal{E}=$1.5~V. (1~поен)}

Вклучете го комбинираниот уред мултиметар да работи како волтметар и измерете го напонот на батеријата 
на нејзините краеви. 
Инструментот може да го регистрира и знакот на напонот.
Запомнете, ако црниот кабел е поврзан на влезот од инструментот означен како COM или со знак заземјување (~\ground), 
а црвениот приклучен кабел за другиот влез за напон V, и ако другиот крај од црниот кабел се поврзи со минусот од батеријата, 
а црвениот кабел со плусот од батеријата тогаш инструментот покажува позитивна вредност, ако се променат местата на каблите ќе покаже негативна вредност. 
Запишете ги покажувањата на волтметарот во двата случаја.

\item \textbf{Измерете го отпорот на големиот бел отпорник. (1 поен)}

Вклучете го мултиметарот како ом-метар и измерете и запишете ја вредноста на отпорот $R_\mathrm{WR}$, (отпор на големиот бел отпорник). 
Работете со точност од 1~$\Omega.$ $R_\mathrm{WR}$=?

\item \textbf{Измерете го отпорот на петте мали отпорници. (3 поени)}

Продолжете да го користите мултиметарот како ом-метар.
Измерете го електричниот отпор на секој од петте мали отпорници дадени во вашиот комплет. 
Залепете жолта етикета што ја имате во комплетот врз секој од нив и запишете $r_1$, $r_2$, $r_3$ итн и запишете ја соодветната вредност. 
Подредете ги по големина како на пример $r_1 < r_2 < r_3 < r_4<r_5$.
Работете со точност до 1~$\Omega.$

\item \textbf{Со помош на петте мали отпорници и батеријата од $\mathcal{E}=$1.5 V измерете ја зависноста помеѓу струјата и напонот на големиот бел отпорник со отпорност RWR во. (7 поени)}

Електрично шема за мерење на зависноста помеѓу струјата и напонот е дадена на Сл.~\ref{MK_resistance-measurement} и Сл.~\ref{MK_R_neg_R_with_crocodiles}. 
Мултиметрот кој го носите употребете го како амперметар поврзете го сериски со белиот отпорник. 
Внимавајте на знаците на поврзување на инструментите со изворот на струја -- струјата си има своја насока! 
Другиот мултиметар поврзете го како волтметар паралелно на проучуваниот бел отпорник и повторно внимавај 
на знаците на поврзување на иснструментот волтметар и поларитетот на вклучувањето. 
За големиот бел отпорник важи законот на Ом $U/I=R_\mathrm{WR}$. 
Ако волтметарот ви покажува позитивна вредност на напонот и амперметарот треба да покажува позитивна вредност 
и се разбира ако напонот е негативен и струјата е негативна. 
Ако знаците на напонот и струјата се спротивни видете каде сте погрешиле во поврзувањето на инструментите.

\begin{figure}[h]
\includegraphics[width=11.5cm]{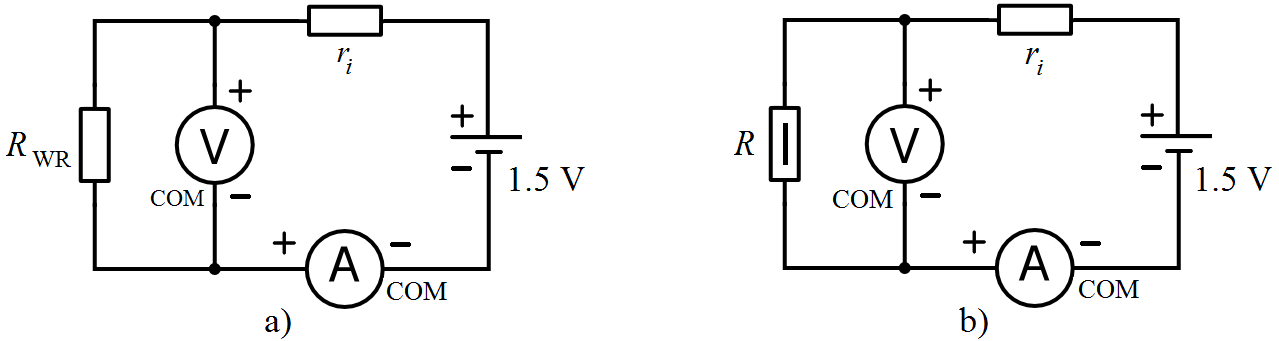}
\caption{Електрични шеми за испитување на дел од волтамперската карактеристика на: 
(a) белиот отпорник и 
(b) ``црната кутија'' и определување на нивните отпори преку волтамперската карактеристика. 
Волтметар (V) е поврзан паралелно со истражуваниот елемент и го мери напонот $U$, a амперметарот (А) е поврзан сериски и ја мери струјата $I$. 
Кога електричното коло се затвора со различни отпори $r_i\in(0,\;600\,\Omega)$ струјата и напонот се различни. Така се добиваат неколку точки од волтамперската карактеристика.
За мали напони односот $R=U/I$ e постојан и тоа е еден можен начин за проверка на законот на Ом и знаците на мерената струја и напон. 
Отпорникот $R_\mathrm{WR}$ на шемата лево е заменет со ``црната кутија'' прикажана на сликата десно и тоа е единственото разлика помеѓу овие две електрични кола.
}
\label{MK_resistance-measurement}
\end{figure}

\begin{figure}[h]
\includegraphics[width=15cm]{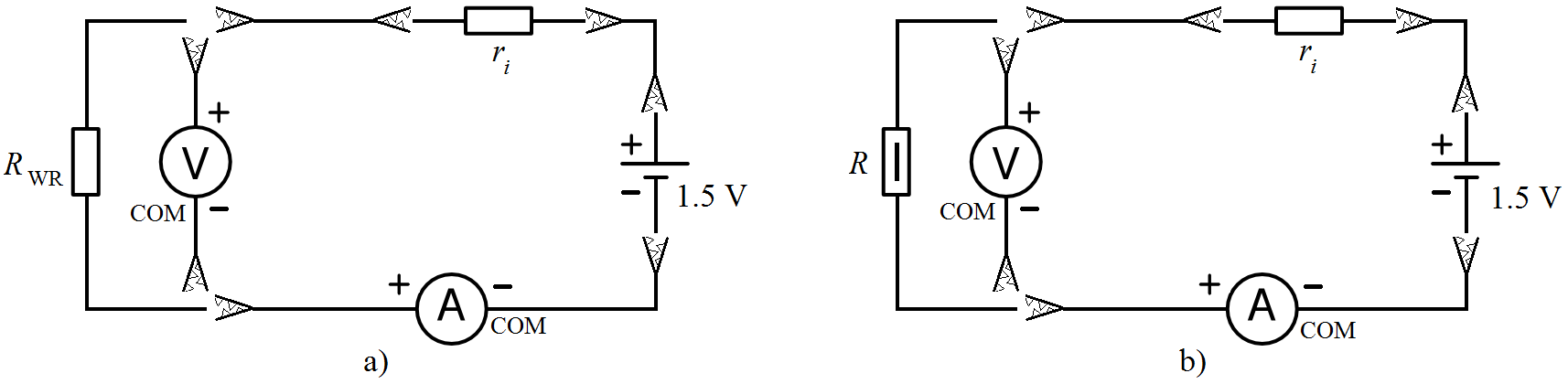}
\caption{Еквивалентна шема на поврзување на електричното коло од Сл.~\ref{MK_resistance-measurement}, преку поврзувачки проводни кабли кои на краевите завршуваат со спојки ``крокодилки''.}
\label{MK_R_neg_R_with_crocodiles}
\end{figure}

Сериски на амперметарот поврзете ја батеријата од 1.5~V. Затворете го струјнотот коло со 0~$\Omega$ и поединечно со секој од 5-те отпорници $r_i.$ 
За секое извршено мерење запишете ги соодветните резултати за напонот и јачината на струјата во табела која има 5 колони и 6 реда исто како табела~\ref{MK_template}: 
1)~број на отпорникот $i$, 
2)~вредноста на отпорникот $r_i$,
3)~Јачината на струјата $I_i$,
4)~напонот на волтметарот $U_i$ и 
5)~пресметаната вредност на отпорот $U_i/I_i$.

\begin{table}[ht]
\caption{
Образец, потребна табела за внесување и обработка на експериментални податоци добиени преку експерименталните задачи покажани на Сл.~\ref{MK_resistance-measurement}. 
Колоните од табелата означуваат: под ред бр 
1~претставува број на отпорникот $i$; 
колоната~2 ја претставува вредноста на отпорникот $r_i$; краток спој значи $r_0=0\;\Omega$; 
колната~3 претставува струјата $I_i$, кој тече низ колото сo различни ri, сериски поврзани со батеријата со напон $\mathcal{E}$; 
колоната~4 претставува $U_i$, напон на краевите од белиот отпорник или ``црната кутија'' и колоната~5 претставува однос помеѓу напонот $U_i$ и струјата $І_i$.
}
\begin{tabular}{| r | r | r | r | r | r | r |}
\tableline
 i& $r_i \, [\Omega]$ & $I_i \,[\mu \mathrm{A}]$ & $U_i\,[\mathrm{V}]$ & $U_i/I_i\,[\Omega]$ \\
\tableline
0 & 0 & & & \\
1 & & & &\\
2 & & & &\\
3 & & & &\\
4 & & & &\\
5 & & & &\\
\tableline
\end{tabular}
\label{MK_template}
\end{table}

\item \textbf{Со помош на петте малецки отпорници и батеријата од $\mathcal{E}=$1.5 V измерете ја зависноста помеѓу струјата и напонот на ``црната кутија (7~поени).}

За таа цел повторете ги истите мерења како и во претходната задача само што белиот отпорник во електричното коло го заменувате со ``црната кутија''.

\item \textbf{Нацртајте ги волтамперскиите карактеристики (ВАК) на ``црната кутија'' и белиот отпорник на еден и ист график, користејќи ги податоците од табелите во претходните две задачи. (7 поени)}

Податоците од табелите претставете ги графички така што на апцисата (\textit{x}-оската) ги нанесувате вредностите на напонот $U_i$ а по ординатата (оската~\textit{y}) струјата $I_i$. 
Ние ви препорачуваме прво да направите мал график во размер $\mathrm{1\,V=1\,cm}$ и $\mathrm{1\,A=1\,cm},$ со координатен почеток $U=0$ и $I=0$.
Доколку ви остане време може да нацртате и график во соодветни координати. 
Вие фактички направивте дел од волт-амперска карактеристика (ВАК) на ``црната кутија'' и на белиот отпорник, 
на зависноста $I(U)$ вклучувајќи само 6 точки, 6 мерења за секој од испитуваните елементи. 
Со помош точките од графикот исцртајте права линија која ги поврзува истите. 
Обележете која од овие прави, за кој испитуван елемент се однесува.

\item \textbf{Одредете го наклонот (коефициентот на правец) на правите изразени преку зависноста $\Delta U / \Delta I$ од графикот на претходната задача. (9~поени)}

Симболот $\Delta$ претставува разлика како помеѓу напоните така и помеѓу струите $\Delta U=U_2-U_1$ и $\Delta I=I_2-I_1$.
Со која карактеристика е поврзан наклонот $\Delta U / \Delta I$? 
Дали пронајдовте нешто необично во однос на карактеристиката на ``црната кутија''?

\subsection{Светот не е идеален. Детално истражување на волтамперската карактеристика (ВАК). Задача за повозрасните ученици кои имаат поголемо практично искуство со волтамперски карактеристики}
\label{MK_the_world_is_not_perfect}
\textit{Повозрасните ученици кои имаат поголемо искуство за испитување на ВАК може да работата на условите во делот~\ref{MK_the_world_is_not_perfect} и да се вратат на почетните услови од делот~\ref{MK_black_box_investigation}, доколку им остане време.}

При мали напони волтамперската карактеристика (ВАК) на ``црната кутија'' е дел од права линија, но како изгледа истата при поголеми вредности на напоните т.е во пошироко мерно напонско подрачје ќе откриете наскоро. 

\begin{figure}[h]
\includegraphics[width=8cm]{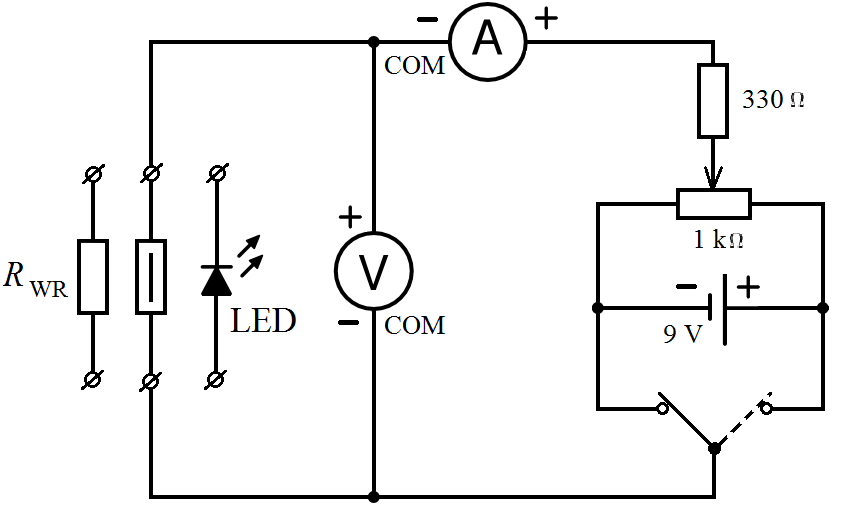}
\caption{
Експериментална постановка за истражување на волтамперската карактеристика (ВАК) $I(U)$. Прво се испитува белиот отпорник $R_\mathrm{WR}$ (крајно лево на сликата), 
потоа тој се заменува со ``црната кутија'' како на шемата и на крај ја вклучувате само лед диодата~(LED)~(прикажана десно). 
Со волтметарот се мери напонот $U$ на испитуваниот елемент, а со амперметарот јачината струјата $I$. Напонот се создава од батерија со електромоторна сила $\mathcal{E}=9\,\mathrm{V}$. 
Кога се врти со рака оската на потенциометар со отпор $1\,\mathrm{k}\Omega$, напонот $U$ ce менува од нула 0 до $+\mathcal{E}$.
При приклучување на ``крокодилката'' од едниот крај на потенциометарот кон другиот, како што е прикажано на сликата (види доле десно на сликата) 
напонот се менува од $-\mathcal{E}$ до 0. Па така напонот може да се менува од $-\mathcal{E}<U<+\mathcal{E}$. 
Отпорот со вредност од $330\,\Omega$ ја ограничува струјата низ LED диодата и ја штити од прегорување.
}
\label{MK_resistance-measurement-I-V-curve}
\end{figure}

\item \textbf{Измерете ја детално зависноста помеѓу струјата и напонот на белиот отпорник. (7~поени)}

Вклучете ја батеријата од 9~V кон соодветните контакти за потенциометарот. 
Таа претставува еден извор на напон за кој вклучувате волтметар и ги мерите вредностите на напонот npu крајните ротации на потенциометарот од $U_\mathrm{min}$ 
до 0 и од 0 до $U_\mathrm{max}$, 
како што е прикажано на Сл.~\ref{MK_resistance-measurement-I-V-curve}. 
Симболично прикажаниот прекинувач се остварува со вклучување на спојка ``крокодилка''. 
Кон белиот отпорник сериски вклучувате амперметар и го поврзувате со изворот на напон. 
Амперметарот ја мери струјата $I_\mathrm{WR}$, низ белиот отпорник, а волтметарот напониот $U$. 
Кога ја вртите оската на потенциометарот напонот се менува и така вие ја истражувате волтамперската карактеристика (ВАК). 

Првин, или на самиот почеток завртетете го потенциометарот помеѓу неговите крајните положби и дознајте во кои интервали на покажувања инструментите треба да работат. 
Потоа испитајте ја зависноста $I_\mathrm{WR}(U)$ како парови струја-напон, $(I_\mathrm{WR},U)$ и запишете ги во табела. 
За белиот отпорник важи Омовиот закон. 
За претставување на линеарната зависност $I_\mathrm{WR}(U)$ cе сосема доволни u 5 точки т.е 5 мерења, од кои двете да бидат при минималните и 
максималните напони соодветно, $U_\mathrm{min}$ и $U_\mathrm{max}$.
Согласно законот на Ом знаците на напонот и струјата треба да бидат исти. 
При разлика во знаците побарајте грешка во шемата.

\item \textbf{Измерете ја детално зависноста помеѓу струјата и напонот на ``црната кутија''. (5 поени)}

Во експерименталната постановка од претходното задача заменете го белиот отпорник со ``црната кутија'' без да правите други промени. 
Вртете ја оската на потенциометарот менувајќи го напонот $U$ помеѓу $U_\mathrm{min}$ и $U_\mathrm{max}$ со промена од 1~$V$ и забележете ги во табела паровите броеви $(I, U)$. 
Запишете ги во табела и напоните, при кои струјата има максимална или минимална вредност.

\item \textbf{Измерете ја детално зависноста помеѓу струјата и напонот на LED диодата. (3 поени)}

Во експерименталната постановка од претходното задача заменете ја ``црната кутија'' со LED диода, повторно без да прават други промени. 
Видете во која крајна положба на потенциометар диодата свети со најјак интензитет. 
Сега вртете ја оската на потенциометарот и следете ги покажувањата наамперметарот. 
При јачини на струи помали од 6~mA запишувајте ги во табела паровите броеви $(I_\mathrm{LED}, U)$ со промена од 1~mA. 
При струи помали од 2~mA=2000~$\mu$А снимајте ги паровите броеви напон-струја со промена од околу 200~$\mu$A додека не дојдете до струја помала од 200~$\mu$A.

\textit{Откога ќе завршете со снимањето на 3-тата волтамперска карактеристика исклучете ја батеријата од 9~V бидејќи брзо се истрошува истата.}

\item \textbf{Претстави ги на заеднички график волтамперските карактеристики (ВАК) на: белиот отпорник $I_\mathrm{WR}(U)$, ``црната кутија'' $I(U)$ u LED диодата $I_\mathrm{LED}(U)$. (10 поени)}

Прво анализирајте ги најмалите u најголемите вредности на струјата и напонот. 
Овие параметри ќе ви го одредат правоаголникот во кој ќе бидат претставени ВАК на трите елементи. 
Ние ви ги препорачуваме следните размери: хоризонтална оска $\mathrm{1\,V=1\,cm}$ и вертикалнa оска $\mathrm{1\,mA=1\,cm}$. 
И за трите елементи (белиот отпорник, црната кутија и лед диодата) BAК $I(U)$ претставуваат континуирани криви. 
Повлечете низ точките прави (или криви) кои максимално добро ги опишуваат експерименталните резултати.

\item \textbf{Зошто се важни волтамперските карактеристики (ВАК)? (5 поени)}

Нацртајте огледална слика на волтамперската карактеристика (ВАК) на LED диодата, при што $I$ ce заменува cо $-І$, односно ја завртуваме ВАК околу хоризонталната оска на напонот. 
Врз ВАК на LED диода означете ги со мало кругче вредностите $(-I^*, U^*)$ од табела~\ref{MK_template_4_setups}. 
Обрнете внимание, дека оваа точка е блиску до пресечната точка на ВАК на ``црната кутија'' u BAК на LED диодата. 
Случајно ли е тоа или не?

\item \textbf{Анализа на ВАК на: белиот отпорник, ``црната кутија'' и LED диодата. (7~поени)}
Ако волтаперските карактеристики BAК $I(U)$ претставуваат одделни прави линии, од нивниот наклон одредете ги соодветните отпори $R=\Delta U/\Delta I.$ 
Симболот $\Delta$ претставува разлика (\textit{difference}), изберете 2 точки од отсечката, 
два пара на вредности напон и струја од графиците и направете ја разликата на напонот $\Delta U= U_2-U_1$ u струјата $\Delta I=I_2-I_1.$ 
Кога разгледуваниот интервал на вредности е мал, кривата е приближно права, па наместо $\Delta$ се пишува $\mathrm{d}.$ 
Отпорот пресметан од наклонот на ВАК се нарекува диференцијална отпор, а реципрочната вредност диференцијална спроводливост $\sigma_\mathrm{diff}=\mathrm{d}I/\mathrm{d}U$.

Пресметај го: a) отпорот на белиот отпорник, b) отпорот од централниот дел од кривата на ``црната кутија'' с) отпорот од левиот дел од кривата на "црната кутија", d) отпорот на десниот дел од кривата на ``црната кутија'' и е) отпорот на LED диоди, коjа свети кога низ неат течат струи во интервалот од 2-5 mA.

\item \textbf{Која карактеристика на волтамперската карактеристика (ВАК) на ``црната кутија'' е од суштинско значење за постојано светкање или трепкање на диодата од двете квалитативни задачи? (4~поени)}

\textit{
За ``црната кутија'' е најважен централниот дел од ВАК. 
На овој дел од ВАК се должи noтојаното светење на диодата и генерирањето на nроменливи струи во осцилаторен круг, 
резонансна контура, кои ги набљудувавте во квалитативните задачи noкажани na Сл.~\ref{MK_LEDNR} и Сл.~\ref{MK_LCNR}, 
како и во многу други можни технички уреди од munот на скриеното во црната кутија.}

Разликата од волтамперските карактеристики (ВАК) на ``црната кутија'' и белиот отпорник ни ја открива која е причината за светењето на LED диодaта. 
Која е оваа разлика?

\item \textbf{Опишете квалитативно како се создава струја, кој предизвикува постојаното светење или трепкање на LED диодата во двете квалитативни задачи. (10~поени)}

\subsection{Мерење на фреквенцијата на електричните осцилации создадени од ``црната кутија''}
\label{MK_dynamics}
Поврзете ја повторно шемата од втората квалитативна задача прикажана на Сл.~\ref{MK_LCNR} кога сите елементи на колото се поврзани паралелно: ``црната кутија'', $LC$ резонаторот, пиезо-плочката (минијатурен звучник, зумер), LED диодата.

Ако наместо зумер (пиезо плочка) се употребат слушалки од мобилен, тогаш кон нив треба сериски да поврземе отпор поголем од 10~k$\Omega$.

\item \textbf{Механичко одредување на фреквенција на осцилации. (7~поени)}

Електричните мерења се полесни и попрецизни. За да ја измериме фреквенцијата на осцилациите без фреквенцметар 
при дадени услови е потребно да се манифестира малку умешност. 
Во приборот располагате со леплива лента. 
Ја прицврстувате LED доиодата на крајот од еластичен ленир со самолеплива трака ``селотејп''. 
Кога ќе го притиснете со една рака ленирот кој е на крајот на масата, а со другата рака предизвикувате осцилации, при што со осцилирањето на ленирот, 
осцилира и лед диодата која свети и е залепена за него, забележувајќи неколку речиси неподвижни светли дамки. 
Ако електричната фреквенција $f_\mathrm{res}$ е множител на механичката $f_\mathrm{mech}$ светлите дамки се неподвижни. 
Пребројте ги светлите петна или само дадете оценка за нивниот број $N$ или ако можете да ги изброете брзо, 
определете ја фреквенцијата на механичките осцилации $f_\mathrm{mech}$. 
Потрудете се ленирот да го заосцилирувате со постојана фреквенција и амплитуда. 
Променувајте ја должината на слободниот крај на ленирот за да постигнете неподвижни светли дамки.
Избројте за 10 секунди колку осцилации прави ленирот и така утврдете ја механичката осцилација $f_\mathrm{mech}$. 
Фреквенцијата на електричното осцилации $f_\mathrm{res}$ изразете ги преку бројот на светлите петна и фреквенцијата на механичките осцилации. 
Колку херци за резонантната фреквенција на електричното осцилации добивате од вашата формула $f_\mathrm{res}(N,f_\mathrm{mech})$?

\item \textbf{Пресметајте ја теоретски фреквенцијата на осцилациите. (2~поени)}

Еден алтернативен метод за одредување на оваа фреквенција е нејзиното теориско пресметување според формулата на $f_\mathrm{res}=1/(2\pi\sqrt{LC}).$ 
Ако на кондензаторот пишува 4.7~$\mu F.$, а на тороидален калем пишува дека имаме 2 намотки по 100~mH и 
при нивно сериско поврзување индуктивност им е $L$=400~mH колку изнесува фреквенцијата? 
Колку пресметаната фреквенција се согласува со експериментално утврдената фреквенција.

\item \textbf{Определување на фреквенција со слушање на осцилациите од зумер. (1 поен)}

Зумерот претставува сигнален елемент ( на германски ``summer'' значи зуење). 
Ако имате музикален слух и се сеќавате на фреквенциите на музичките тонови може да се определи на кој музички тон одговара фреквенцијата што ја емитира зумерот. 
Кај шемата број 3 имате вклучено и зумер кој емитува некоја фреквенција. 
Колку е фреквенцијата што ја емитува? 
Не се очекува голема точност и 50\% грешка би било задоволително добра согласност на овие 3 методи за одредување на фреквенција без фреквенцметар.

\end{enumerate}%

\section{Теориска задача. (35~поени)}
\label{MK_TheoreticalProblem}

\subsection{Услов на теориската задача}

\begin{figure}[h]
\includegraphics[width=9cm]{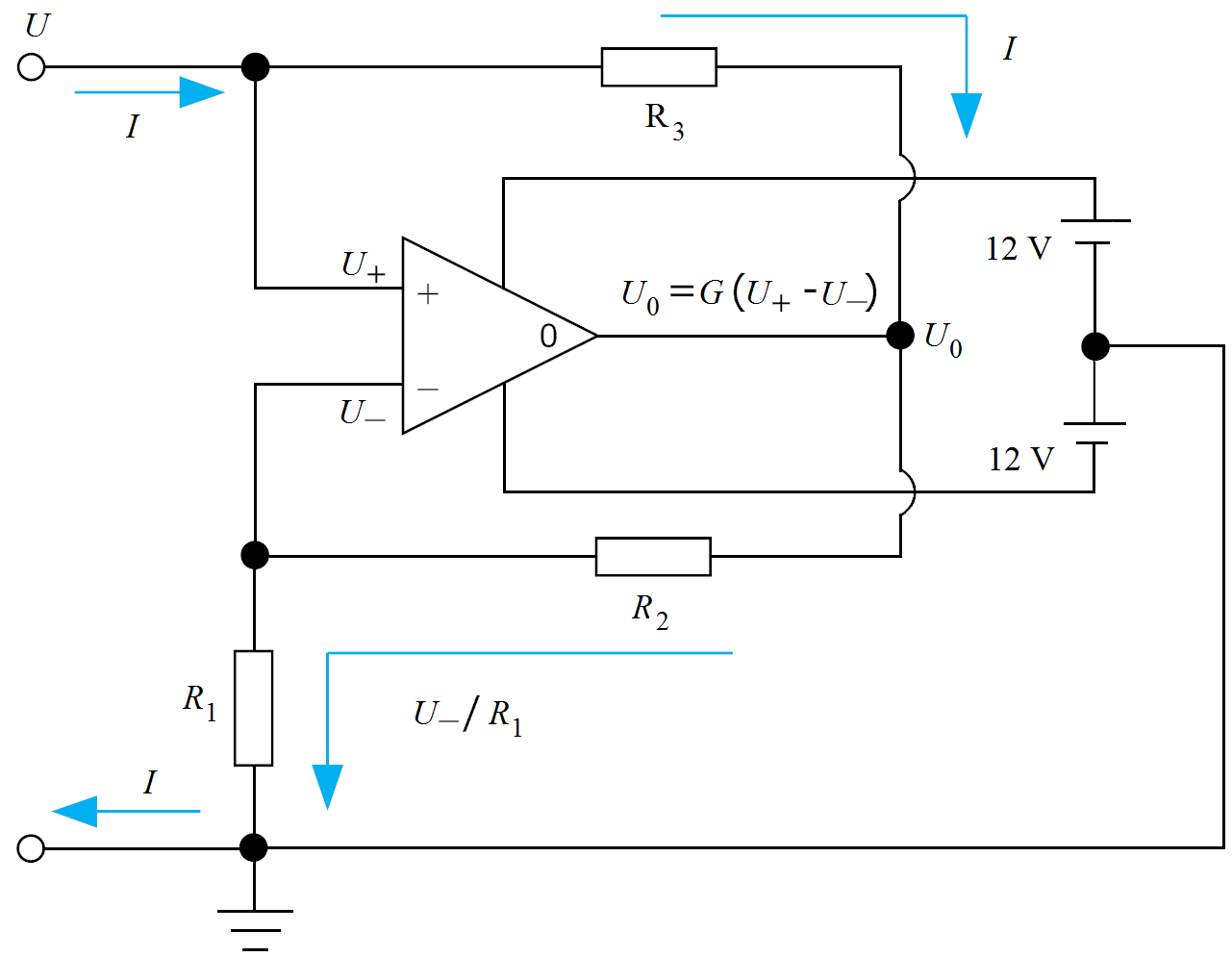}
\caption{Пресметајте го со 1\% процентна точност ефективниот отпор $R=U/I$ на колото во зависност од трите отпорници $R_1,$ $R_2$ и $R_3$ на означениот 
со триаголник напонски засилувач со коефициент на засилување $G=10^5,$ кој се напојува со две батерии со напон $V_S$. Напоните $U_0$, $U_{-}$, како и струјата $I$ cе непознати. 
Со $U_0$ е означен напонот изведен од $U_\mathrm{output}=(U_+-U_-)G.$}
\label{MK_circuit}
\end{figure}

На електричното коло прикажано на Сл.~\ref{MK_circuit}, влезните струи во точките (+) и (--) се нула, а излезната струја во точката (0) е таква што соодветните напони се поврзани со релацијата $U_0=(U_+ - U_-)G$, каде што $G$ е коефициентот на засилување и има многу голема вредност, $G=10^5 \gg 1$. 
Електронскиот елемент означен со триаголник (засилувач) се напојува со две батерии и во условот на задачата се претпоставува дека споменатите напони се помали од напонот на напојување на батериите $\mathcal{E}_\mathrm{B}=12\;\mathrm{V}.$
Точката помеѓу двете батерии е поврзана со проводник со крајот на отпорникот $R_1$ и со една од влезните електроди на шемата. 
Погодно е напонот во то таа заедничка точка да се избере за нула $U_\mathrm{CP}=0$ или како електроехничарите велат тоа е ``земја'' (\ground). 
Индексот $\mathrm{CP}$ доаѓа од англиските зборови Common Point u на мултиметрите се означува со COM. 
Во спротивeнслучај, ако $U_\mathrm{CP} \neq 0$, за равенката за засилувањето на nапонот имаме $U_0=(U_+ - U_-)G+U_\mathrm{CP}$. 
Струјата, кој истекува во ``земјата'', е нула. 
Пресметајте го со точност од 1\% (три значајни цифри) односот помеѓу влезниот напон $U$ и струјата $I$ што тече во колото и изразете го тој ефективен отпор $R=U/I=R(R_1,R_2, R_3)$
во зависност од трите отпори на шемата. 
За да поедноставите, може да претпоставите дека коефициентот на засилување се стреми кон бесконечност $G \rightarrow \infty$. 
Во добиениот израз заменете $R_1=R_2=10\;\mathrm{k} \Omega,\; R_3=1.5 \;\mathrm{k} \Omega$. 
На кратко: 
(1) Се бара крајната формула за отпорот на шемата и 
(2) дасе пресмета според неа нумеричката вредност на отпорот, со точност од 1\%. 
Каков е знакот на отпорот $R=U/I$ u колку ома е неговиот модул. 
Како тој отпор ја запалува LED диодата?

%

%
\subsection{Гатанка (2~поени)} 
Црвенo и цврсто a лета. Што е тоа?

\section{Задача за домашна работа, $\mathbf{137\,\$}$}
\label{MK_Homework}
\textit{По завршувањето на Олимпијадата, најдете шрафцигер и извадете ги шрафовите од каnакот на ``црната кутија''. 
Извадете ги батериите или nоставете еден од прекинувачите oдnоложба On на Оff.}

За мали напони волт-амперската карактеристика (ВАК) на ``црната кутија'' е права линија со постојан однос $U/I$ согласно како и законот на Ом. 
Обидте се да го измерите отпорот на ``црната кутија'' со омметар. 
Обидете се да ги споредите покажувањата на омметарот со отпорот добиен преку испитување на ВАК.
Објаснете зошто отпорот на црната кутија дефинитивно не може да се измери со омметар? 
Какви промени од шемата на сл.~\ref{MK_circuit} треба да се реализират во ``црната кутија'' за да го направи можно мерењето на отпорот со омметар, на пример со мултиметар DT-830B od 20~k$\Omega$ со кој располагавте на олимпијадата? 
Првиот кој ќе одговори на барем на едно од двете прашања и ro испрати од адресатта, со која се регистрирал за Олимпијадата на \texttt{epo@bgphysics.eu} до 07:00 на 1 ноември 2015 година ќе освои паричната награда од 137~\$. 
Можете да работите во тим, да користите литература, да користите \textit{Google} и преку интернет да се консултирате со радиоинженери и универзитетски професори по електроника насекаде по светот. 
Кога во Куманово е полноќ, во Калифорнија доцна попладе, а Јапонија започнува ден – секогаш во светот има колеги што работат.

\newpage

\section{Translation into Serbian: Услови задатка}
\begin{figure}[h]
\includegraphics[width=8.8cm]{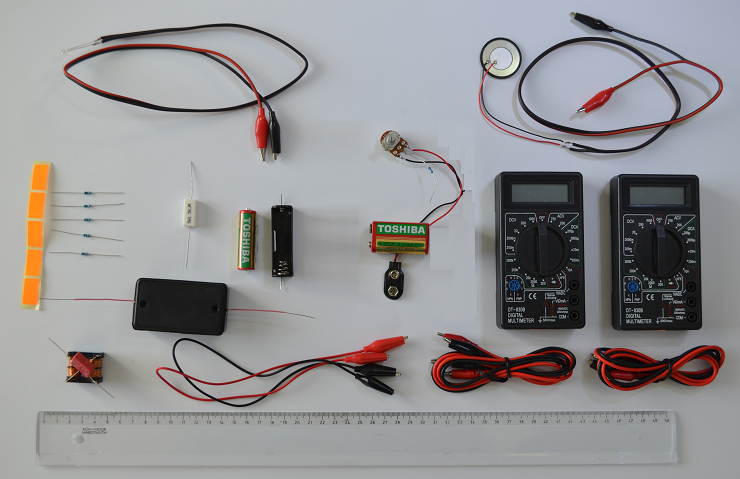}
\caption{Opis eksperimentalnog seta.
Vi nosite jedan multimetar sa priključnim kablovima i dobićete još jedan
od organizatora na privremeno korišćenje.Tako da sada trebate imati dva multimetra, jednu bateriju od 1.5~V~(AA), dve 
baterije od 9~V, jedan beli keramički otpornik sa otporom od 1.5~k$\Omega$, 
druge otpornike,žute etikete, jedan kondenzator sa 
piezo-pločicom koji igra ulogu minijaturnog zvučnika, LED diodu povezanu na dugačke žice, plastični lenjir, četiri 
prikčjučna kabla za multimetre (crveni i crni), potenciometar povezan na kontakte za baterije od 9~V, paralelno 
povezane kondenzator i kalem (predstavljajući rerzonantni $LC$ krug) i najvažniji deo~--~jednu crnu kutiju.}
\label{SR_setup}
\end{figure}

Pomoću datog kompleta eksperimentalnih uređaja (set uređaja) prikazanog na slici~\ref{SR_setup}. ispitajte zavisnost između 
struje i napona $I(U)$ ili kako se još naziva voltamperska karakteristika (VAK) ``crne kutije'', koja se ne sme otvoriti za 
vreme Olimpijade.

\subsubsection{Opis zadataka na Olimpijadi}
Slede dva kvalitativna zadadtka opisanih u delu~\ref{SR_quatity_tasks}, čiji je cilj da proverite dali uređaj sakriven u ``crnoj kutiji'' radi. U poglavlju~\ref{SR_black_box_investigation} su detaljno opisani eksperimenti koje trebate izvršiti datim kompletom da bi istražili volt-ampersku 
karakteristiku (VAK) ``crne kutije''. Dalje, u poglavlju~\ref{SR_the_world_is_not_perfect} uradićete detaljnije istraživanje i uvideti da su fizička svojstva ovog objekta koja ste do sad otkrili validna u određenim granicama. Nakon statične analize sledi poglavlje~\ref{SR_dynamics} u kom ćete mehaničkim opitima istaživati dinamička svojstva ``crne kutije''. Poglavlje~\ref{SR_TheoreticalProblem} ima i čisto teoretski deo, 
povezan teoretskim opisom predloženih eksperimenata. Učenicima koji nisu mnogo sigurni u eksperiment, ali se bolje 
nose sa matematikom predlažemo da se skoncentrišu na teoretski deo zadatka. Za neumorne je domaći zadatak opisan u 
poglavlju~\ref{SR_Homework}, uz novčanu premiju od 137~\$ -- krajnji rok za predavanje (01.11.2015) sutra u 7:00.

Nakon završetka Olimpijade komplet dobijate na poklon kabinetu fizike. 
Molimo vas da ga predate vašem 
nastavniku da bi pokazali eksperiment đacima iz vaše škole.
Organizatorima jedino trebate vratiti primljeni multimetar. 
Želimo vam dobro raspoloženje, interesantno istaživanje, zabavu i uspeh.

\section{Dva kvalitativna zadatka}

\label{SR_quatity_tasks}
\subsection{А. Palenje LED diode ``crnom kutijom''. Ako dioda ne svetli, obratite se}
U kompletu prikazanom na sl.~\ref{SR_setup} imate ``crnu kutiju'' i LED diodu povezanu na krajevima dužim kablovima sa
štipaljkama. Povežite LED diodu na ``crnu kutiju'' i ako ne zasvetli premenite mesta štipaljkama. 
U jednoj kombinaciji za pvezivanje dioda zasveti.
Pri daljnjih kvantitativnih zadataka potrebno je mjerenje struje i napona na LED (Light-emitting Diode) i objasniti kako su ove varijable povezane s karakteristikama struje napona od elemenata.

\begin{figure}[h]
\includegraphics[width=3.3cm]{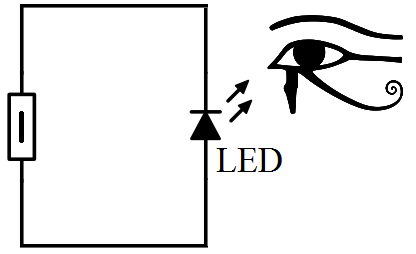}
\caption{Kod jedne od dve moguće kombinacije polariteta dioda zasvetli kada je povezana na electrode ``crne kutije''.
Koje su karakteristike te kutije i šta je unutra je praktično zadatak Olimpijade}
\label{SR_LEDNR}
\end{figure}

\subsection{Pobuđivanje (generisanje) električnih oscilacija ``crnom kutijom''.
Ako LED dioda ne trepće ili piezo-pločica ne zuji obratite se nastavnicima}
\begin{figure}[h]
\includegraphics[width=8.8cm]{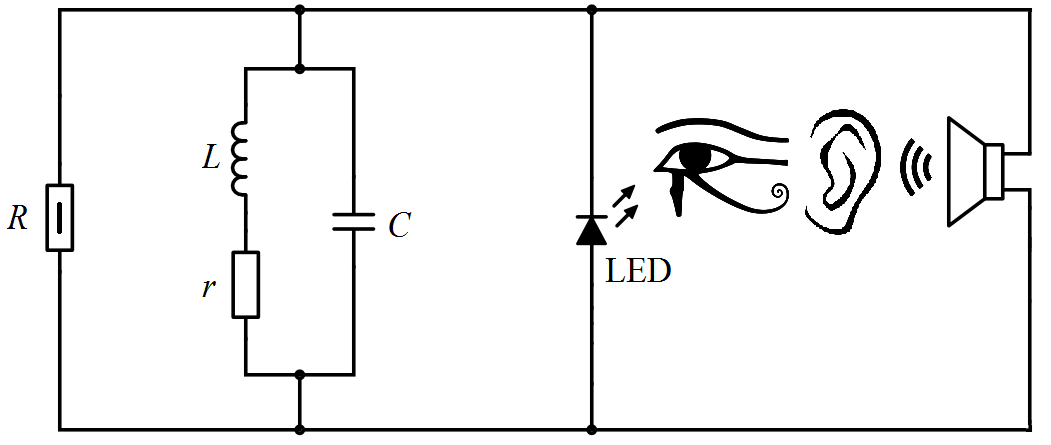}
\caption{Pobuđivanje električnih oscilacija u $LC$ rezonantnom oscilatzornom krugu pomoću ``crne kutije''. 
Ako pomerimo LED diodu u ovom slučaju vidimo da svetlost pulsira, a čujemo zujanje piezo-pličice izazvano naizmeničnim naponom.}
\label{SR_LCNR}
\end{figure}

Paralelno na crnu kutiju i LED diodu prikčjučite $LC$ konturu sastavljenu od paralelno povezanog kondenzatora 
kapaciteta $C$ i kalema induktivnosti $L$ i unutrašnjeg otpora $r$. 
Kalem je namotan na feritno jezgro. 
Pomerite LED diodu i vedećete da svetlost pulsira. 
Ako paralelno uključite i piezo-pločicu, kako je prikazano na slici~\ref{SR_LCNR}, čućete i slabo zujanje. 
Proverite dali LED dioda treperi i piezo-pločica zuji. 
Naredni zadaci se odnose na detaljno kvantitativno istrživanje i teoretsko objašnjenje svetlosti i zujanja.

\begin{figure}[h]
\includegraphics[width=8.8cm]{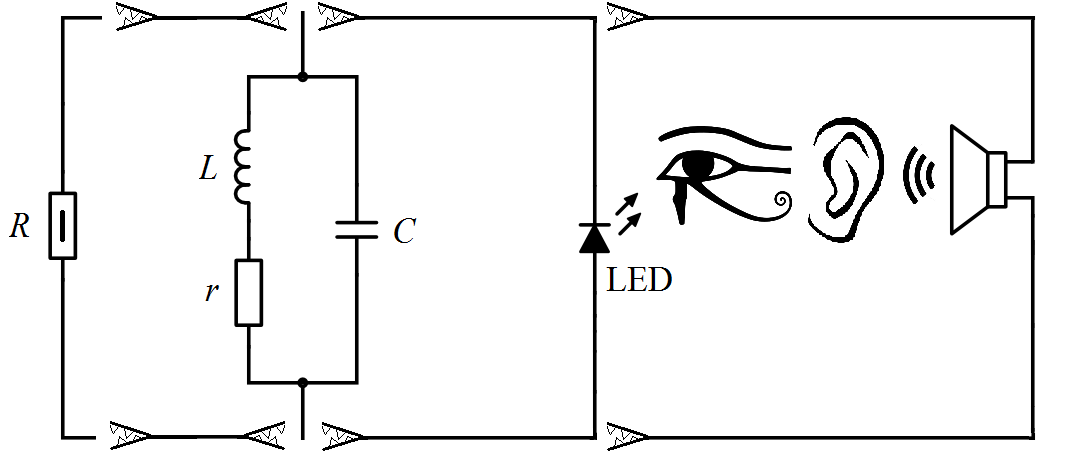}
\caption{Šema za povezivanje električnog kola iz slike~\ref{SR_LCNR}, pomoću priključnih kablova koji završavaju štipaljkama.}
\label{SR_LCNR-with-crocodiles}
\end{figure}

Kod kog od ova dva kvalitativna zadatka dioda svetli jače ’ u slučaju kada je dioda povezana sama na ``crnu kutiju'' ili 
kada je priključen i rezonantni oscilatzorni krug, t.j. kod postojane ili pulsirajuće svetlosti?

\pagebreak
\section{Eksperimentalni zadatak 100~poena}
\subsection{Istraživanje statičkog ponašanja ``crne kutije``. Odsjek za mlađe učenike}
\label{SR_black_box_investigation}


\begin{enumerate}%
\item \textbf{Merenje napona i struje koja teče kroz ``crnu kutiju'' povezanu sa LED diodom i bez nje. (7~poena)}

Povežite šeme kao na sl.~\ref{SR_Hypothesis_reject} i izmerite napon U na ``crnoj kutiji'' i struju $I$ koja teče kroz nju.
Ako LED dioda iz prve ili druge šeme ne svetli, povežite je obrnuto, t.j. promenite joj polaritet. 
Rezultate merenja unesite u tabelu identičnu Tabeli~\ref{SR_template_4_setups}, kako je priazana niže.
Usled čega se pojavljuje mala razlika između napona $U_a$ i $U^*$?

\begin{figure}[h]
\includegraphics[width=16.2cm]{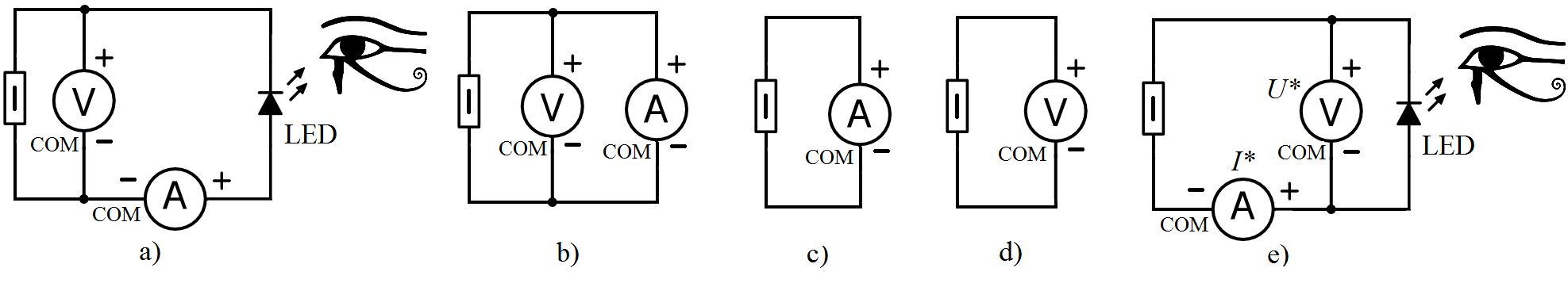}
\caption{Pet načina za istraživanje ``crne kutije'': 
(a) Kod postrojenja iz sl.~\ref{SR_LEDNR} dodaju se još ampermetar i voltmetar. 
Na njima se mogu učitati vrednosti $U_a$ u $I_a$. 
(b) Zamenimo LED diodu provodnikom i zabeležimo vrednosti na uređajima $U_b$ i $I_b$. 
(c) Uklonimo voltmetar i učitamo jačinu koju pokazuje ampermetar $I_c$.
(d)Zamenite ampermetar voltmetrom i upišite vrednost napona $U_d$. 
(e) Izmerite napon $U^*$ jačinu $I^*$ po poslednjoj šemi.Upišite merenja u tabelu kao u primeru Tabela~\ref{SR_template_4_setups}.}
\label{SR_Hypothesis_reject}
\end{figure}

\begin{table}[h]
\caption{Model tabele za obradu eksperimentalnih podataka za eksperiment prikazan na slici~\ref{SR_Hypothesis_reject}.}
\begin{tabular}{| c| c | c | }
\tableline
$\#$ & $I$ [$\mu A$] & $U$ [$V$ ] \\
\tableline
a) & $I_a=\qquad \qquad$ & $U_a=\qquad \qquad$ \\
b) & $I_b=\qquad \qquad$ & $U_b=\qquad \qquad$ \\
c) & $I_c=\qquad \qquad$ & \\
d) & & $U_d=\qquad \qquad$ \\
e) & $I^*=\qquad \qquad$ & $U^*=\qquad \qquad$ \\
\tableline
\end{tabular}
\label{SR_template_4_setups}
\end{table}

\item \textbf{Izmerite napon baterije iz kompleta sa ems $\mathcal{E}=$1.5~V. (1 poen)}

Uključite kombinovani uređaj multimer da radi kao voltmetar i izmerite napon baterije na njenim krajevima. 
Instrument može da registruje i znak napona.
Upamtite,ako je crni kabel povezan na ulaz instrumenta označenim kao COM ili znakom uzemljenja (~\ground), 
a crveni kabel za drugi ulaz za napon~$V$, i ako se drugi kraj crnog kabla poveže na minus baterije a 
crveni kabel na plus baterije instrument pokazuje pozitivnu vrednost.
Ako se promene mesta kablova pokazaće negativnu vrednost.
Upišite vrednosti na voltmetru u oba slučaja.

\item \textbf{Izmerite otpor velikog belog otpornika. (1~poen)}

Uključite multimetar kao om-metar i izmerite i upišite vrednost otpora $R_\mathrm{WR}$, (otpor velikog belog otpornika). 
Radite sa tačnošću od 1~$\Omega.$ $R_\mathrm{WR}$=?

\item \textbf{Izmerite otpor pet malih otpornika. (3 poena)}

Nastavite da koristite multimetar kao om-metar.Izmerite električni otpor svakog od pet malih otpornika u vašem kompletu.
Zalepite žutu etiketu iz kompleta na svaki od otpornika i upišite $r_1 < r_2 < r_3 < r_4<r_5$ i upišite odgovarajuću vrednost. 
Radite sa tačnošću od 1~$\Omega.$

\item \textbf{Pomoću pet malih otpornika i baterije od од $\mathcal{E}=$1.5 V izmerite zavisnost između struje i napona velikog belog otpornika sa otporom $R_\mathrm{WR}$. (7 poena)}

Električna šema za merenje zavisnosti između struje i napona data je na Sl.~\ref{SR_resistance-measurement} i Sl.~\ref{SR_R_neg_R_with_crocodiles}. 
Multimetar upotrebite kao ampermetar, povežite ga serijski sa belim otpornikom.Pazite na znakove za povezivanje instrumenta sa izvorom struje -- struja ima svoj smer! 
Drugi multimetar povežite kao voltmetar paralelno na proučavani beli otpornik i ponovo pazite na znakove za povezivanje voltmetra i polaritet uključivanja.
Za veliki beli otpornik važi Omov zakon $U/I=R_\mathrm{WR}$. 
Ako voltmetar pokazuje pozitivnu vrednost i ampermetar treba da pokazuje pozitivnu vrednost, i obrnuto.
Ako su znaci napona i struje suprotni proverite gde ste pogrešili kod povezivanja instrumenata.

\begin{figure}[h]
\includegraphics[width=11.5cm]{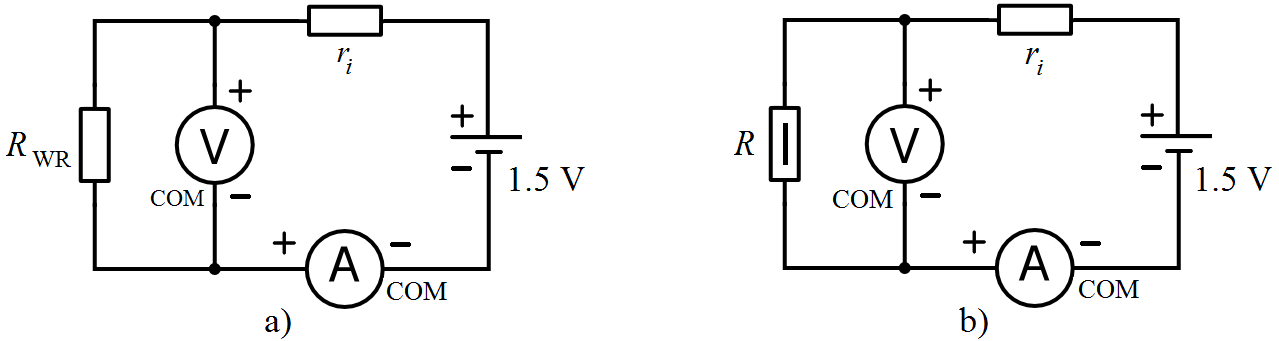}
\caption{Električna šema za ispitavanje dela voltamperske karakteristike (a) belog otpornika (b) ``crne kutije'', i
određivanje njihovih otpora kroz voltampersku karakteristiku.
Voltmetar (V) je povezan paralelno na istrživani element i meri napon $U$, 
a ampermetar (A) je povezan serijski i meri struju $I$. 
Kada se električno kolo zatvara različitim 
otporima $r_i\in(0,\;600\,\Omega)$ struja i napon su različiti. Tako se dobija nekoliko tačaka voltamperske karakteristike. Za male 
napone odnos $R=U/I$ je postojan i to je jedan mogući način provere Omovog zakona i znakova izmerene struje i 
napona.
Otpornik $R_\mathrm{WR}$ na šemi levo zamenjen je ``crnom kutijom'' na šemi desno i to je jedina razlika između ova dva 
električna kola.}
\label{SR_resistance-measurement}
\end{figure}

\begin{figure}[h]
\includegraphics[width=15cm]{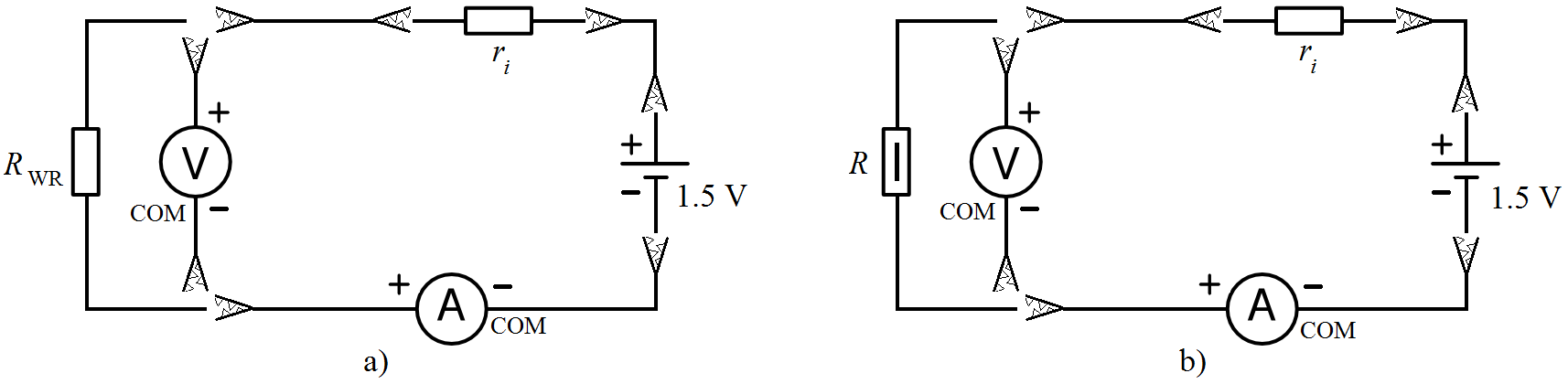}
\caption{Ekvivalentna šema povezivanja električnog kola iz Sl.~\ref{SR_resistance-measurement} pomoću kablova koji završavaju štipaljkama.}
\label{SR_R_neg_R_with_crocodiles}
\end{figure}

Serijski povežite na ampermetar bateriju od 1.5~V.
Zatvorite strujno kolo sa 0~$\Omega$ i pojedinačno svakim od pet otpornika $r_i.$ 
Rezultate svakog pojedinačnog merenja upišite u odgovarajuća polja tabele sa 5 kolona i 6 redova, kao kod Tabele~\ref{SR_template}:
1)~broj otpornika $i$, 
2)~vrednost otpornika $r_i$,
3)~jačinu struje $I_i$,
4)~napon voltmetra $U_i$,
5)~izračunatu vrednost otpora $U_i/I_i$.

\begin{table}[ht]
\caption{Obrazac, potrebna tabela za unos i obradu eksperimentalnih podataka dobijenih kroz eksperimentalne zadatke prikazane na Sl.~\ref{SR_resistance-measurement}. 
Kolone tabele predstavljaju: 
Kolona 1 broj otpornika $i$, 
Kolona 2 vrednost otpornika $r_0=0\;\Omega$ znači kratak spoj, 
Kolona 3 struju $I_i$ koja teče kroz kolo sa različitim $r_i$, serijski povezani na bateriju napona $\mathcal{E}$, 
Kolona 4 napon $U_i$, napon na krajevima belog otpornika ili ``crne kutije'' i 
Kolona 5 odnos između napona $U_i$ i struje $I_i$.}
\begin{tabular}{| r | r | r | r | r | r | r |}
\tableline
 i& $r_i \, [\Omega]$ & $I_i \,[\mu \mathrm{A}]$ & $U_i\,[\mathrm{V}]$ & $U_i/I_i\,[\Omega]$ \\
\tableline
0 & 0 & & & \\
1 & & & &\\
2 & & & &\\
3 & & & &\\
4 & & & &\\
5 & & & &\\
\tableline
\end{tabular}
\label{SR_template}
\end{table}

\item \textbf{Pomoću pet malih otpornika i baterije od $\mathcal{E}=$1.5~V izmerite zavisnost između struje i napona ``crne kutije''. (7~poena)}

Ovde ponovite ista merenja kao i u prethodnom zadatku s time da beli otpornik zamenite ``crnom kutijom''.

\item \textbf{Nacrtajte voltamperske karakteristike (VAK) ``crne kutije'' i belog otpornika na jednom grafikonu,
koristeći podatke iz tabela iz prethodna dva zadatka. (7~poena)}

Podatke iz tabela predstavite tako da na apcisi (\textit{x}-osa) nanesete vrednosti napona $U_i$ a na ordinati (\textit{y}-osa) struje $I_i$.
Preporučujemo da prvo uradite manji grafikon razmera $\mathrm{1\,V=1\,cm}$ i $\mathrm{1\,A=1\,cm},$ sa koordinatnim početkom $U=0$ i $I=0$.
Ukoliko vam preostane vremena možete nacrtati i grafikon sa odgovarajućim koordinatama.Faktički ste uradili deo voltamperske karakteristike (VAK) ``crne kutije'' i belog otpornika, zavisnosti $I(U)$ uključujući samo 6 tačaka, 
6 merenja svakog od ispitinanih elemenata.
Pomoći tačaka na grafikonu iscrtajte pravu liniju koja ih povezuje.
Označite prave na koji se ispitani element odnose.

\item \textbf{Odredite naklon (koeficijent pravca) pravih linija izražen kroz zavisnost $\Delta U / \Delta I$ iz grafikona prethodnog zadatka. (9~poena)}

Simbol $\Delta$ predstavlja razliku kako napona tako i struja $\Delta U=U_2-U_1$, $\Delta I=I_2-I_1$. 
Kojom karakteristikom je povezan naklon $\Delta U / \Delta I$?
Jeste li otkrili nešto neobično u odnosu na karakteristiku ``crne kutije''?

\subsection{Svet nije idealan.
Detaljno ispitivanje voltamperske karakterisitke (VAK).
Zadatak za starije učenike koji imaju veće praktično iskustvo sa voltamperskim karakterisitkama}

\label{SR_the_world_is_not_perfect}
\textit{Stariji učenici koji imaju veće praktično iskustvo sa voltamperskim karakterisitkama mogu raditi na uslovima u delu~\ref{SR_the_world_is_not_perfect} i da se vrate na početne uslove iz dela~\ref{SR_black_box_investigation}.}

Kod malih napona voltamperska karakteristika (VAK) ``crne kutije'' je deo prave linije, 
ali kako izgleda VAK kod većih vrednosti napona t.j. u širem mernom naponskom području otkrićete uskoro.

\begin{figure}[h]
\includegraphics[width=8cm]{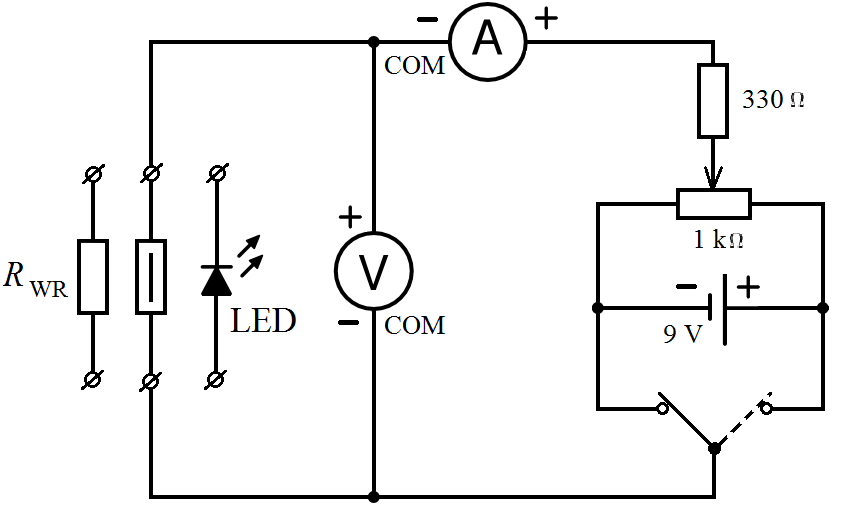}
\caption{Eksperimentalna šema za istraživanje voltamperske karakteristike (VAK) $I(U)$. 
Prvo se ispita beli otpornik $R_\mathrm{WR}$ (sasvim levo na slici), zatim se on menja ``crnom kutijom'' kao na šemi i na kraju uključite samo LED diodu (prikazanu desno). 
Voltmetrom se meri napon $U$ ispitivanog elementa, a ampermetrom jačina struje $I$. 
Napon se stvara iz baterije elektromotorne snage $\mathcal{E}=9\,\mathrm{V}$. 
Kada se rukom okreće osovina potenciometra sa otporom $1\,\mathrm{k}\Omega$ napon $U$ se menja od nule 0 do $+\mathcal{E}$. 
Priključivanjem štipaljke sa jednog na drugi kraj potenciometra,kako je prikazano na slici (vidi dole desno na slici) napon se menja od $-\mathcal{E}$ до 0.
Tako se napon može menjati od $-\mathcal{E}<U<+\mathcal{E}$. 
Otpor od $330\,\Omega$ ograničava struju kroz LED diodu i štiti je od pregorevanja.}
\label{SR_resistance-measurement-I-V-curve}
\end{figure}

\item \textbf{Izmerite detaljno zavisnost između struje i napona belog otpornika. (7~poena)}
 
Priključite bateriju od 9~V na odgovarajuće kontakte potenciometra.
Baterija je izvor napona na koji uključite voltmetar i 
merite vrednosti napona kod krajnjih rotacija potenciometra od $U_\mathrm{min}$ do 0 i od 0 do $U_\mathrm{max}$, kako je prikazano na sl.~\ref{SR_resistance-measurement-I-V-curve}.
Simbolično prikazani prekidač se realizuje uključivanjem štipaljke.
Na beli otpornik serijski uključite ampermetar i povežete ga na izvor napona.
Ampermetar meri strju $I_\mathrm{WR}$ kroz beli otpornik,a voltmetar napon $U$.
Kada okrećete osovinu potenciometra napon se menja i tako istražujete voltampersku karakteristiku (VAK).
Najpre okrenite potenciometar od jedne do druge krajnje vrednosti i saznajte u kojim intervalima otčitavanja trebaju raditi instrumenti.
Zatim ispitajte zavisnost $I_\mathrm{WR}(U)$ kao parovi struja-napon, $(I_\mathrm{WR},U)$ i upišite ih u tabelu.
Za beli otpornik važi Omov zakon.
Za predstavljanje linearne zavisnosti $I_\mathrm{WR}(U)$ sasvim je dovoljno pet tačaka odnosno pet merenja,od kojih da dva budu kod minimalnog i maksimalnog napona $U_\mathrm{min}$ i $U_\mathrm{max}$.
U skladu sa Omovim zakonom znaci napona i struje trebaju biti isti.ako se pojavi razlika u znacima,potražite grešku u šemi.

\item \textbf{Izmerite detaljno zavisnost između struje i napona ``crne kutije''. (5~poena).}

U eksperimentalnom postrojenju iz prethodnog zadatka zamenite beli otpornik ``crnom kutijiom'' bez bilo kakvih drugih promena.
Okretanjem osovine potenciometra menjajte napon $U$ od $U_\mathrm{min}$ do $U_\mathrm{max}$ promenama od 1~V i zabeležite u tabeli parove brojeva $(I,U)$. 
U tabelu upišite i napone kod kojih struja ima minimalnu i maksimalnu vrednost.

\item \textbf{Izmerite detaljno zavisnost između struje i napona LED diode. (3~poena)}

U eksperimentalnom postrojenju iz prethodnog zadatka zamenite ``crnu kutijiu'' LED diodom bez bilo kakvih drugih promena.
Vidite u kom krajnjem položaju potenciometra dioda svetli najjačim intenzitetom. Zatim okrećite osovinu potenciometra i pratite vrednosti na ampermetru.Kod jačina struje manjim od 6~mA upišite u tabelu parove brojeva $(I_\mathrm{LED}, U)$ promenama od 1~mA. 
Od struja manjih od 2~mA=2000~$\mu$A snimite parove brojeva (ILED,U) promenamo od oko 200~$\mu$A dok ne dođette do struje manje od 200~$\mu$А.

\textit{Nakon završetka snimanja treće voltamperske karakteristike isključite bateriju od 9~V jer se brzo troši.}

\item \textbf{Predstavite na zajedničkom grafikonu voltamperske karakteristike (VAK): belog otpornika $I_\mathrm{WR}(U)$, ``crne kutije'' $I(U)$ i LED diode $I_\mathrm{LED}(U)$. (10~poena).}

Najpre analizirajte najmanje i najveće vrednosti struje i napona.Ovi parametri će vam odrediti pravougaonik u kom će biti predstavljene VAK triju elemenata.
Mi vam preporučujemo sledeće razmere:horizontalna osa $\mathrm{1\,V=1\,cm}$ i vertikalna osa $\mathrm{1\,mA=1\,cm}$.
Za sva tri elementa (beli otpornik, ``crna kutija'' i LED dioda) VAK $I(U)$ predstavljaju kontinuirane krive. 
Povucite kroz tačke prave (ili krive) koje maksimalno dobro opisuju eksperimentalne rezultate.

\item \textbf{Zašto su važne voltamperske karakteristike (VAK). (5~poena)}

Nacrtajte sliku u ogledalu voltamperske karakteristike (VAK) LED diode, zamenom $I$ sa $–I$, 
odnosno okrenemo VAK oko horizontalne ose napona. 
Na VAK-u LED diode malim kružićem označite vrednosti $(-I^*,U^*)$ iz tabele I. 
Obratite pažnju dali je ova tačka blizu presečne tačke VAK-a ``crne kutije'' LED diode. 
Dali je to slučajno ili nije?

\item \textbf{Analiza VAK-a: belog otpornika, ``crne kutije'' i LED diode. (7~poena)}

Ako voltamperske karakteristike (VAK) $I(U)$ predstavljaju pojedine prave linije, iz njihovih naklona odredite 
odgovarajuće otpore$R=\Delta U/\Delta I.$ 
Simbol $\Delta$ predtsavlja razliku (difference): Izaberite dve tačke duži, dva para vrednosti 
napona i struje iz grafikona i napravite razliku napona $\Delta U= U_2-U_1$ i struje $\Delta I=I_2-I_1.$ 
Kada je posmatrani interval mali, kriva je približno prava, pa se umesto $\Delta$ piše $\mathrm{d}.$
Otpor izračunat iz naklona VAK-a se naziva diferencijalni otpor, a recipročna vrednost diferencijalna sprovodljivost $\sigma_\mathrm{diff}=\mathrm{d}I/\mathrm{d}U$.

Izračunajte: a) otpor belog otpornika, b) otpor centralnog dela krive ``crne kutije'' c) otpor levog dela krive ``crne kutije'' d) otpor desnog dela krive ``crne kutije'', i e) otpor LED diode kada kroz nju teku struje u intervalu od 2 do 5~mA.

\item \textbf{Koja karakteristika voltamperske karakteristike (VAK) ``crne kutije'' je od suštinskog značaja za stalno svetljenje ili treperenje diode iz oba kvalitativna zadatka? (4~poena).}

\textit{Za ``crnu kutiju'' je najvažniji centralnog deo VAK-a. 
Ovom delu VAK-a dioda duguje stalno svetljenje i generisanje promenljivih struja u oscilatornom krugu, rezonansnoj konturi, koje posmatramo u kvalitativnim zadacima pokazanim na Sl.~\ref{SR_LEDNR} i Sl.~\ref{SR_LCNR}, kao i drugim brojnim mogućim tehničkim uređajima tipa sakrivenog u ``crnoj kutiji''.}

Razlika voltamperskih karakteristika ``crne kutije'' i belog otpornika nam otkriva razlog svetljenja LED diode.
Koja je ta razlika?

\item \textbf{Opišite kvalitativno kako se stvara struja koja izaziva stalno svetljenje ili treperenje LED diode iz oba kvalitativna zadatka (10~poena).}

\subsection{Merenje frekvencije električnih oscilacija stvorenim od ``crne kutije''}
\label{SR_dynamics}
Povežite ponovi šemu iz drugog kvalitativnog zadatka prikazanog na Sl.~\ref{SR_LCNR},
kada su svi elementi povezani paralelno: ``crna kutija'', $LC$ rezonator, piezo-pločica (minijaturni zvučnik, zumer) LED dioda.

Ako se, umjesto zujalicu (piezo pločice) koristi slusalice za mobilni telefon, mora se seriski povezati, otpor veći od 10 k $ ~ \ Omega $

\item \textbf{Mehaničko određivanje frekvencije oscilacija. (7~poena)}

Električna merenja su lakša i preciznija. Da bi izmerili frekvenciju oscilacija bez frekvencmetra kod datih uslova 
potrebno je manifestovati malo umeća. U priboru imate lepljivu traku. Pričvrstite LED diodu na kraj elestičnog lenjira 
samolepljivom trakom-selotejpom. Kada jednom rukom pritisnete lenjir postavljen na kraju stola a drugom rukom 
izazivate oscilacije na način da oscilovanjem lenjira osciluje i LED dioda koja svetli zalepljena na njemu, primetićete 
nekoliko skoro nepokretnih svetlih tačaka. Ako je električna frekvencija $f_\mathrm{res}$ množitelj mehaničke $f_\mathrm{mech}$ svetle tačke su 
nepokretne. Prebrojte svetle mrlje ili samo dajte ocenu njihovog broja $N$, ili ako ih možete brzo izbrojiti, odredite 
frekvenciju mehaničkih oscilacija $f_\mathrm{mech}$. Potrudite se da lenjir zaoscilujete postojanom frekvencijom i amplitudom. Menjajte 
dužinu slobodnog dela lenjira da bi postigli nepokretne svetle tačke. Izbrojte koliko oscilacija pravi lenjir za deset 
sekundi i tako utvrdite mehaničku oscilaciju $f_\mathrm{mech}$. Frekvenciju električnih oscilacija $f_\mathrm{res}$ izrazite kroz broj svetlih mrlja i 
frekvenciju mehaničkih oscilacija.Koliko herci rezonantne frekvencije električnih oscilacija dobijate iz vaše formule $f_\mathrm{res}(N,f_\mathrm{mech})$?

\item \textbf{Izračunajte teoretski frekvenciju oscilacija. (2~poena)}

Alternativni metod za određivanje ove frekvencije je njeno teoretsko računanje po formuli $f_\mathrm{res}=1/(2\pi\sqrt{LC}).$
na kondenzatoru piše 4.7~$\mu F$, a na toroidalnom kalemu piše da imamo dva namotaja po 100~mH, i pri njihovom 
serijskom povezivanju induktivnost im je $L$=400~mH, koliko iznosi frekvencija. Koliko se izračunata frekvencija 
poklapa sa eksperimentalno utvrđenom frekvencijom.

\item \textbf{Određivanje frekvencije slušanjem oscilacija na zumeru. (1~poen)}

Zumer je signalni element (nemački ``summer''= zujanje). Ako ste muzikalni i sećate se frekvencija muzičkih tonova
možete odrediti kom muzičkom tonu odgovara frekvencija koju emituje zumer. Kod šeme broj 3 imate uključen zumer 
koji emituje neku frekvenciju. Kolika je frekvencija koju emituje?
Ne očekuje se visoka tačnost i 50\% greške se smatra zadovoljavajućom kod ova tri metoda određivanja frekvencije bez 
frekvencmetra.

\end{enumerate}%

\section{Teoretski zadatak, 35~poena}
\label{SR_TheoreticalProblem}

\subsection{Uslovi teoretskog zadatka}
\begin{figure}[h]
\includegraphics[width=9cm]{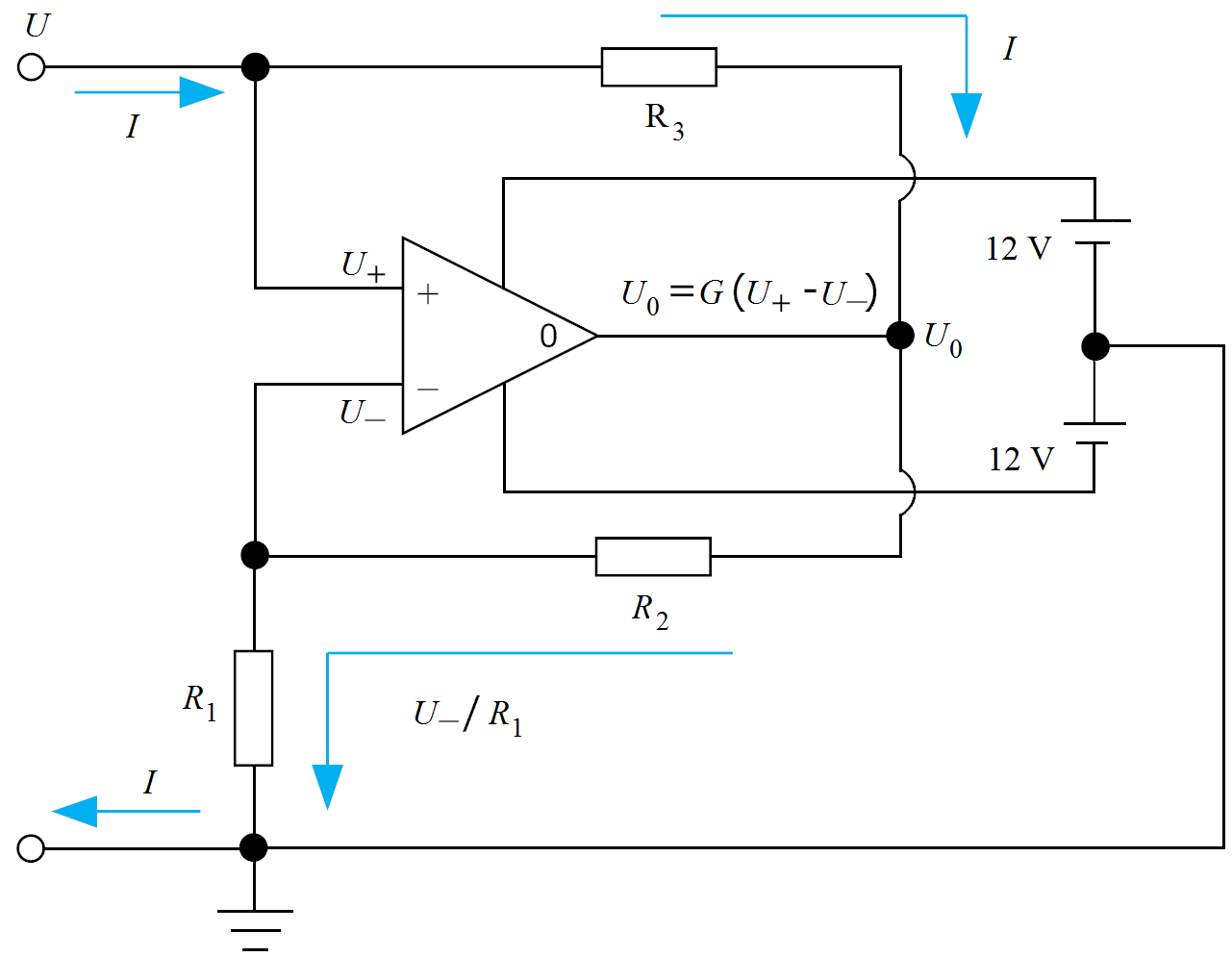}
\caption{Izračunajte uz tačnost od 1\% efektivni otpor $R=U/I$ kola u zavisnosti od tri otpornika $R_1,$ $R_2$ i $R_3$ na 
trouglom označenim naponskim pojačivačem sa koeficijentom pojačavanja $G=10^5,$ koji se napaja sa dve baterije sa 
naponom $V_S$.
Naponi $U_0$, $U_{-}$, kao i struja I su nepoznati. Sa $U_0$ je označen napon izveden iz $U_\mathrm{output}=(U_+-U_-)G.$}
\label{SR_circuit}
\end{figure}

U električnom kolu prikazanom na Sl.~\ref{SR_circuit} ulazne struje u tačkama (+) i (-) su jednake nuli, a izlazna struja u tački (0) je 
takva da odgovarajući naponi su povezani relacijom $U_0=(U_+ - U_-)G$, gde je G koeficijent pojačavanja i ima jako 
veliku vrednost, $G=10^5 \gg 1$. 
Elektronski element označen trouglom (pojačivač) se napaja iz dve baterije i u 
uslovu zadatka se pretpostavlja da su pomenuti naponi manji od napona napajanja baterija $\mathcal{E}_\mathrm{B}=12\;\mathrm{V}.$
Tačka između dve baterije je povezana provodnikom sa krajem otpornika $R_1$ i jednom od ulaznih elektroda na 
šemi.
Pogodno je da se napon u toj zajedničkoj tački izabere za nulu, $U_\mathrm{CP}=0$, ili kako elektrotehničari kažu to 
je „uzemljenje“ (\ground). Indeks $\mathrm{CP}$ dolazi iz engleskog Common Point i na multimetrima se označava kao COM. 
U suprotnom slučaju,ako je $U_\mathrm{CP} \neq 0$, za jednačinu pojačavanja napona imamo $U_0=(U_+ - U_-)G+U_\mathrm{CP}$.
Struja koja ističe u ``uzemljenje'' je nula.Izračunajte uz tačnost od 1\% (tri značajna broja) odos između ulaznog 
napona U i struje I koja teče kroz kolo i izrazite taj efektivni otpor $R=U/I=R(R_1,R_2, R_3)$ u zavisnosti od tri 
otpora na šemi.
Da bi uprostili,možete pretpostaviti da koeficijent pojačavanja stremi ka beskonačnosti
$G \rightarrow \infty$.
U dobijenom izrazu zamenite $R_1=R_2=10\;\mathrm{k} \Omega,\; R_3=1.5 \;\mathrm{k} \Omega$. 
Ukratko: 
1) Traži se krajnja formula za otpor šeme i
2) izračunati po njoj numeričku vrednost otpora uz tačnost od 1\%. Kakav je znak otpora $R=U/I$ i koliko oma ima njegov modul? 
Kako taj otpor pali LED diodu?

\subsection{Zagonetka, 2~poeni} 
Crveno i čvrsto, a leti. Šta je to?

\section{Zadatak za domaći rad, $\mathbf{137\,\$}$}
\label{SR_Homework}
\textit{Po završetku Olimpijade, nađite šrafciger i odvrnite šrafove sa poklopca ``crne kutije''. 
Izvadite baterije ili promenite položaj jednog od prekidača sa ``On'' na ``Off''.}

Za male napone voltamperska karakteristika (VAK) ``crne kutije'' je prava linija sa postojanim
odnosom $U/I$ u skladu sa Omovim zakonom.
Pokušajte da izmerite otpor ``crne kutije'' ommetrom. 
Pokušajte da uporedite vrednosti na ommetru sa otporom dobijenim kroz ispitivanje VAK-a.
Objasnite zašto se otpor ``crne kutije'' definitivno ne može izmeriti ommetrom.
Kakve promene iz šeme na Sl.~\ref{SR_circuit} treba realizovati u ``crnoj kutiji'' da bi bilo moguće merenje otpora ommetrom, na 
primer multimetrom DT-830B od 20~k$\Omega$ kakvim raspolažete na Olimpijadi?
Prvi koji odgovori na bar jedno od dva pitanja i pošalje odgovor sa adrese kojom se registrovao za Olimpijadu na 
\texttt{epo@bgphysics.eu}s do 07:00 1.novembra 2015 osvojiće novčanu nagradu od 137~\$. 
Možete raditi u timu, koristiti 
literaturu, \textit{Google}, i da se na internetu konsultujete sa radioinženjerima i univerzitetskim profesorima elektronike svuda 
po svetu. Kada je ponoć u Kumanovu, u Kaliforniji je kasno popodne, u Japanu počinje dan -- svuda na svetu postoje 
kolege koje rade.

\end{document}